\RequirePackage{fix-cm} % Fix LaTeX2e bugs.
\documentclass[a4paper, twoside, 12pt, dvips, reqno]{amsart}

\usepackage{fixltx2e}     % Fix LaTeX2e bugs.

\usepackage{amssymb}
\usepackage{amsmath}

\usepackage{a4wide}

\usepackage{indentfirst}
\usepackage{graphicx}
\usepackage{psfrag}

\usepackage[usenames,dvipsnames]{pstricks}
\usepackage{epsfig}
\usepackage{pst-grad} % For gradients
\usepackage{pst-plot} % For axes

\usepackage{subfigure}

\usepackage{ctable, caption}

\usepackage[english]{babel}
\usepackage[latin1]{inputenc}

\usepackage{color}
\definecolor{oneblue}{rgb}{0,0.0,0.75}
\usepackage[colorlinks,
            urlcolor=oneblue,
            linkcolor=oneblue,
            citecolor=oneblue,
            bookmarksopen=false,
            pagebackref]{hyperref}

\numberwithin{equation}{section}

\newtheorem{rem}{Remark}

\newcommand{\R}{\mathbb{R}}
\newcommand{\Z}{\mathbb{Z}}
\newcommand{\Ah}{\mathcal{A}}
\newcommand{\Th}{\mathcal{T}}
\newcommand{\Fh}{\mathcal{F}}
\newcommand{\Gh}{\mathcal{G}}
\newcommand{\Hh}{\mathcal{H}}
\newcommand{\Sh}{\mathcal{S}}
\newcommand{\ip}{{i+\frac12}}
\newcommand{\im}{{i-\frac12}}
\newcommand{\xim}{x_{i-\frac12}}
\newcommand{\xip}{x_{i+\frac12}}

\newcommand{\Fip}{\Fh_{i+\frac12}}
\newcommand{\Fim}{\Fh_{i-\frac12}}
\newcommand{\Gip}{\Gh_{i+\frac12}}
\newcommand{\Gim}{\Gh_{i-\frac12}}
\newcommand{\dx}{\Delta x}
\newcommand{\dt}{\Delta t}

\newcommand{\diag}{\mathop{\mathrm{diag}}}
\makeatletter
\newcommand{\sech}{\mathop{\operator@font sech}}
\newcommand{\sign}{\mathop{\operator@font sign}}
\makeatother

\usepackage{acronym}
\acrodef{FV}{Finite Volumes}
\acrodef{FD}{Finite Differences}
\acrodef{DG}{Discontinuous Galerkin}
\acrodef{PDEs}{Partial Differential Equations}
\acrodef{SWs}{Solitary Waves}
\acrodef{NSWE}{Nonlinear Shallow Water Equations}

\begin{document}

\title[Finite volume schemes for dispersive waves]{Finite volume schemes for dispersive wave propagation and runup}

\author[D. Dutykh]{Denys Dutykh$^*$}
\address{LAMA, UMR 5127 CNRS, Universit\'e de Savoie, Campus Scientifique,
73376 Le Bourget-du-Lac Cedex, France}
\email{Denys.Dutykh@univ-savoie.fr}
\urladdr{http://www.lama.univ-savoie.fr/~dutykh/}
\thanks{$^*$ Corresponding author}

\author[Th. Katsaounis]{Theodoros Katsaounis}
\address{Department of Applied Mathematics, University of Crete, Heraklion, 71409 Greece \\ Inst. of App. and Comp. Math.(IACM), FORTH, Heraklion, 71110, Greece}
\email{thodoros@tem.uoc.gr}
\urladdr{http://www.tem.uoc.gr/~thodoros/}

\author[D. Mitsotakis]{Dimitrios Mitsotakis}
\address{UMR de Math\'ematiques, Universit\'e de Paris-Sud, B\^atiment 425, P.O. Box, 91405 Orsay, France}
\email{Dimitrios.Mitsotakis@math.u-psud.fr}
\urladdr{http://sites.google.com/site/dmitsot/}

\begin{abstract}
Finite volume schemes are commonly used to construct approximate solutions to conservation laws. In this study we extend the framework of the finite volume methods to dispersive water wave models, in particular to Boussinesq type systems. We focus mainly on the application of the method to bidirectional nonlinear, dispersive wave propagation in one space dimension. Special emphasis is given to important nonlinear phenomena such as solitary waves interactions, dispersive shock wave formation and the runup of breaking and non-breaking long waves.
\end{abstract}


\maketitle

%%%%%%%%%%%%%%%%%%%%%%%%%%%%%%%%%%%%%%%%%%%%%%%
%%%%% SECTION
%%%%%%%%%%%%%%%%%%%%%%%%%%%%%%%%%%%%%%%%%%%%%%%

\section{Introduction}

The simulation of water waves in realistic and complex environments is a very challenging problem. Most of the applications arise from the areas of coastal and naval engineering, but also from natural hazards assessment. These applications may require the computation of the wave generation \cite{Dutykh2006, Kervella2007}, propagation \cite{Titov1997}, interaction with solid bodies, the computation of long wave runup \cite{TS94, TS} and even the extraction of the wave energy \cite{Simon1981}. Issues like wave breaking,  robustness of the numerical algorithm in wet-dry processes along with the validity of the mathematical models in the near-shore zone are some  basic problems in this direction \cite{Hibberd1979}. During past years, the classical shallow water equations have been employed to solve some of  these problems \cite{Anastasiou1999, Dutykh2009a}:
\begin{equation}\label{eq:E1.1}
	\begin{array}{l}
		H_t+(Hu)_x=0,\\
		(Hu)_t + \bigl(Hu^2 + \frac{g}{2} H^2 \bigr)_x = g H D_x,
	\end{array}
\end{equation}
where $H(x,t) := \eta(x,t) + D(x)$ is the total water depth, $D(x)$ describes the depth below the mean sea level while $\eta(x,t)$ is the free surface elevation, $u(x,t)$ denotes the depth-averaged fluid velocity and $g$ is the gravity acceleration constant. Mathematically, equations \eqref{eq:E1.1} represent a system of conservation laws describing the propagation of infinitely long waves with a hydrostatic pressure assumption. The wave breaking phenomenon is commonly assimilated to the formation of shock waves (or hydraulic jumps) which is a common feature of hyperbolic p.d.e's.  Consequently, the finite volume(FV) method has become the method of choice for these problems due to its excellent intrinsic conservative and shock-capturing properties \cite{Anastasiou1999, DeKa, DeKaKa, Dutykh2009a}. Furthermore the shallow water equations have been proven in practice to predict accurately the maximum runup of long waves \cite{Heitner1970, Synolakis1987, TS94, TS, Kanoglu1998, Cea2004}.

On the other hand, various studies have shown that the inclusion of dispersive effects is beneficial for the description of long wave propagation and runup processes \cite{Zelt1991, Wei1995, Li2002a}. Moreover, J.A.~Zelt \cite{Zelt1991} reported a divergence in the prediction of the rundown and in the prediction  of the reflected wave-train after the wave climbing on the shore when a dispersionless model is employed. According to J.A.~Zelt,  the results of the nonlinear dispersive model considered in \cite{Zelt1991} showed better performance compared to \eqref{eq:E1.1}. During the last fifty years numerous dispersive models have been proposed for the simulation of long waves \cite{Serre1953, Peregrine1967, BS, Nwogu1993, Kennedy2000, Madsen03, Mitsotakis2007}. 

In this work we will study numerically  bidirectional water wave models. Specifically, we consider the following family of Boussinesq type systems of water wave theory, introduced in \cite{BCS},  written in nondimensional, unscaled variables
\begin{equation}\label{E1.4i}
\begin{aligned}
& \eta_t+u_x+(\eta u)_x+a\, u_{xxx}-b\,\eta_{xxt}=0,\\
& u_t+\eta_x+uu_x+c\,\eta_{xxx}-d\,u_{xxt}=0,
\end{aligned}
\end{equation}
where $a,\, b, \, c, \,d\in\R$, $\eta=\eta(x,t)$, $u=u(x,t)$ are real functions defined for $x\in \mathbb{R}$ and $t\geq 0$.

For more realistic situations we introduce a new Boussinesq type system with variable bottom topography based on Peregrine's system, \cite{Peregrine1967}. The new system incorporates a very important property --- the invariance under vertical translations, thus more appropriate for practical applications such as wave runup on non-uniform  shores. In dimensional variables the model reads
\begin{equation}\label{E.Per5i}
\begin{array}{l}
H_t+Q_x=0,\\
Q_t+(\frac{Q^2}{H}+\frac{g}{2}H^2)_x -\left[\frac{H^2}{3}Q_{xxt}-(\frac{1}{3}H_x^2-\frac{1}{6}HH_{xx})Q_t+\frac{1}{3}HH_xQ_{xt}\right]=gHD_x , 
\end{array}
\end{equation}
where $H(x,t)=\eta(x,t)+D(x)$, $Q(x,t)=H(x,t) u(x,t)$. 

There is a wide range of numerical methods in the literature for computing approximate solutions to these models. Finite difference (FD) schemes \cite{Fun, Jamois2006, Hamidou2009}, finite element methods \cite{Bona2007, Mitsotakis2007, DMII, Avilez-Valente2009} and spectral methods \cite{PD, Nguyen2008a} have been proposed.  More contemporary discontinuous Galerkin (DG)  schemes have  also been adapted with some success to dispersive wave equations \cite{YS, Levy2004, Eskilsson2006, Eskilsson2006a} while  the application of \acf{FV} or hybrid FV/FD methods remain most infrequent for this type of problems. To our knowledge, only a few very recent works are in this direction \cite{Bellotti2001, EIK, Benkhaldoun2008, TP, SMi}.

Finite volume method is well known for its accuracy, efficiency and robustness for approximating solutions to conservation laws and in particular to nonlinear shallow water equations \eqref{eq:E1.1}. The aforementioned bidirectional models \eqref{E1.4i} and \eqref{E.Per5i} are rewritten in a conservative  form and discretization by the finite volume method follows. Three different numerical fluxes are employed
\begin{itemize}
\item a simple \emph{average flux} (m-scheme),
\item a \emph{central flux}, (KT-scheme) \cite{NT, KT}, as a representative of central schemes,
\item  a \emph{characteristic flux} (CF-scheme), as a representative of the linearized Riemann solvers, \cite{Ghidaglia1996}, 
\end{itemize}
along with \emph{TVD}, \emph{UNO} and \emph{WENO} reconstruction techniques, \cite{Sweby1984, HaOs, LOC}. Time discretization is based on Runge-Kutta (RK) methods which preserve the total variation diminishing(TVD) property of the finite volume scheme, \cite{Shu1988a, Gottlieb2001, Spiteri2002}. We use explicit RK methods since we work with BBM type systems \eqref{E1.4i} and not with KdV-type systems  which is well known to be notoriously stiff, \cite{Mitsotakis2007}. These methods have been studied thoroughly in the case of nonlinear conservation laws. The average flux although is known to be unstable for conservation laws is proved to be very accurate for nonlinear dispersive waves. On the other hand finite volume methods based on the central flux as well as on characteristic flux work equally well for the numerical simulation  of waves even in realistic environments. 

The performance of the finite volume method applied to models \eqref{E1.4i} and to the new system \eqref{E.Per5i} is studied in a systematic way through a series of numerical experiments.  Our main focus is the evaluation, in terms of accuracy, efficiency and robustness of second order finite volume methods compared to high order schemes. In particular, in this study we take up on the following points 
\begin{itemize}
\item  accuracy of the finite volume method in the propagation of solitary waves with very satisfactory results. 
\item  conservation of various invariant quantities during the formation of dispersive shocks is studied numerically. The finite element as well as spectral methods  break down for these experiments. The finite volume method provides very accurate results.
\item interactions of solitary waves are computed with high accuracy. It is shown numerically that Boussinesq type systems describe better overtaking collisions of solitary waves than unidirectional models like KdV-BBM. We compare our results, whenever possible, with experimental measurements with very good agreement.
\item finite volume method allows to use appropriate techniques to treat the transition from wet to dry regions and vice versa. These techniques are applied successfully to systems with dispersive terms modeling runup of long waves. On the other hand, when the model fails  due to wave breaking, the method allows to use locally the nonlinear shallow water system, thus enabling us to resolve a wide spectrum of hydrodynamic phenomena using a single computational framework.
\item it is shown numerically the advantage of using dispersive models over standard nonlinear shallow water equations in computing the wave runup and, in particular, in capturing the reflected wave. It's also illustrated by an example the importance  of the system being invariant under vertical translations. 
\end{itemize}

The paper is organized as follows. In Section \ref{sec:models} Boussinesq type systems are presented along with some of their basic properties. A new system with uneven bottom and invariant under vertical translations is derived.  In Section \ref{sec:num} the finite volume method is presented for a general framework incorporating all models.

Section \ref{sec:interact} presents a series of numerical experiments  for the Boussinesq systems \eqref{E1.4i}. In this mathematical setting we validate the finite volume method and measure its accuracy. We study the propagation as well as the interaction of solitary waves: we consider in particular head-on and overtaking collisions, but also we present results concerning the small dispersion effect. Finally, in Section \ref{sec:bottom} the new system with variable bottom, \eqref{E.Per5i} is studied. Numerical simulations of non-breaking and breaking long wave runup are presented and compared with experimental data. 

%%%%%%%%%%%%%%%%%%%%%%%%%%%%%%%%%%%%%%%%%%%%%%%
%%%%% SECTION
%%%%%%%%%%%%%%%%%%%%%%%%%%%%%%%%%%%%%%%%%%%%%%%

\section{Mathematical models}\label{sec:models}

We present briefly the mathematical models being considered and some of their main properties.

\subsection{Dispersive models with flat bottom}
We consider the following family of Boussinesq type systems of water wave theory, introduced in \cite{BCS}, which may be written in nondimensional, unscaled variables
\begin{equation}\label{E1.4}
\begin{array}{l}
\eta_t+u_x+(\eta u)_x+a u_{xxx}-b\eta_{xxt}=0,\\
u_t+\eta_x+uu_x+c\eta_{xxx}-du_{xxt}=0,
\end{array}
\end{equation}
where $\eta=\eta(x,t)$, $u=u(x,t)$ are real functions defined for $x\in \R$ and $t\geq 0$. Coefficients $a$, $b$, $c$ and $d$ are defined as
%\begin{equation}\label{E1.5}
%\begin{array}{l}
%a=\frac{1}{2}(\theta^2-\frac{1}{3})\nu,\,\,\,\,\,\,\,
%b=\frac{1}{2}(\theta^2-\frac{1}{3})(1-\nu),\\
%\\
%c=\frac{1}{2}(1-\theta^2)\mu,\,\,\,\,\,\,\,
%d=\frac{1}{2}(1-\theta^2)(1-\mu),
%\end{array}
%\end{equation}
\begin{equation}\label{E1.5}
a=\frac{1}{2}(\theta^2-\frac{1}{3})\nu,\ 
b=\frac{1}{2}(\theta^2-\frac{1}{3})(1-\nu),\ 
c=\frac{1}{2}(1-\theta^2)\mu,\ 
d=\frac{1}{2}(1-\theta^2)(1-\mu),
\end{equation}
where $0\le\theta\le 1$ and $\mu, \nu\in\R$. The variables in (\ref{E1.4}) are non-dimensional and unscaled: $x$ and $t$ are proportional to position along the channel and time, respectively, while $\eta(x,t)$ and $u(x,t)$ are proportional to the deviation of the free surface above an undisturbed level and to the horizontal velocity of the fluid at a height $y=-1+\theta(1+\eta(x,t))$, respectively. In terms of these variables the channel bottom is located at $y=-1, (\theta=0)$, while the free surface corresponds to $\theta=1$.  Boussinesq systems (\ref{E1.4}) with $b=d$ conserve the energy functional:
\begin{equation}\label{EInv}
I_1(t)=\int_{\R} (\eta^2(x,t)+(1+\eta(x,t))u^2(x,t)-c\,\eta_x^2(x,t)-a\,u_x^2(x,t))\; dx,
\end{equation}
i.e. $I_1(t)=I_1(0)$ for $t\ge 0$. System (\ref{E1.4}) is derived under the following assumptions:
$$
\varepsilon:=A/h_0\ll 1 \ , \quad \sigma:=h_0/\lambda\ll 1\, \quad S:=\frac{A\lambda^2}{h_0^3} = O(1)\, ,
$$ 
where $S$ is the Stokes (or Ursell) number, $A$ is a typical wave amplitude of fluid of depth $h_0$ and $\lambda$ is a characteristic wavelength. If one takes $S=1$ and switches to scaled, dimensionless variables, one may derive from  Euler equations a scaled version of (\ref{E1.4}) by appropriate asymptotic expansion in powers of $\varepsilon$, cf. \cite{BCS}:
\begin{equation}\label{E1.4s}
\begin{array}{l}
\eta_t+u_x+\varepsilon(\eta u)_x+\varepsilon[a u_{xxx}-b\eta_{xxt}]=O(\varepsilon^2),\\
u_t+\eta_x+\varepsilon uu_x+\varepsilon[c\eta_{xxx}-du_{xxt}]=O(\varepsilon^2),
\end{array}
\end{equation}
from which we obtain (\ref{E1.4}) by unscaling and neglecting higher order terms $O(\varepsilon^2)$.

We list several examples of particular Boussinesq systems of the form (\ref{E1.4}) that we will refer to in the sequel. The initial-value problem for all these systems has been shown to be at least nonlinearly well-posed locally in time, cf. \cite{Bona2004}.
\begin{enumerate}
\item[(i)] The 'classical' Boussinesq system ($\mu=0$, $\theta^2=1/3$, i.e. $a=b=c=0$, $d=1/3$ in \eqref{E1.4}), whose initial-value problem is globally well-posed, \cite{Am, Scho},
\begin{equation}\label{E1.6}
\begin{array}{l}
\eta_t+u_x+(\eta u)_x=0,\\
u_t+\eta_x+uu_x-\frac{1}{3}u_{xxt}=0.
\end{array}
\end{equation}
\item[(ii)] The BBM-BBM system ($\nu=\mu=0$, $\theta^2=2/3$, i.e. $a=c=0$, $b=d=1/6$ in \eqref{E1.4}), whose initial-value problem is locally well-posed, \cite{BC}, 
\begin{equation}\label{E1.7}
\begin{array}{l}
\eta_t+u_x+(\eta u)_x-\frac{1}{6}\eta_{xxt}=0,\\
u_t+\eta_x+uu_x-\frac{1}{6}u_{xxt}=0.
\end{array}
\end{equation}
\item[(iii)] The Bona-Smith system ($\nu=0$, $\mu=(4-6\theta^2)/3(1-\theta^2)$, i.e. $a=0$,
$b=d=(3\theta^2-1)/6$, $c=(2-3\theta^2)/3$, $2/3<\theta^2<1$ in \eqref{E1.4}), whose initial-value problem is globally well-posed, cf. \cite{BS}. The limiting form of this system as $\theta\rightarrow 1$, corresponding to $a=0$, $b=d=1/3$, $c=-1/3$, is the system actually studied by Bona and Smith, \cite{BS}. These systems are given by
\begin{equation}\label{E1.8}
\begin{array}{l}
	\eta_t+u_x+(\eta u)_x-\frac{3\theta^2-1}{6}\eta_{xxt}=0,\\
	u_t+\eta_x+uu_x+\frac{2-3\theta^2}{3}\eta_{xxx}-\frac{3\theta^2-1}{6}u_{xxt}=0.
\end{array}
\end{equation}
\end{enumerate}
The existence of solitary wave solutions to the above systems, in some cases the uniqueness also, has been proved in \cite{C1, Chen1998, Dougalis2004} and in the case of the Bona-Smith type systems (\ref{E1.8}),  for each $\theta^2\in (7/9,1)$,  there exists one solitary wave in closed form, \cite{Chen1998}
\begin{equation}\label{E1.9}
\begin{array}{l}
\eta(\xi)=\eta_0\, {\sech}^2(\lambda\xi),\\
u(\xi)=B\,\eta(\xi),
\end{array}
\end{equation}
with
\begin{equation}\label{E1.10}
\textstyle{
\begin{array}{cc}
  \eta_0=\frac{9}{2}\cdot\frac{\theta^2-7/9}{1-\theta^2},&
  c_s=\frac{4(\theta^2-2/3)}{\sqrt{2(1-\theta^2)(\theta^2-1/3)}},\\
  \lambda=\frac{1}{2}\sqrt{\frac{3(\theta^2-7/9)}{(\theta^2-1/3)(\theta^2
  -2/3)}},& B=\sqrt{\frac{2(1-\theta^2)}{\theta^2-1/3}}.
  \end{array}}
\end{equation}

\subsection{Dispersive models with variable bottom}
For more realistic applications one should consider Boussinesq systems with variable bottom. In his  pioneering work Peregrine, \cite{Peregrine1967}, derived the following Boussinesq type system
\begin{equation}\label{E.Per0}
\begin{array}{l}
\eta_t+[(D+\eta)u]_x=0,\\
u_t + g\eta_x + uu_x - \frac{D}{2}(Du)_{xxt} + \frac{D^2}{6}u_{xxt}=0,
\end{array}
\end{equation}
where $\eta(x,t)$ and $u(x,t)$ are defined as before and $D(x)$ describes the water depth below its rest position.  Many other systems have been derived also, including systems with improved dispersion characteristics \cite{Nwogu1993}, high-order Boussinesq systems \cite{Madsen03} and other generalizations of (\ref{E1.4}), cf. \cite{Mitsotakis2007}. Most of these systems break Galilean invariance and the invariance under vertical translations. This is a restrictive property especially in the studies of realistic problems like the water wave runup on non-uniform shores. We note also that the complete water wave problem possesses these symmetries, \cite{Benjamin1982}.

To overcome this deficiency we develop a new system, analogous to the original Peregrine's system, \cite{Peregrine1967},  which is invariant under vertical translations. To derive the system we begin with \eqref{E.Per0} written in dimensionless scaled variables (in analogy with (\ref{E1.4s}))
\begin{equation}\label{E.Per1}
\begin{array}{l}
\eta_t+[(D+\varepsilon\eta)u]_x=0,\\
u_t+\eta_x+\varepsilon uu_x-\sigma^2\left[\frac{D}{2}(Du)_{xxt}-\frac{D^2}{6}u_{xxt}\right]=O(\varepsilon^2, \varepsilon\sigma^2).
\end{array}
\end{equation}
Then by setting $H=D+\varepsilon \eta$, we obtain
\begin{equation}\label{E.Per2}
\begin{array}{l}
H_t+\varepsilon(Hu)_x=0,\\
(Hu)_t+(\varepsilon Hu^2+\frac{1}{2\varepsilon}H^2)_x\\ -\sigma^2\left[\frac{HD}{2}(Du)_{xxt}-\frac{HD^2}{6}(\frac{Du}{D})_{xxt}\right] =\frac{1}{\varepsilon}HD_x+O(\varepsilon^2, \varepsilon\sigma^2).
\end{array}
\end{equation}
Observing that $\left(\frac{Du}{D}\right)_{xx} = [2\frac{D_x^2}{D^3} - \frac{D_{xx}}{D^2}](Du) - 2\frac{D_x}{D^2}(Du)_x + \frac{1}{D}(Du)_{xx}$ and that $H=D+O(\varepsilon)$ we have that
\begin{equation}\label{E.Per3}
\begin{array}{l}
H_t + \varepsilon(Hu)_x=0, \\
(Hu)_t + (\varepsilon Hu^2 + \frac{1}{2\varepsilon}H^2)_x  \\ 
- \sigma^2\left[\frac{D^2}{3}(Du)_{xxt}-(\frac{1}{3}D_x^2-\frac{1}{6}DD_{xx})Du_t + \frac{1}{3}DD_x(Du)_{xt}\right] \\ = \frac{1}{\varepsilon}HD_x + O(\varepsilon^2, \varepsilon\sigma^2).
\end{array}
\end{equation}
By setting $Q=Hu$, and using again the relation $H=D+O(\varepsilon)$ we have
\begin{equation}\label{E.Per4}
\begin{array}{l}
H_t+\varepsilon Q_x=0,\\
Q_t+(\varepsilon Q^2/H+\frac{1}{2\varepsilon}H^2)_x \\ -\sigma^2\left[\frac{H^2}{3}Q_{xxt}-(\frac{1}{3}H_x^2-\frac{1}{6}HH_{xx})Q_t+\frac{1}{3}HH_xQ_{xt}\right]  = \frac{1}{\varepsilon}HD_x+O(\varepsilon^2, \varepsilon\sigma^2).
\end{array}
\end{equation}
In dimensional variables, neglecting the higher order terms at the right-hand side we obtain
\begin{equation}\label{E.Per5}
\begin{array}{l}
H_t+Q_x=0,\\
Q_t+(Q^2/H+\frac{g}{2}H^2)_x - P(H,Q) = gHD_x. \\ 
P(H,Q) = \frac{H^2}{3}Q_{xxt}-(\frac{1}{3}H_x^2-\frac{1}{6}HH_{xx})Q_t+\frac{1}{3}HH_xQ_{xt}
\end{array}
\end{equation}
where $H(x,t)=\eta(x,t)+D(x)$, $Q(x,t)=H(x,t) u(x,t)$. We underline that system (\ref{E.Per5}) is invariant under vertical translations and therefore more appropriate for studying  long wave runup. Moreover, the linearization of the system (\ref{E.Per5}) coincides with the original Peregrine's system (\ref{E.Per1}) and therefore, inherits all its linear dispersive characteristics. On the other hand system (\ref{E.Per5}) cannot be regarded as a correct asymptotic model to the Euler equations since it contains terms of the order $O(\varepsilon\sigma^2)$. However, such terms considered in the correct (small amplitude and long wave) regime are negligible, hence their contribution will be negligible. Finally we note that ignoring the dispersive terms $P(H,Q)$ of system (\ref{E.Per5}) we obtain the shallow water equations \eqref{eq:E1.1}.

We also note that even though Boussinesq systems are not valid in the near-shore region, in practice they appear to predict well the behavior of small amplitude waves from moderately deep to shallower waters and for smooth flows, cf. \cite{Zelt1991}. Of course,  more accurate systems in the near-shore zone have been derived such as the S\'{e}rre equations (sometimes referred also as Green-Naghdi equations), cf. \cite{Serre1953, Lannes2009, Dias2010}. These systems appeared in practice to model better the breaking phenomena in the near shore zone but recent numerical studies of the S\'{e}rre system showed that unphysical oscillations might appear in analogy with the Boussinesq equations during the wave breaking and the runup process, \cite{CBB1, CBB2}. 

\subsection{Source terms}

Nonlinear shallow water model \eqref{eq:E1.1} and Boussinesq system \eqref{E.Per5i} may be completed to take into account some dissipative or friction effects which are very beneficial in describing the wave breaking phenomena. Usually this is accomplished by including appropriate source or dissipative terms into momentum conservation equations \eqref{eq:E1.1} or \eqref{E.Per5i}. Possible choices are the following :
\begin{align}
& \text{  Friction: }  & F(u,H) = -c_m\, g\,\frac{u|u|}{H^{1/3}},  \label{SFri}\\
& \text{ Viscosity: }  & V(u,H) = \mu\frac{\partial}{\partial x}\left( H \frac{\partial u}{\partial x}\right), \label{SVis}
\end{align} 
where $c_m$ is the Manning roughness coefficient and $\mu$ denotes the kinematic viscosity of the fluid. The particular form of the source terms is suggested by empirical laws, which are generally obtained for steady state flows. Similar models have been derived from  then Navier-Stokes system for incompressible flows with a free surface. More complex friction laws can be also formulated to model bottom rugosity effects, etc.

%%%%%%%%%%%%%%%%%%%%%%%%%%%%%%%%%%%%%%%%%%%%%%%
%%%%% SECTION
%%%%%%%%%%%%%%%%%%%%%%%%%%%%%%%%%%%%%%%%%%%%%%%

\section{Numerical schemes}\label{sec:num}

In the present section we generalize the finite volume method to systems \eqref{E1.4i} and \eqref{E.Per5i} of dispersive PDEs. In our work we rely on corresponding schemes for conservation laws.  Next we present briefly the finite volume framework for conservation laws. Based on this framework we introduce finite volume schemes for the dispersive models. 

\subsection{Finite volume method for conservation laws}
We  consider the initial value problem
\begin{equation}\label{CL0}
\begin{aligned}
 & w_t + F(w)_x = S(w), \ x\in\R,  t >0\\
 & w(x,0) = w_0(x), 
 \end{aligned}
\end{equation}
where $w(x,t)$ is the state variable, $F$ denotes the flux and $S$ is the source term.  Let $\Th= \{x_i\}, \ i\in\Z$ denotes a partition of $\R$ into cells $C_i= [\xim,\xip]$ where $x_i = (x_{\ip}+x_{\im})/2$ denotes the midpoint of $C_i$. Let $\dx_i= \xip-\xim$ be the length of  the cell $C_i$,  $\dx_{\ip}=x_{i+1}-x_i$.  Without loss of generality we assume a uniform partition $\Th$, that is $\dx_i=\dx_{\ip}=\dx, \ i\in\Z$. Let $w_i$ denotes the cell average of $w$ on $C_i$ i.e $w_i(t) = \frac{1}{\dx}\int_{C_i} w(x,t)\,dx$. Then a simple integration of \eqref{CL0} over a cell $C_i$ yields 
\begin{equation}\label{FV0}
\frac{d}{dt}w_i(t) +\frac{1}{\dx}\left(F(w(\xip,t))-F(w(\xim,t))\right) = \frac{1}{\dx}\int_{C_i} S(w(x,t))\, dx.
\end{equation}

\subsubsection{Semidiscrete schemes}

We now define the semidiscrete finite volume approximation of $w(x,t)$. Let $\chi_{C_i}$ denotes the characteristic function of the cell $C_i$, we seek a piecewise constant function $w_h(x,t)=\sum_{i\in\Z} W_i(t)\chi_{C_i}(x)$ with
\begin{equation}\label{FV1}
\begin{array}{l}
\frac{d}{dt}W_i(t) +\frac{1}{\dx}\left(\Fip-\Fim\right) = \Sh_i,\quad i\in\Z,  \\
W_i(0)=\frac{1}{\dx}\int_{C_i}w(x,0)\,dx , \quad i\in\Z , 
\end{array}
\end{equation}
where $\Fip = \Fh(W_{\ip}^L,W_{\ip}^R)$ is an approximation to $F(w(\xip,t))$ while  $\Sh_i$ approximates the source term $\Sh_i= \Sh_i(W_i)\approx \frac{1}{\dx}\int_{C_i} S(w(x,t))\, dx$. The values $W_{\ip}^L,W_{\ip}^R $ are approximations to the point value $w(\xip,t)$ from cells $C_i, \ C_{i+1}$ respectively and  $\Fh$ is a numerical flux function which is consistent and monotone.  The values $W_{\ip}^L,W_{\ip}^R $ are computed by a reconstruction process described below (see Section \ref{sec:reconstruct}).

\subsubsection{The numerical fluxes}
There are many possible choices for the numerical flux function $\Fh$. In the present study we choose to work with three following fluxes
\begin{align}
& \Fh^m(W,V) = F\left(\frac{W+V}{2}\right),  \label{AVFlux}\\
& \Fh^{KT}(W,V) = \frac12\left\{\left[ F(V) + F(W)\right] - \Ah(W,V)\left[V-W\right]\right\}, \label{KTFlux} \\
& \Fh^{CF}(W,V)  = \frac12\left\{\left[ F(V) + F(W)\right] - \Ah(W,V)\left[F(V)-F(W)\right]\right\}. \label{CFFlux}
\end{align}
The \emph{average} flux \eqref{AVFlux} is the simplest one. It is well known that although this flux  is unstable for nonlinear conservation laws, it is proven very stable and accurate for nonlinear dispersive models.

The \emph{central} flux \eqref{KTFlux} is a Lax-Friedrichs type flux and is a representative of central schemes \cite{KT, NT}. The operator $\Ah$ is related to the characteristic speeds of the flow and is defined as 
\begin{equation}\label{KTA}
\Ah(W,V) = \max\left[\rho\left(DF(W)\right), \rho\left(DF(V)\right)\right], 
\end{equation}
where $DF$ denotes the Jacobian matrix and $\rho(A)$ is the spectral radius of $A$. 

The \emph{characteristic} flux function \eqref{CFFlux}, \cite{Ghidaglia1996, Ghidaglia2001}, is similar to the upwind  flux and the operator $\Ah$ in this case is defined by
\begin{equation}\label{CFA}
\Ah(W,V) = \sign\left(DF\left(\frac{W+V}{2}\right)\right).
\end{equation}

\subsubsection{The reconstruction process}\label{sec:reconstruct}
The values $W_{\ip}^L,W_{\ip}^R$ are approximations to $w(\xip,t)$ from cells $C_i$ and $C_{i+1}$ respectively. The simplest possible choice is to take the piecewise constant approximation in each cell, 
\begin{equation}\label{RC0}
W_{\ip}^L = W_i, \quad W_{\ip}^R = W_{i+1}. 
\end{equation}
The resulting semidiscrete finite volume scheme is formally first order accurate in space. 

To construct a higher order  scheme in space, the piecewise constant data is replaced by a piecewise polynomial representation. The main idea here is to construct higher order approximations to $w(\xip,t)$ using the computed cell averages $W_i$. For this purpose the classical MUSCL type (TVD2) linear reconstruction \cite{Kolgan1975, Leer1979} as well as UNO2, \cite{HaOs} or WENO type reconstructions, \cite{LOC}, have been developed.

The classical TVD2  linear reconstruction is given by the  following formulas:
\begin{equation}\label{RCTVD}
W_{\ip}^L = W_i + \frac12 \phi(r_i) (W_{i+1}-W_i), \quad W_{\ip}^R = W_{i+1} - \frac12 \phi(r_{i+1}) (W_{i+2}-W_{i+1}),
\end{equation}
where $r_{i} = \frac{W_{i} - W_{i-1}}{W_{i+1} - W_{i}}$, and $\phi$ is an appropriate slope limiter, \cite{Sweby1984}. There  are many options for a limiter function. Some of the most usual choices are
\begin{itemize}
\item MinMod (MM) limiter: $ \phi(\theta)=\max(0,\min(1,\theta))$,
\item VanLeer (VL)  limiter: $ \phi(\theta)= \frac{\theta+|\theta|}{1+|\theta|}$, 
\item Monotonized Central (MC) limiter: $ \phi(\theta)=\max(0,\min((1+\theta)/2,2,2\theta))$, 
\item Van Albada (VA) limiter: $ \phi(\theta)= \frac{\theta+\theta^2}{1+\theta^2}$. 
\end{itemize}
The last three limiters have been shown to exhibit sharper resolution of discontinuities since  they do not reduce the slope as severely as (MM) near a discontinuity. The TVD2 reconstruction is second order accurate except at the local extrema where it reduces to the first order. A remedy is  to consider the UNO2 type reconstruction.

The UNO2 reconstruction  is a linear interpolation which is second order accurate even at local extrema, \cite{HaOs}. The values $W_{\ip}^L,W_{\ip}^R$ are defined as 
\begin{equation}\label{RCUNO}
W_{\ip}^L = W_i + \frac12 S_i, \quad W_{\ip}^R = W_{i+1} - \frac12 S_{i+1},
\end{equation}
where
\begin{align*}
 & S_i = m(S_i^+, S_i^-), \quad S_i^{\pm} = d_{i\pm\frac12}W \mp \frac12 D_{i\pm\frac12}W, \\
 & d_{\ip}W = W_{i+1}-W_i , \quad D_{\ip}W = m(D_iW, D_{i+1}W), \\
 & D_iW = W_{i+1}-2W_i+W_{i-1}, \quad m(x,y) = \frac12(\sign(x)+\sign(y))\min(|x|,|y|).
\end{align*}
Using either (TVD2) or (UNO2) reconstructions the semidiscrete finite volume scheme \eqref{FV1} is formally second order accurate. 

In order to achieve higher order accuracy we also employ WENO type reconstructions for the values $W^R_{i\pm \frac{1}{2}}$, $W^L_{i\pm \frac{1}{2}}$.  We implemented 3rd and 5th order accurate WENO methods (also referred to as WENO3 and WENO5, respectively) as they are described in \cite{LOC}. For the sake of simplicity we only present the WENO3 case.  In order to compute the approximations $W^L_{i+\frac{1}{2}}$ and $W^R_{i-\frac{1}{2}}$, we first compute the 3rd order reconstructed values 
\begin{align*}
&W^{(0)}_{i+\frac{1}{2}}=\frac{1}{2}(W_i+W_{i+1}), \qquad W^{(1)}_{i+\frac{1}{2}}=\frac{1}{2}(-W_{i-1}+3W_i),\\
&W^{(0)}_{i-\frac{1}{2}}=\frac{1}{2}(3W_i-W_{i+1}), \qquad W^{(1)}_{i-\frac{1}{2}}=\frac{1}{2}(W_{i-1}+W_i).
\end{align*}
We define the smoothness parameters
\begin{align*}
 & \beta_0=(W_{i+1}-W_i)^2,\qquad  \beta_1=(W_i-W_{i-1})^2,
 \end{align*}
and the parameters $d_0=\frac{2}{3}$, $d_1=\frac{1}{3}$ and $\tilde{d}_0=d_1$, $\tilde{d}_1=d_0$, along with the weights
\begin{equation*}
  \omega_0=\frac{\alpha_0}{\alpha_0+\alpha_1},  \quad  \omega_1=\frac{\alpha_0}{\alpha_0+\alpha_1}, \quad
  \tilde{\omega}_0=\frac{\tilde{\alpha}_0}{\tilde{\alpha}_0+\tilde{\alpha}_1}, \quad    \tilde{\omega}_1=\frac{\tilde{\alpha}_1}{\tilde{\alpha}_0+\tilde{\alpha}_1},
 \end{equation*}
where $\alpha_i=\frac{d_i}{\epsilon+\beta_i}$, $\tilde{\alpha}_i=\frac{\tilde{d}_i}{\epsilon+\beta_i}$ and $\epsilon$ to be a small, positive number (in our computations we set $\epsilon=10^{-15}$).
Then the reconstructed values are given by the following formulas
\begin{equation}\label{RCWENO} 
W^L_{i+\frac{1}{2}}=\sum_{r=0}^{1}\omega_r W^{(r)}_{i+\frac{1}{2}}, \qquad  W^R_{i-\frac{1}{2}}=\sum_{r=0}^{1}\tilde{\omega}_r W^{(r)}_{i-\frac{1}{2}}. 
 \end{equation}

\subsubsection{Discretization of source terms}
The finite volume discretization of the source term $S(w)$ in \eqref{CL0} depends on the particular choice.  On the other hand the resulting approximation should preserve the upwind nature and the overall scheme should be well balanced. One possible discretization of the source term $S(w)$ is given by: 
\begin{equation}\label{FVS}
\frac{1}{\dx}\int_{C_i} S(w)\, dx \approx \frac{\Sh_{\im}+\Sh_{\ip}}{2}, \quad \Sh_{\ip}=S\left(\frac{W_{\ip}^L+W_{\ip}^R}{2}\right).
\end{equation}

\subsubsection{Fully discrete schemes}

Equation \eqref{FV1} is an initial value problem and can be discretized by various methods. In our case we use a special class of Runge-Kutta methods which ensure the TVD property of the finite volume scheme, \cite{Shu1988a, Gottlieb2001, Spiteri2002}.

Let $\dt$ be the time step and let $t^{n+1}=t^n+\dt, \ n\ge0$ be discrete time levels. Assuming at  $t^n$ the approximations $\{W_i^n\}, \ i\in\Z$ are known then $W_i^{n+1}$ are defined by
\begin{equation}\label{FV2}
\begin{aligned}
& W_i^{n+1} = W_i^n - \frac{\dt}{\dx}\sum_{j=1}^s b_j\left(\Fip^{n,j}-\Fim^{n,j}\right) + \dt  \sum_{j=1}^s b_j\
 \Sh_i^{n,j},  \\
& W_i^{n,j} = W_i^n - \frac{\dt}{\dx}\sum_{\ell=1}^s a_{j\ell}\left(\Fip^{n,\ell}-\Fim^{n,\ell}\right) + \dt \sum_{\ell=1}^s a_{j\ell}\,  \Sh_i^{n,\ell},
\end{aligned}
\end{equation}
where $\Fip^{n,j} = \Fh(W_i^{n,j}, W_{i+1}^{n,j})$, $ \Sh_i^{n,j}=  \Sh(W_i^{n,j})$. The set of constants $A=(a_{j\ell}), \ b=(b_1,\dots,b_s)$ define an $s-$stage Runge-Kutta method. The following \emph{tableau} are examples of explicit TVD RK-methods which are of 2nd and 3rd order respectively
\begin{equation}\label{RKM}
\begin{tabular}{c c | c}
0 & 0 & 0 \\
1 & 0 & 1 \\ \hline
$\frac12$ & $\frac12$ & 
\end{tabular}
\qquad
\begin{tabular}{c c c | c}
0 & 0 & 0 &  0 \\
1 & 0 & 0 &  1 \\ 
$\frac14$ & $\frac14$ & 0 & $\frac12$ \\ 
\hline
$\frac16$ & $\frac16$ & $\frac23$ \\
\end{tabular}
\end{equation}
In our computations we mainly use the three stage 3rd order method. 

\subsection{Finite volume schemes for dispersive models}

To construct the finite volume schemes for the dispersive models the main idea is to rewrite the governing equations or systems in a conservative like form and discretize the resulting conservation laws using the aforementioned framework. One can use any of the numerical fluxes, $\Fh^m, \ \Fh^{KT}, \ \Fh^{CF}$ and  reconstruction techniques TVD2, UNO2 or WENO. Temporal discretization is based on the TVD-Runge-Kutta methods, \eqref{RKM}.  

\subsubsection{Boussinesq systems with flat bottom}

Boussinesq systems \eqref{E1.4i}  can be rewritten in a conservative like form as follows:
\begin{equation}\label{E2.2}
({\bf I}-{\bf D}){\bf v}_t+\left[{\bf F}({\bf v})\right]_x+\left[{\bf G}({\bf v})\right]_x=0, 
\end{equation}
where ${\bf v}=(\eta,u)^T$, ${\bf F}({\bf v})=((1+\eta)u,\eta+\frac{1}{2}u^2)^T$, ${\bf G}({\bf v})=(a\,u_{xx},c\,\eta_{xx})^T$, and ${\bf D}=\diag\,(b\,\partial^2_x,d\,\partial^2_x)$.  The simplest discretization is based on the average fluxes $\Fh^m$ for ${\bf F}$ and $\Gh^m$ for ${\bf G}$. For the other two choices of the numerical flux $\Fh$ the evaluation of a Jacobian is needed.  Let  $A$ denotes the Jacobian of ${\bf F}$, then 
$$
A=\left(
\begin{array}{cc}
u & 1+\eta\\ 
1 & u 
\end{array} \right),
$$ 
with eigenvalues  $\lambda_i=u\pm \sqrt{1+\eta}$, $i=1,2$. It is readily seen, since ${\bf F}$ is a hyperbolic flux, that $A$ can be decomposed as $A=L\Lambda R$ thus for the characteristic flux $\Fh^{CF}$ we have with $\mu=\frac{W+V}{2}$, $s_i=\sign(\lambda_i), \ i=1,2$
$$
\Ah(W,V)=\left(
 \begin{array}{cc} 
 \frac{1}{2}(s_1+s_2) & \frac{1}{2}\sqrt{1+\mu_1}(s_1-s_2)\\ 
\frac{s_1-s_2}{2\sqrt{1+\mu_1}} & \frac{1}{2}(s_1+s_2) 
\end{array} \right).
$$ 
For evaluating the numerical fluxes $\Fh, \ \Gh$ simple cell averages or higher order approximations such as UNO2 \eqref{RCUNO} or WENO \eqref{RCWENO} can be used. 

\begin{rem}
The discretization of the elliptic operator ${\bf D}$ is based on the standard centered difference. This is a second order accurate approximation and it is compatible with the TVD2 and UNO2 reconstructions. For higher order interpolation we modify the elliptic and flux discretization. Indeed, the finite volume scheme is modified as 
\begin{multline*}\label{FV1a}
\frac{d}{dt}\left[\frac{{\bf V}_{i-1}+10{\bf V}_i+{\bf V}_{i+1}}{12}-(b, d)\frac{{\bf V}_{i+1}-2{\bf V}_i+{\bf V}_{i-1}}{\dx^2}\right]  \\ + \frac{\Hh_{i-1}+10\Hh_i+\Hh_{i+1}}{12}=0, 
\end{multline*}
where $\Hh_i=\frac{1}{\dx}(\Fip-\Fim) +\frac{1}{\dx}(\Gip-\Gim)$, resulting in a high order accurate approximation.  Thus in the WENO3 case a global third order accuracy is observed, while for  WENO5 interpolation, we profit only locally by the 5th order accuracy of the reconstruction,  cf. Section \ref{sec:accu}.
\end{rem}

\begin{rem}
In the sequel for the discretization of the dispersive term ${\bf G}$ we use mainly the average numerical flux $\Gh^m$ defined as $\Gh^m_{i+\frac{1}{2}}=(a,c)\frac12({\bf Y}_i+{\bf Y}_{i+1})$, where ${\bf Y}_i=\frac{1}{\Delta x^2}({\bf V}_{i+1}-2{\bf V}_i+{\bf V}_{i-1})$. In case of higher order WENO reconstructions we use the average numerical flux based on the reconstructed values of ${\bf Y}_i$ i.e. the flux $\Gh^{lm}_{i+\frac{1}{2}}=(a,c)\frac12({\bf Y}^L_{i+\frac{1}{2}}+{\bf Y}^R_{i+\frac{1}{2}})$, where ${\bf Y}^L_{i+\frac{1}{2}} $ and ${\bf Y}^R_{i+\frac{1}{2}}$ are reconstructed values of ${\bf Y}_i$.
\end{rem}

\subsubsection{Boussinesq system with variable bottom}

We write system (\ref{E.Per5}) in terms of dependent variables ${\bf v}:=(H, Q)^T$ in the following conservative  form 
\begin{equation}\label{E5.1}
[{\bf D}({\bf v}_t)] + [{\bf F}({\bf v})]_x = {\bf S}({\bf v}),
\end{equation}
where
\begin{align} 
& {\bf D}({\bf v}_t)=
\begin{pmatrix} H_t \\  (1+\frac{1}{3}H_x^2-\frac{1}{6}HH_{xx})Q_t-\frac{1}{3}HH_x Q_{xt} -\frac{H^2}{3}Q_{xxt}
\end{pmatrix},  \label{E5.2} \\
&  {\bf F}({\bf v})=
\begin{pmatrix}
Q\\
\frac{Q^2}{H} + \frac{g}{2}H^2
\end{pmatrix},\qquad 
{\bf S}({\bf v})=\begin{pmatrix} 0\\ gHD_x \end{pmatrix}. \label{E5.3}
\end{align}

We consider a uniform mesh and we denote by $H_i$, $U_i$ and $D_i$ the corresponding cell averages. To discretize the dispersive terms in \eqref{E5.2} we consider the following approximations:
\begin{multline*}\label{E5.4}
\frac{1}{\dx}\int_{x_{i-\frac{1}{2}}}^{x_{i+\frac{1}{2}}}\left[1+\frac{1}{3}(H_x)^2-\frac{1}{6}HH_{xx}\right]\;Q\;dx \approx \\
\left(1+\frac{1}{3}\left(\frac{H_{i+1}-H_{i-1}}{2\dx} \right)^2 - \frac{1}{6}H_i \;\frac{H_{i+1}-2H_i+H_{i-1}}{\dx^2} \right) Q_i, 
\end{multline*}

\begin{equation}\label{E5.5}
\frac{1}{\dx}\int_{x_{i-\frac{1}{2}}}^{x_{i+\frac{1}{2}}} \frac{1}{3}HH_xQ_x \;dx \approx 
\frac{1}{3}H_i\;\frac{H_{i+1}-H_{i-1}}{2\dx}\frac{Q_{i+1}-Q_{i-1}}{2\dx}, 
\end{equation}

\begin{equation}\label{E5.6}
\frac{1}{\dx}\int_{x_{i-\frac{1}{2}}}^{x_{i+\frac{1}{2}}} \frac{1}{3}H^2 Q_{xx}\;dx \approx 
\frac{1}{3}H_i^2\;\frac{Q_{i+1}-2Q_i+Q_{i-1}}{\dx^2}.
\end{equation}
The aforementioned  discretizations lead to a linear system with tridiagonal matrix denoted by ${\bf L}$ that can be inverted efficiently by a variation of Gauss elimination for tridiagonal systems with computational complexity $O(d)$, d-being the dimension of the system.  We note that on the dry cells the matrix becomes diagonal since $H_i$ is zero on dry cells. For the time integration the explicit third-order TVD-RK method, \eqref{RKM} is used. In the numerical experiments we observed that the fully discrete scheme is stable and preserves the positivity of $H$ during the runup under a mild restriction on the time step $\dt$.

Therefore, the semidiscrete problem of (\ref{E5.2}) - (\ref{E5.3}) is written as a system of o.d.e's in the form 
\begin{equation}\label{E5.7}
{\bf L}_i{{\bf v}_i}_t+\frac{1}{\dx}({\Fh}_{i+\frac{1}{2}}-{\Fh}_{i-\frac{1}{2}})=\frac{1}{\dx}{\bf S_i}, 
\end{equation}
where ${\bf L}_i$ is the $i-$th row of matrix ${\bf L}$ and ${\Fh}_{i+\frac{1}{2}}$ can be chosen as one of the numerical flux functions mentioned  in the previous sections. In the sequel we will use the KT and the CF numerical fluxes. In this case the Jacobian of ${\bf F}$ is given by the matrix
$$A=\begin{pmatrix} 0 & 1 \\ gH-(Q/H)^2 & 2Q/H \end{pmatrix},$$ and the eigenvalues
are $\lambda_{1,2}=Q/H\pm\sqrt{gH}$.  Therefore, the CF numerical flux takes the form
\begin{equation}\label{E5.8}
{\Fh}_{i+\frac{1}{2}}=\frac{{\bf F}({\bf V}_{i+\frac{1}{2}}^L)+{\bf F}({\bf V}_{i+\frac{1}{2}}^R)}{2}-{\bf U}({\boldsymbol \mu})\frac{{\bf F}({\bf V}_{i+\frac{1}{2}}^R)-{\bf F}({\bf V}_{i+\frac{1}{2}}^L)}{2}, 
\end{equation}
where ${\boldsymbol \mu}=(\mu_1,\mu_2)^T$ are the Roe average values,
$$
\mu_1=\frac{H_{i+\frac{1}{2}}^L+H_{i+\frac{1}{2}}^R}{2}, \quad 
\mu_2=\frac{\sqrt{H_{i+\frac{1}{2}}^L}U_{i+\frac{1}{2}}^L+\sqrt{H_{i+\frac{1}{2}}^R}U_{i+\frac{1}{2}}^R} {\sqrt{H_{i+\frac{1}{2}}^L}+\sqrt{H_{i+\frac{1}{2}}^R}}
$$ 
 and
\begin{equation}\label{E5.9}
{\bf U}({\boldsymbol \mu})=\begin{pmatrix}\frac{s_2(\mu_2+c)-s_1(\mu_2-c)}{2c} & \frac{s_1-s_2}{2c}\\
\frac{(s_2-s_1)(\mu_2^2-c^2)}{2c} & \frac{s_1(\mu_2+c)-s_2(\mu_2-c)}{2c} \end{pmatrix}, \ c=\sqrt{g \mu_1},\ s_i=\sign(\lambda_i) . 
\end{equation}
In order to guarantee the positivity of the reconstructed values $H_{i+\frac{1}{2}}$ on the cell  interfaces  we employ the well balanced hydrostatic reconstruction algorithm,  \cite{Audusse2004}. Here we briefly recall the great lines of this reconstruction algorithm.

In the cell $C_i$ we compute first the reconstructions ${\bf V}_{i,r}$ and ${\bf V}_{i,l}$ at $(i+\frac{1}{2})^-$ and $(i-\frac{1}{2})^+$, respectively using  either TVD2 or UNO2 with MinMod limiter. Moreover, we compute in the same way the values $\eta_{i,l}$ and $\eta_{i,r}$ of the free surface elevation $\eta_i=H_i-D_i$. Now we can deduce the values $D_{i,l}=H_{i,l}-\eta_{i,l}$ and $D_{i,r}=H_{i,r}-\eta_{i,r}$. Letting $D_{i+\frac{1}{2}}=\min (D_{i,r},D_{i,l})$ we compute
\begin{equation}\label{E5.10}
H_{i+\frac{1}{2}}^R=\max (0,H_{i,r}+D_{i,r}-D_{i+\frac{1}{2}}), \quad H_{i+\frac{1}{2}}^L=\max (0,H_{i+1,l}+D_{i+1,l}-D_{i+\frac{1}{2}}),
\end{equation}
and we deduce conservative reconstructed variables
\begin{equation}\label{E5.11}
{\bf V}_{i+\frac{1}{2}}^L=\begin{pmatrix} H_{i+\frac{1}{2}}^L\\ H_{i+\frac{1}{2}}^L u_{i,r} \end{pmatrix}, \qquad {\bf V}_{i+\frac{1}{2}}^R=\begin{pmatrix} H_{i+\frac{1}{2}}^R \\ H_{i+\frac{1}{2}}^Ru_{i+1,l}\end{pmatrix}.
\end{equation}
Then the term ${\bf S}_i$ can be written as ${\bf S}_i={\bf S}_{i+\frac{1}{2}}^L+{\bf S}_{i+\frac{1}{2}}^R+{\bf S}_{ci}$, where
\begin{equation*}
{\bf S}_{i+\frac{1}{2}}^L=\begin{pmatrix} 0\\ \frac{g}{2}\left[(H_{i+\frac{1}{2}}^L)^2-(H_{i,r})^2\right] \end{pmatrix},
\qquad
{\bf S}_{i+\frac{1}{2}}^R=\begin{pmatrix} 0\\ \frac{g}{2}\left[(H_{i,l})^2 -(H_{i+\frac{1}{2}}^R)^2\right] \end{pmatrix}
\end{equation*}
and 
\begin{equation*}
{\bf S}_{ci}=\begin{pmatrix} 0\\ g\frac{H_{i,l}+H_{i,r}}{2}(D_{i,l}-D_{i,r}) \end{pmatrix}.
\end{equation*}
Numerical experiments show that the resulting scheme is well-balanced even for Boussinesq system of equations.

\subsubsection{Boundary conditions}

In the case of Bona-Smith type systems with flat bottom we consider herein only the initial-periodic boundary value problem which is known to be well-posed \cite{ADM1}.

In case of the modified Peregrine's system with an uneven bottom we use reflective boundary conditions. We note that for the classical Boussinesq system posed in a bounded domain $I=[b_1,b_2]$, one needs to impose boundary conditions only in one of the two dependent variables, cf. \cite{FP}.  In the case of reflective boundary conditions it is sufficient to take $u(b_1,t)=u(b_2,t)=0$ cf. \cite{AD3}. In \cite{AD3} it was also observed that during solitary waves reflection the derivatives $\eta_x(b_1,t)=\eta_x(b_2,t)\to 0$, while for other wave types these derivatives remained very small.

In our case we consider analogous reflective boundary conditions taking the cell averages  of  $u$ on the first and the last cell  to be $u_0=u_{N+1}=0$.  We don't impose explicitly boundary conditions on $H$. The reconstructed values on the first and the last cell are computed using neighboring ghost cells and taking odd and even extrapolation for $u$ and $H$ respectively. These specific boundary conditions appeared to reflect incident waves on the boundaries while conserving the mass.

%%%%%%%%%%%%%%%%%%%%%%%%%%%%%%%%%%%%%%%%%%%%%%%
%%%%% SECTION
%%%%%%%%%%%%%%%%%%%%%%%%%%%%%%%%%%%%%%%%%%%%%%%
\section{Interactions of solitary waves}\label{sec:interact}

For the Boussinesq system \eqref{E1.4i} we present initially results demonstrating  the accuracy of the finite volume scheme. We study the propagation as well as the interaction of solitary waves. In particular we consider head-on and overtaking collisions.

\subsection{Accuracy test, validation} \label{sec:accu}
We consider the initial value problem with periodic boundary conditions for the Bona-Smith  systems (\ref{E1.8}) with known solitary wave solutions (\ref{E1.9}) -- (\ref{E1.10}) to study the accuracy of the finite volume method. We fix $\theta^2=8/10$ in the system and an analytic solitary wave of amplitude $\eta_0=1/2$ is used as the exact solution in $[-50, 50]$  computed up to $T=200$. The error is measured with respect to discrete $L^2$ and $L^{\infty}$ scaled norms $E_h^2, \  E_h^{\infty}$, namely 
\begin{align*}
& E_h^2(k)=\|u-U^k\|_h/\|U^0\|_h, \quad  \|u-U^k\|_h=\left(\sum_i \dx |u(x_i)-U^k_i|^2\right)^{1/2} , \\
& E_h^{\infty}(k)=\|u-U^k\|_{h,\infty}/\|U^0\|_{h,\infty}, \quad \|u-U^k\|_{h,\infty}=\max_i |u(x_i)-U^k_i|,
\end{align*} 
where $U^k=\{U^k_i\}_i$ denotes the solution of the fully-discrete scheme at the time $t^k=k\, \dt$. The expected theoretical order of convergence was confirmed for all finite volume methods presented above.Three indicative cases, demonstrating the order of convergence, are reported in Table \ref{ROC} :  a) for the average flux, b) for the KT flux with TVD2 reconstruction using the minmod limiter and c) for CF flux with WENO3 reconstruction.  The order of convergence for the WENO5 method cannot be obtained since a 4th order discretization is used for the elliptic operator.
\begin{table}%
\centering
\subtable[Average Flux]{
\begin{tabular}{|c|c|c|} 
\toprule%
$\dx$ & Rate($E_h^2$) &  Rate($E_h^{\infty}$) \\ \midrule
0.5          & 1.910 & 1.978 \\ \hline
0.25        & 1.910 & 1.954 \\ \hline
0.125     & 1.923 &  1.937 \\ \hline
0.0625   & 1.936 &  1.941 \\ \hline
0.03125 & 1.946 &  1.948 \\ 
\bottomrule%
\end{tabular}}
\subtable[KT-TVD2(MinMod)]{
\begin{tabular}{|c|c|c|} 
\toprule%
$\dx$ & Rate($E_h^2$) &  Rate($E_h^{\infty}$) \\ \midrule
0.5          & 2.042  & 2.032 \\ \hline
0.25        & 2.033  & 2.029 \\ \hline
0.125     & 2.026  &  2.023 \\ \hline
0.0625   & 2.021  &  2.019 \\ \hline
0.03125 & 2.017 &  2.016 \\ 
\bottomrule%
\end{tabular}}\\
\subtable[CF-WENO3]{
\begin{tabular}{|c|c|c|} 
\toprule%
$\dx$ & Rate($E_h^2$) &  Rate($E_h^{\infty}$) \\ \midrule
0.5          & 2.976  & 2.975 \\ \hline
0.25        & 3.017  & 3.022 \\ \hline
0.125     & 3.031  &  3.044 \\ \hline
0.0625   & 3.042  &  3.059 \\ \hline
0.03125 & 3.051 &  3.073 \\ 
\bottomrule%
\end{tabular}}

\caption{Rates of convergence.}%
\label{ROC}%
\end{table}

\begin{figure}%
\centering
\subfigure[Evolution of $\eta$ amplitude (Bona-Smith)]%
{\includegraphics[width=0.49\textwidth]{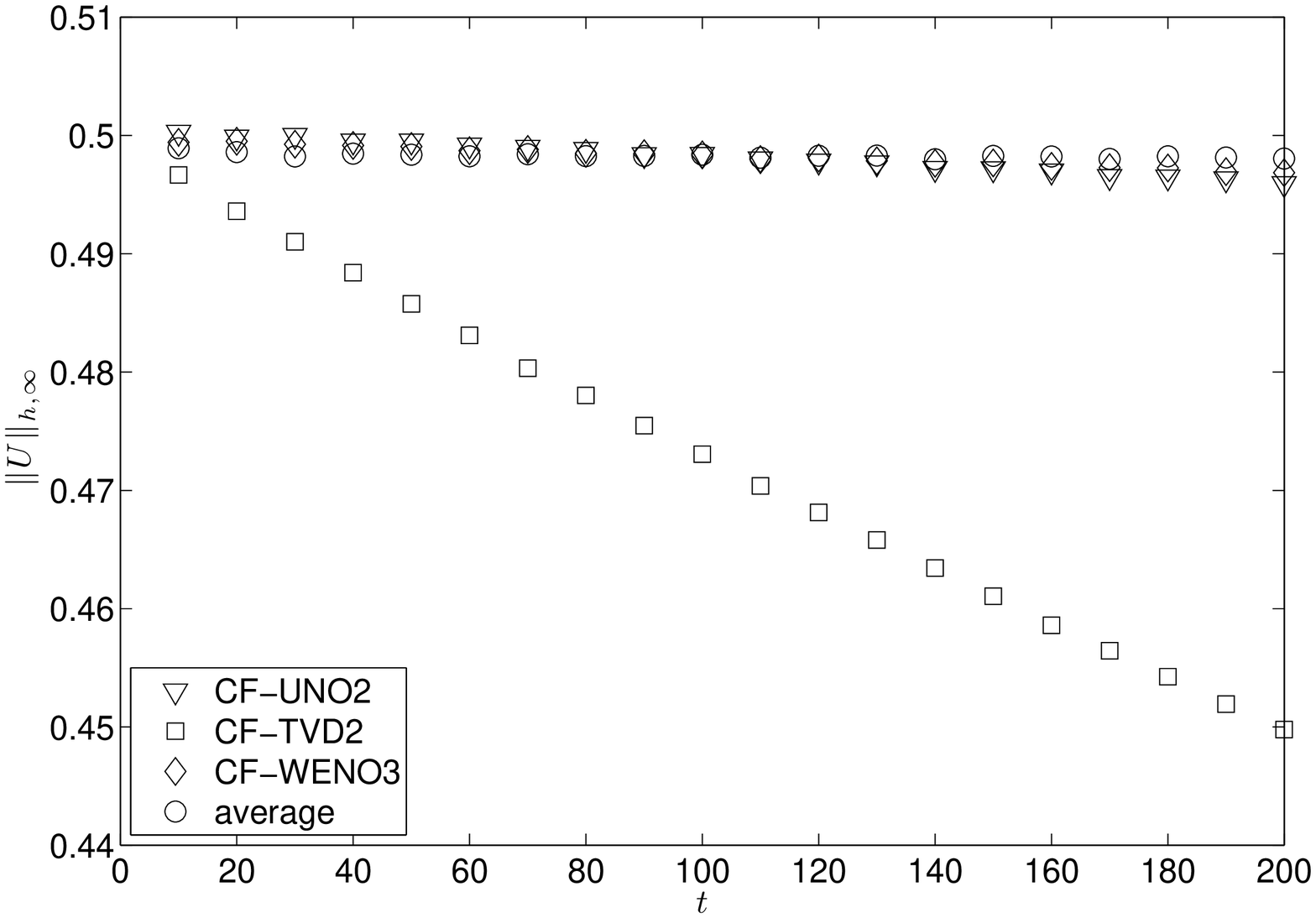}}
\subfigure[Evolution of $I_1^h$ (Bona-Smith)]%
{\includegraphics[width=0.49\textwidth]{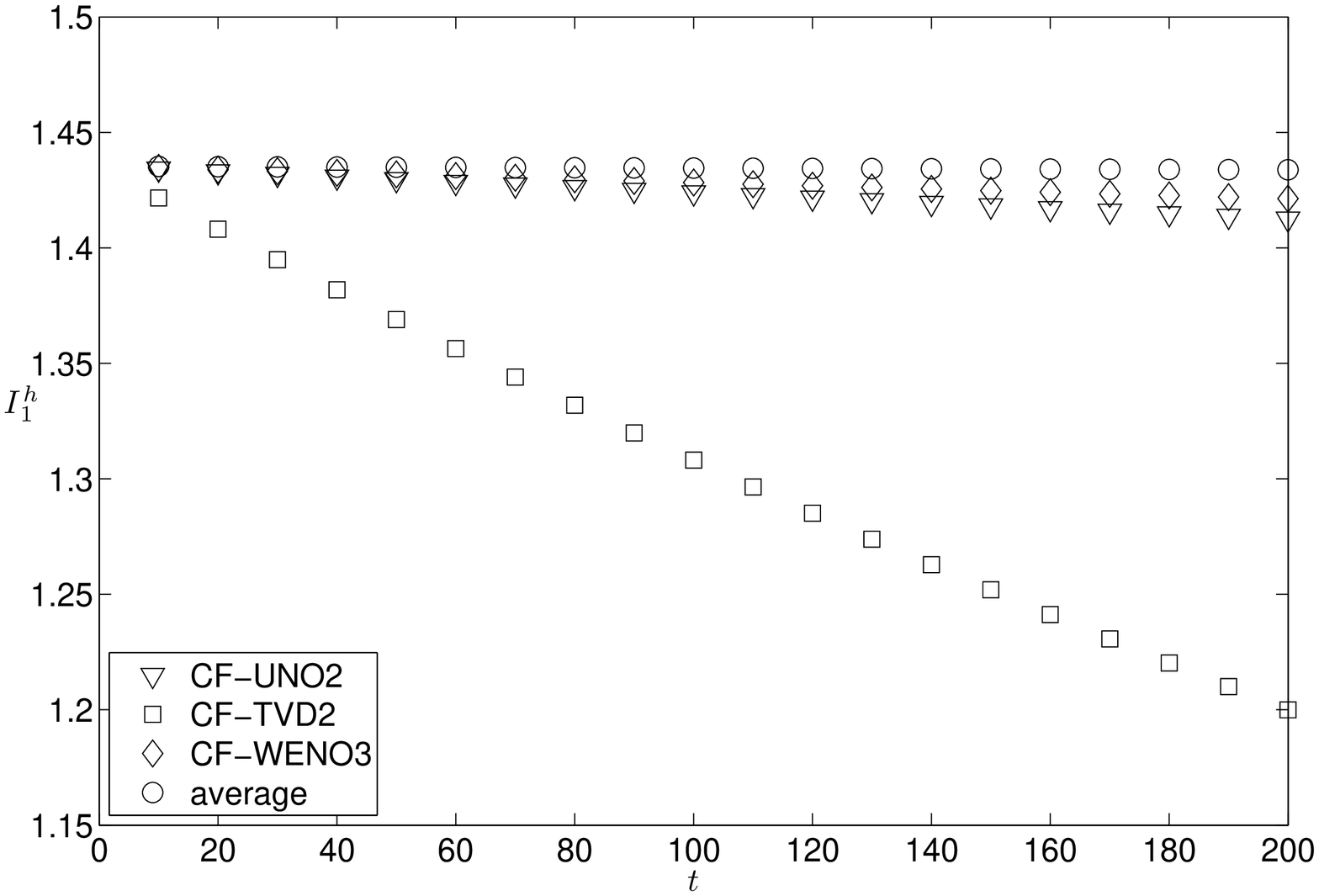}}
\subfigure[Evolution of $\eta$ amplitude (classical Boussinesq)]%
{\includegraphics[width=0.6\textwidth]{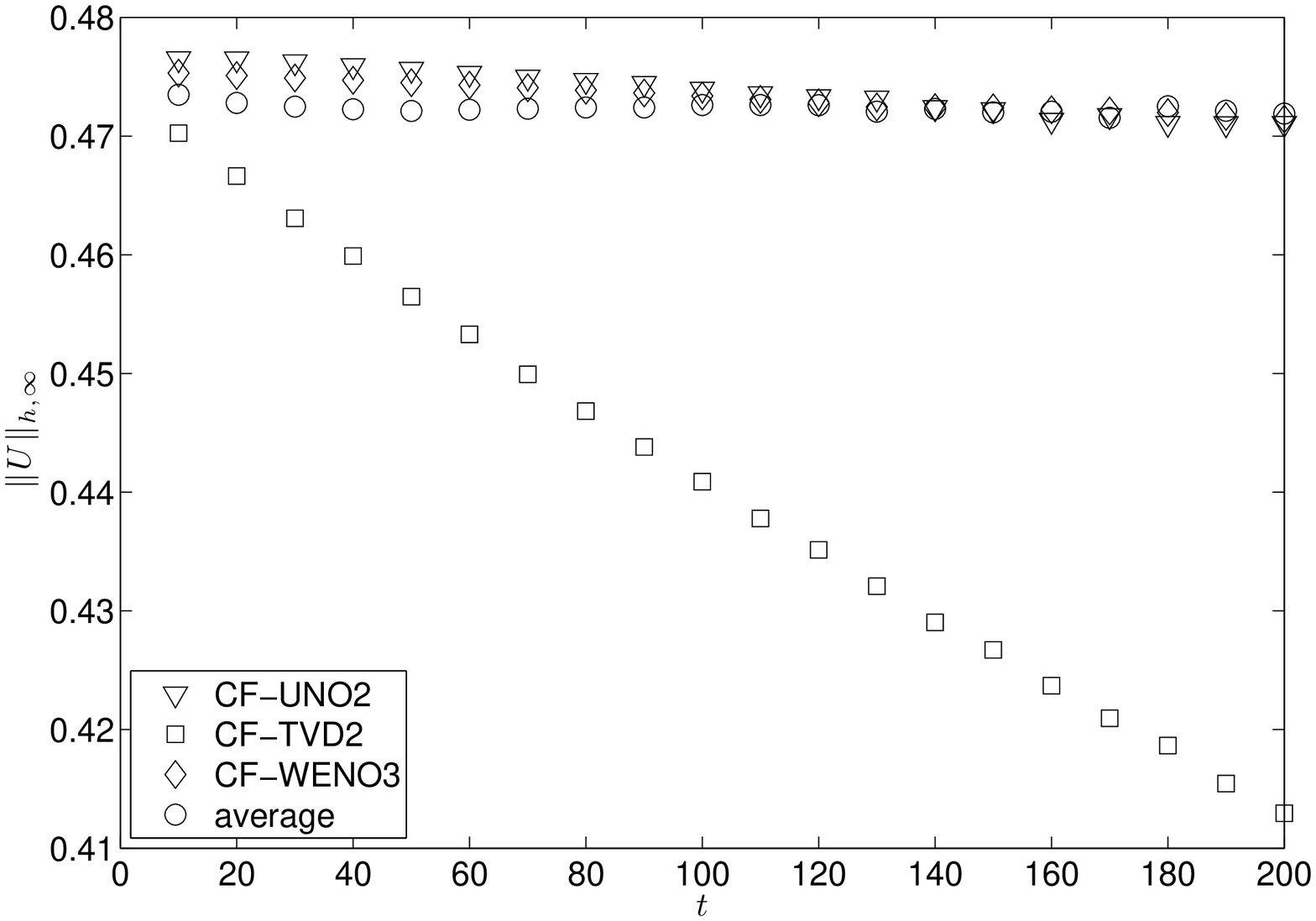}}
\caption{Preservation of the solitary wave amplitude and conservation of the invariant $I_1^h$: $G^{m}$ flux with Minmod limiter}%
\label{F12}%
\end{figure}

We also check the preservation of the invariant \eqref{EInv} by computing its discrete counterpart:
\begin{equation}\label{EInv2}
I_1^h=\sum_i \dx\left(\eta_i^2+[(1+\eta_i)u_i]^2-c\left[\frac{\eta_{i+1}-\eta_i}{\dx}\right]^2-
a\left[\frac{u_{i+1}-u_i}{\dx}\right]^2\right),
\end{equation}
as well as the discrete mass $I_0^h=\dx\sum_i \eta_i$.  Figure \ref{F12} shows the evolution of the amplitude and the invariant $I_1^h$ of the  solitary wave up to $T=200$. The comparison of various methods is performed. We observe that the UNO2 reconstruction is more accurate  while KT and the CF schemes show comparable performance. (We note that the invariant $I_0^h = 1.932183566158$ conserved the digits shown for all numerical schemes. In this experiment we took $\dx=0.1, \ \dt = \dx/2$.)  Figure \ref{F12} (a) and (c) show the evolution of the amplitude of the analytical solitary wave of the Bona-Smith system ($\theta^2=8/10$) and of the solitary wave produced by the solution of the analogous  ordinary differential equations  system of the classical Boussinesq system respectively. In the case of the classical Boussinesq system we took $c_s=1.2$ and we used the method described in \cite{DMII}.

%%%%%%%%%%%%%%%%%%%%%%%%%%%%%%%%%%

\begin{figure}%
\centering
\subfigure[$t=100$]{\includegraphics[scale=.43]{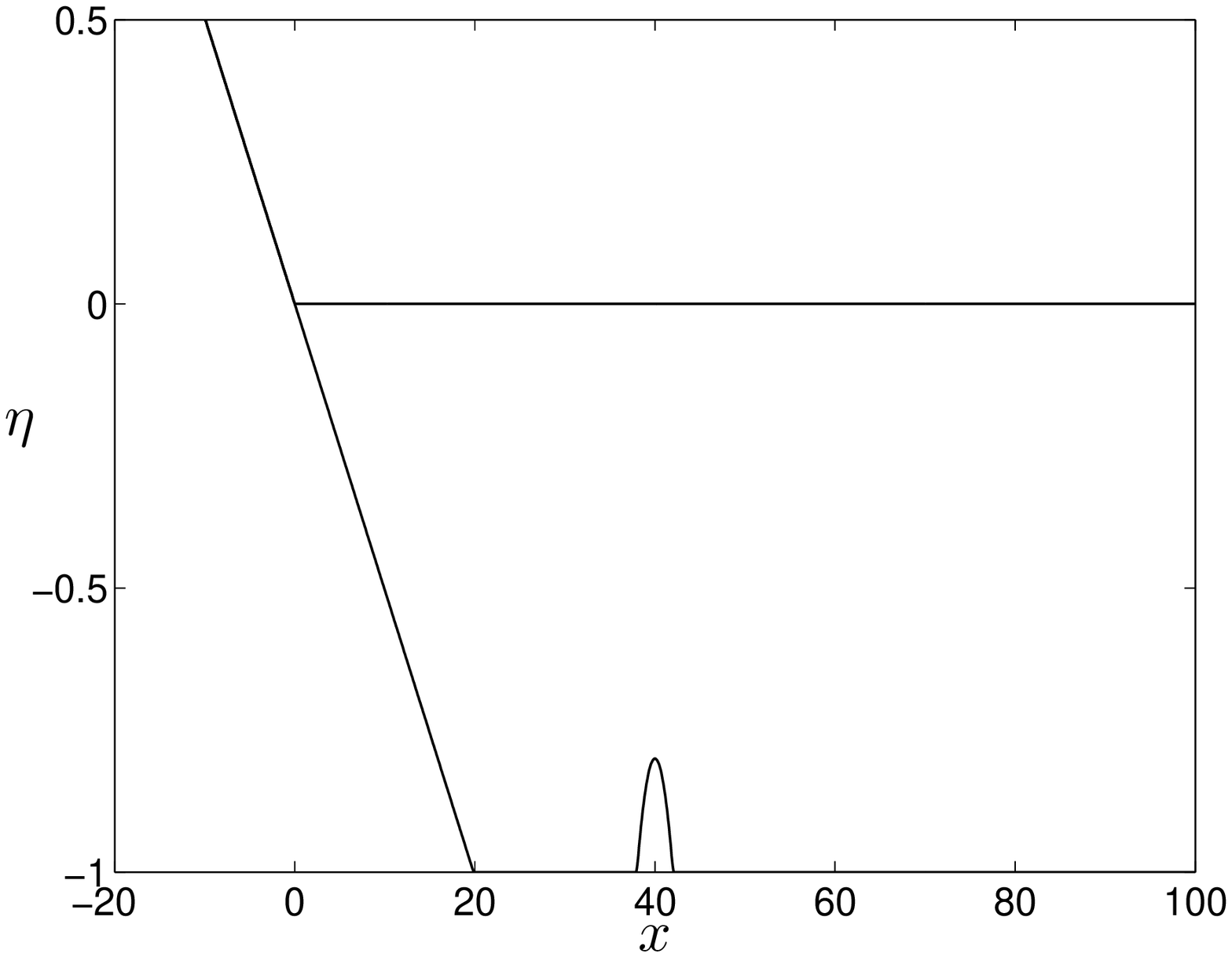}}
\subfigure[magnification of the solution at $t=100$]%
{\includegraphics[scale=.43]{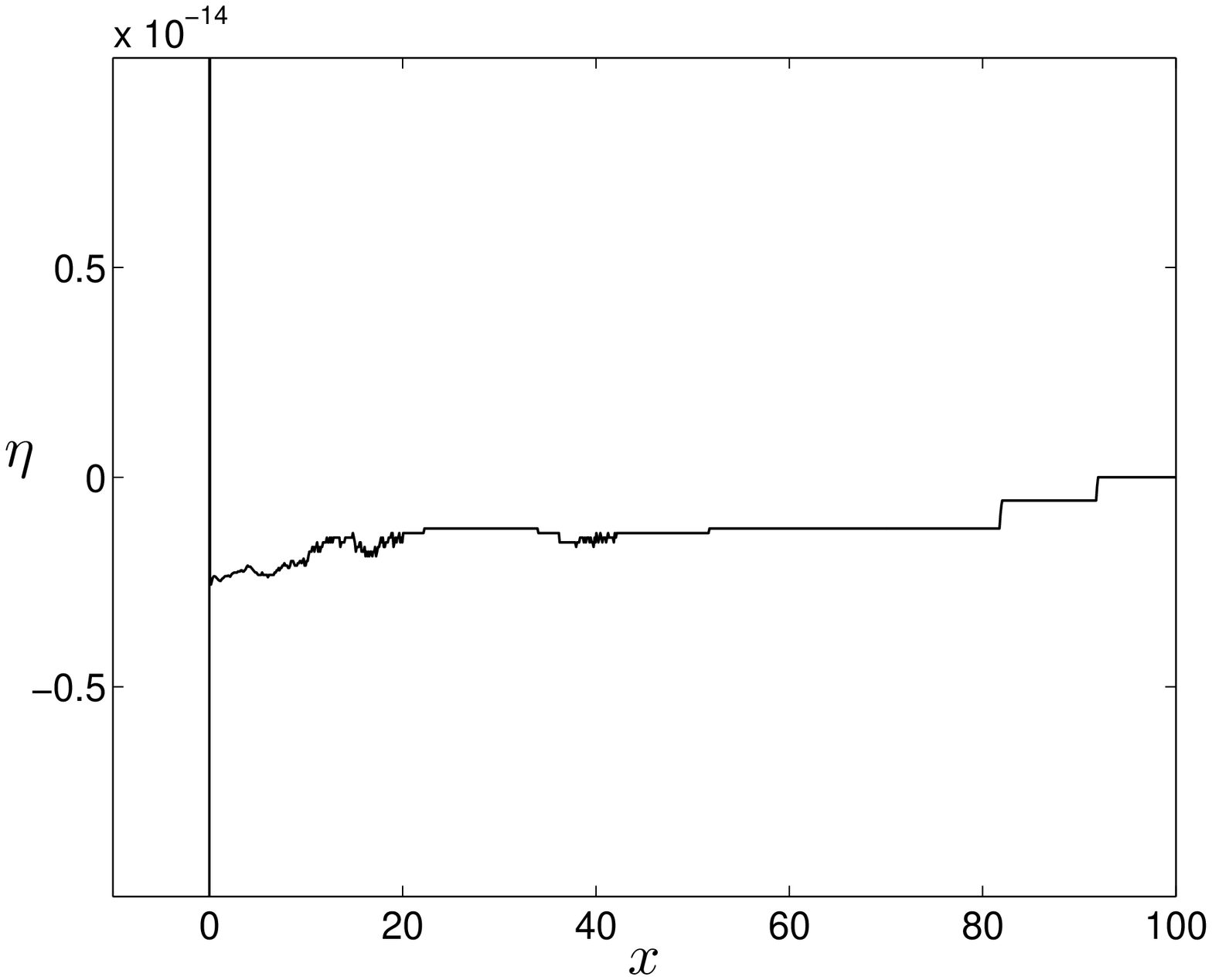}}
\caption{Steady state preservation}%
\label{wellb}%
\end{figure}

The well balanced property of the finite volume schemes is also verified numerically.  We consider a uniform shore, cf. Section \ref{sec:bottom},  including a wet and dry region. The bottom is also modified by adding a small parabolic type hump located at $x=40$. We tested the steady state preservation of all fluxes and possible reconstruction techniques.  The results are similar in all cases.  In Figure \ref{wellb}  we present the case of FC flux along with UNO2. 

%%%%%%%%%%%%%%%%%%%%%%%%%%%%%%%%%%%%%%
%%%%%%%%%%%%%%%%%%%%%%%%%%%%%%%%%%%%%%

\subsection{Head-on collisions}

The head-on collision of two counter-propagating solitary waves is characterized by the change of the shape along with a small phase-shift of the waves as a consequence of the nonlinearity and dispersion. These effects have been studied extensively  before by numerical means using high order numerical methods such as finite differences, \cite{BC}, spectral and finite element methods \cite{ADM2, DDMM, PD} and experimentally in \cite{CGHHS}. In Figure \ref{F15a}  we present the numerical solutions of the BBM-BBM system (\ref{E1.7}) and the Bona-Smith system (\ref{E1.8}) with $\theta^2=9/11$ (in dimensional and unscaled variables) along with the experimental data from \cite{CGHHS}. The spatial variable is expressed in centimeters while the time  in seconds. The solutions were obtained using the CF-scheme with UNO2 and WENO3  reconstruction using $\dx = 0.05$ cm and $\dt=0.01$ s. For this experiment we constructed solitary waves for  Boussinesq systems by solving the respective o.d.e's system in the spirit of \cite{Bona2007} such that they fit to experimentally generated solitary waves before the collision. The speeds of the right and left-traveling solitary waves are $c_{r,s}=0.854 \mbox{ m/s}$ and $c_{l,s}=0.752 \mbox{ m/s}$ respectively.

%%%%%%%%%%%%%%%%%%%%%%%%%%%%%%%%%%%%%%

\begin{figure}%
\centering
\subfigure[$t=18.29993 s$]{\includegraphics[scale=.33]{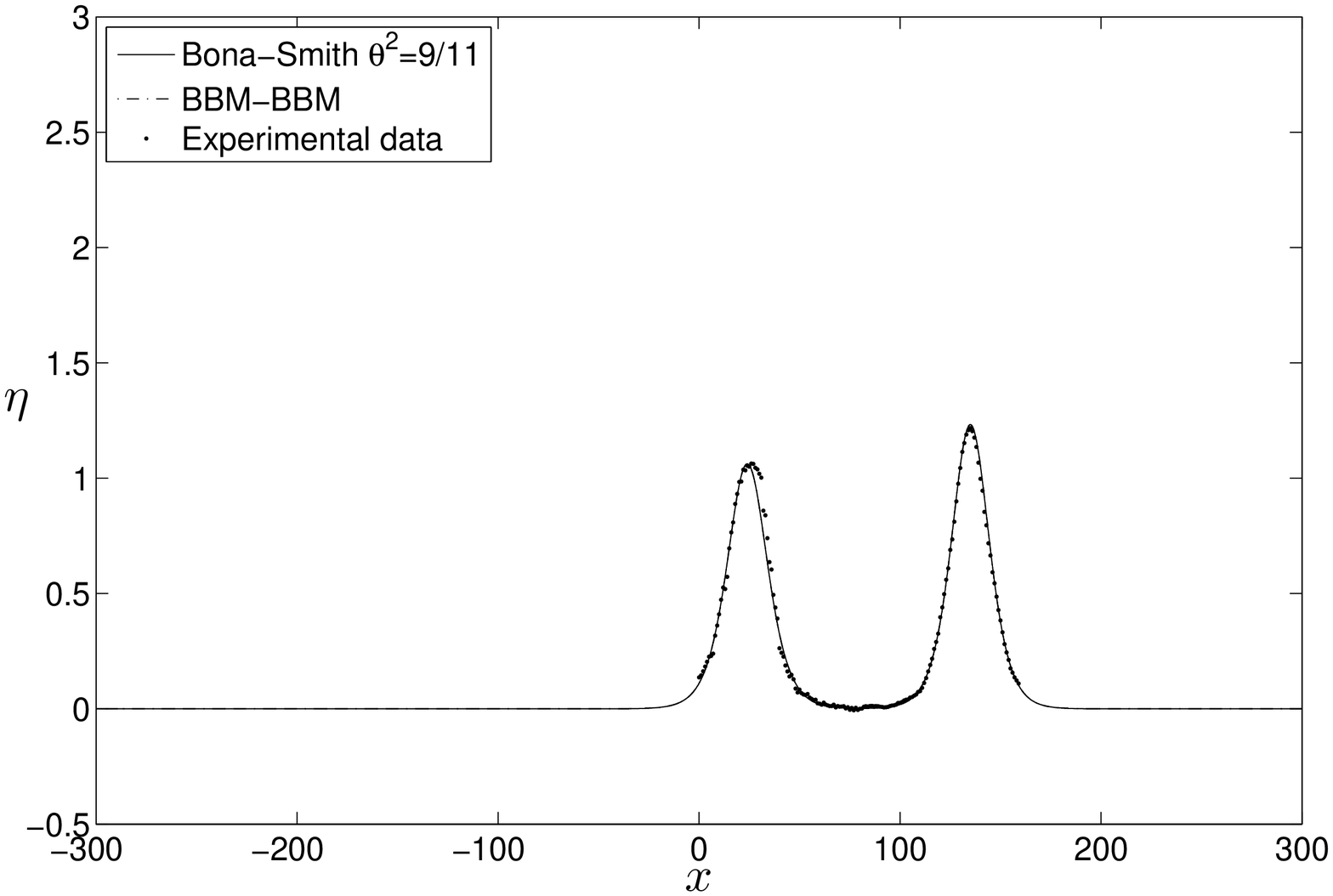}}
\subfigure[$t=18.80067 s$]{\includegraphics[scale=.33]{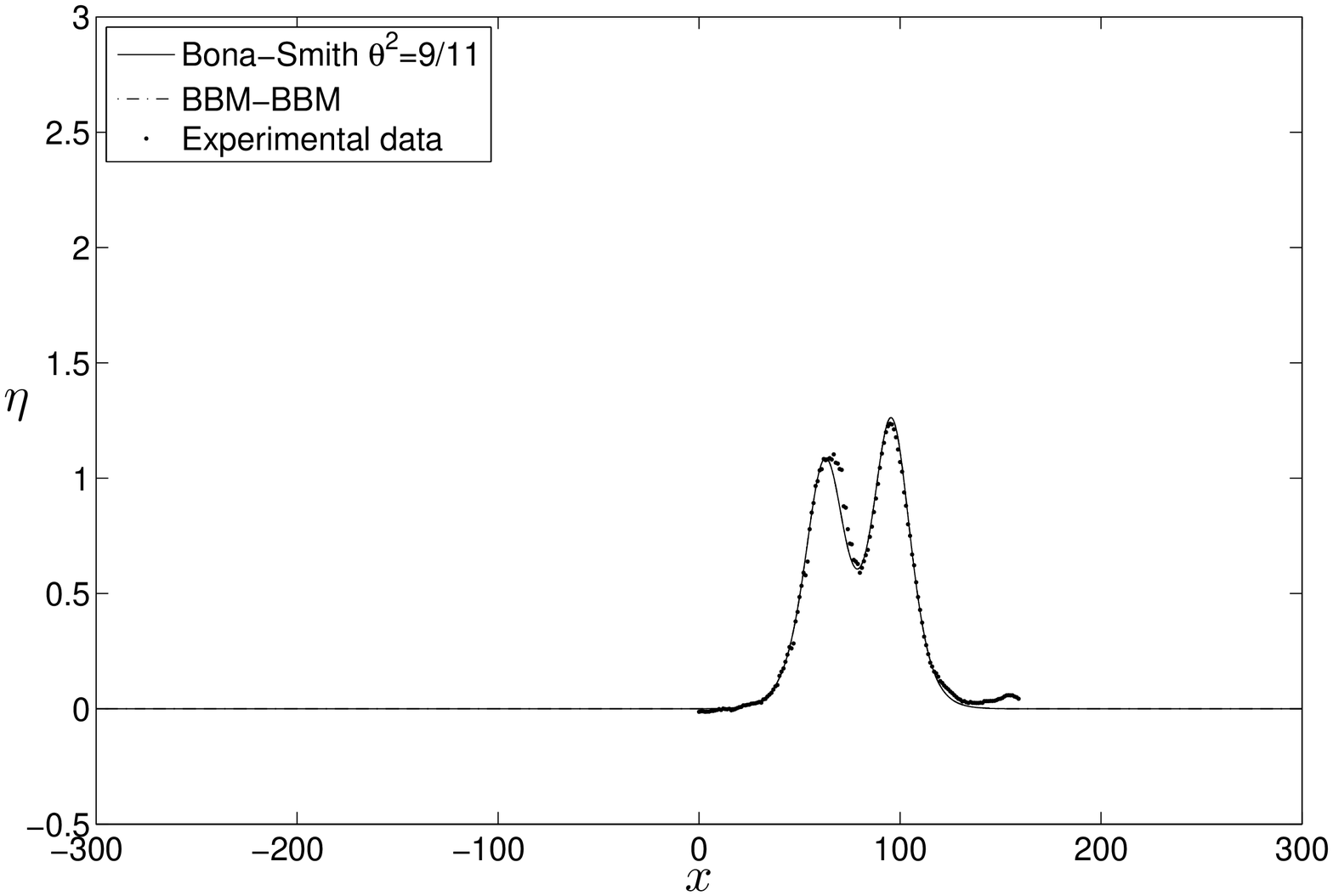}} 
\subfigure[$t=19.00956 s$]{ \includegraphics[scale=.33]{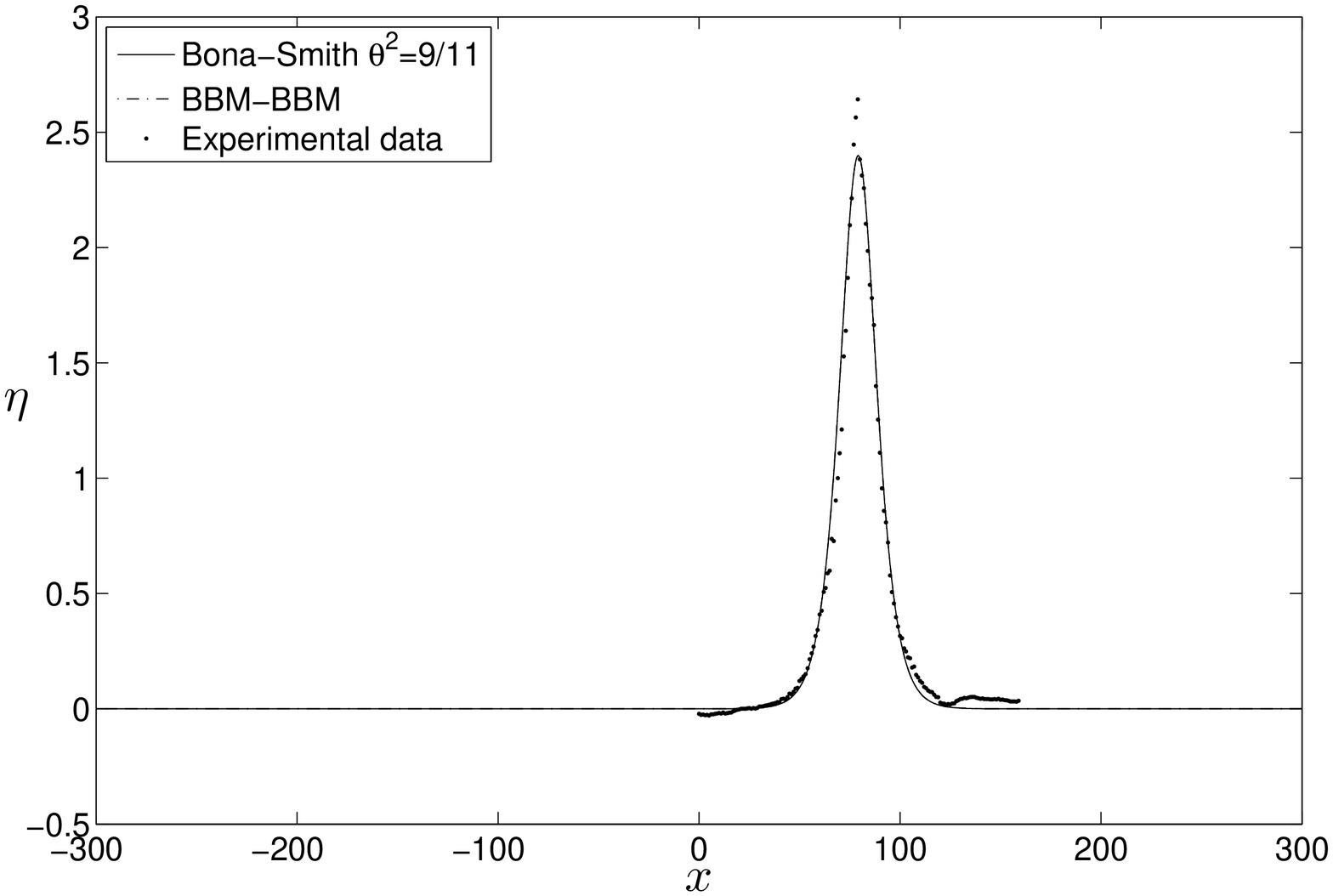}} 
\subfigure[$t=19.15087 s$]{\includegraphics[scale=.33]{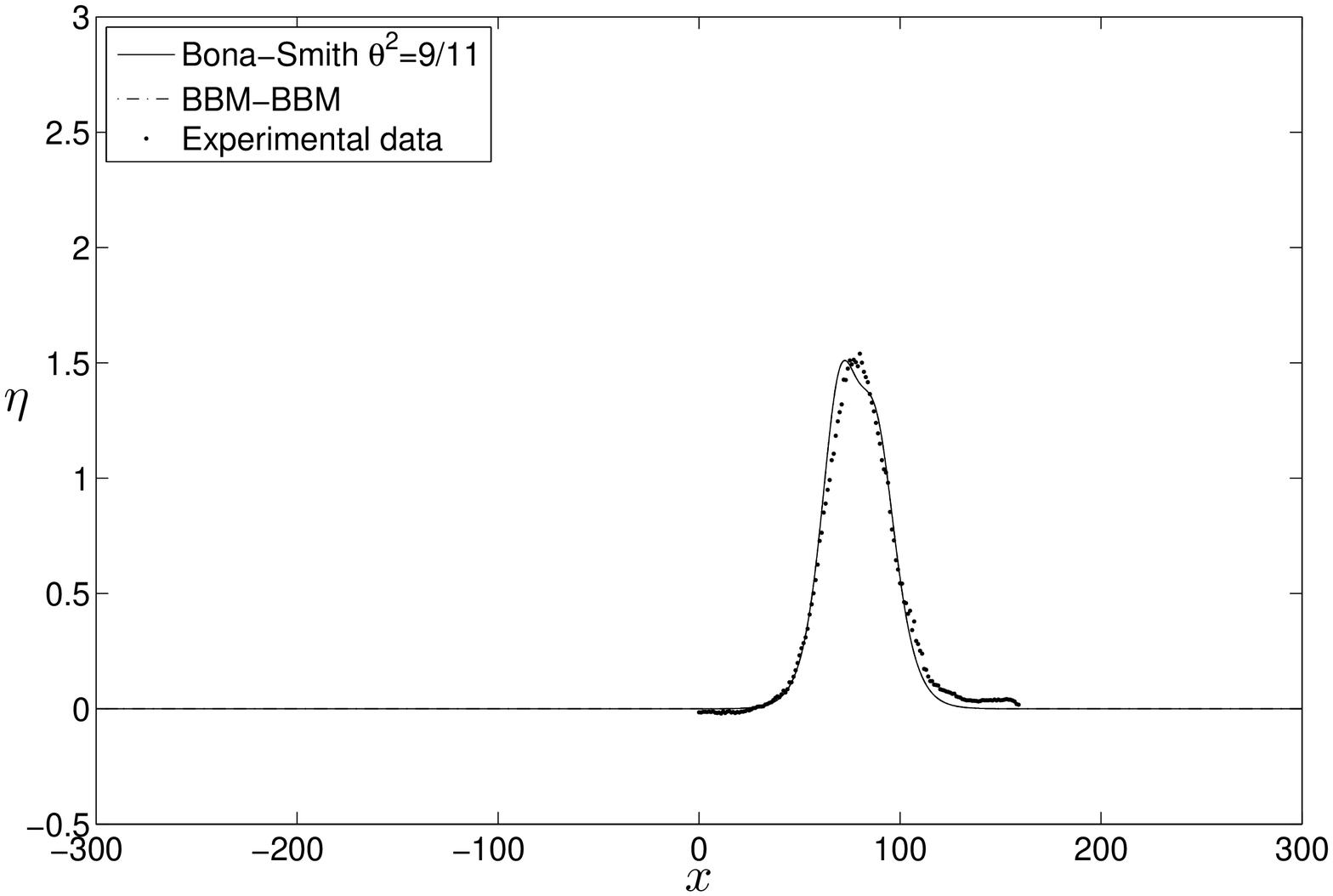}} 
\subfigure[$t=19.19388 s$]{\includegraphics[scale=.33]{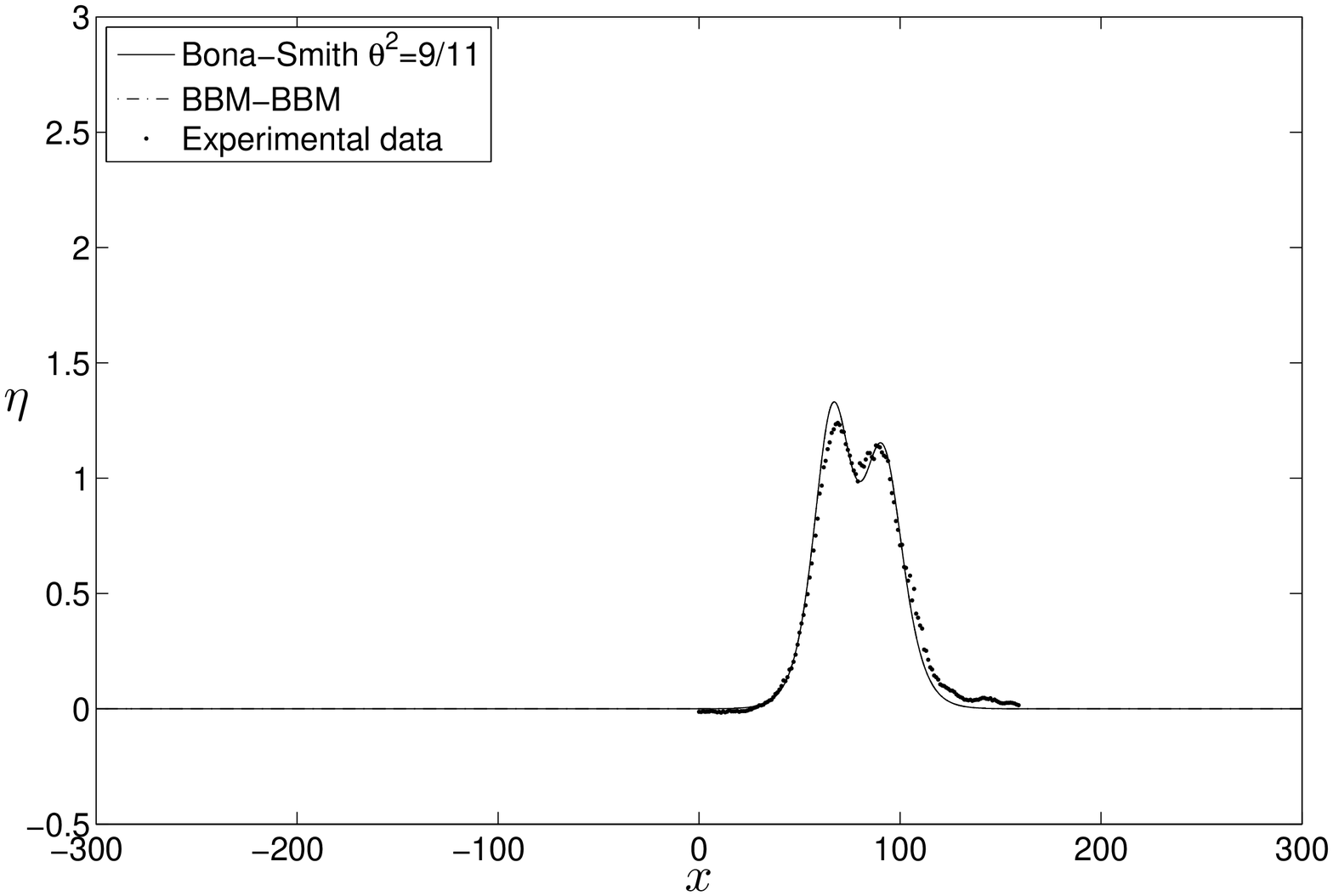}} 
\subfigure[$t=19.32904 s$]{\includegraphics[scale=.33]{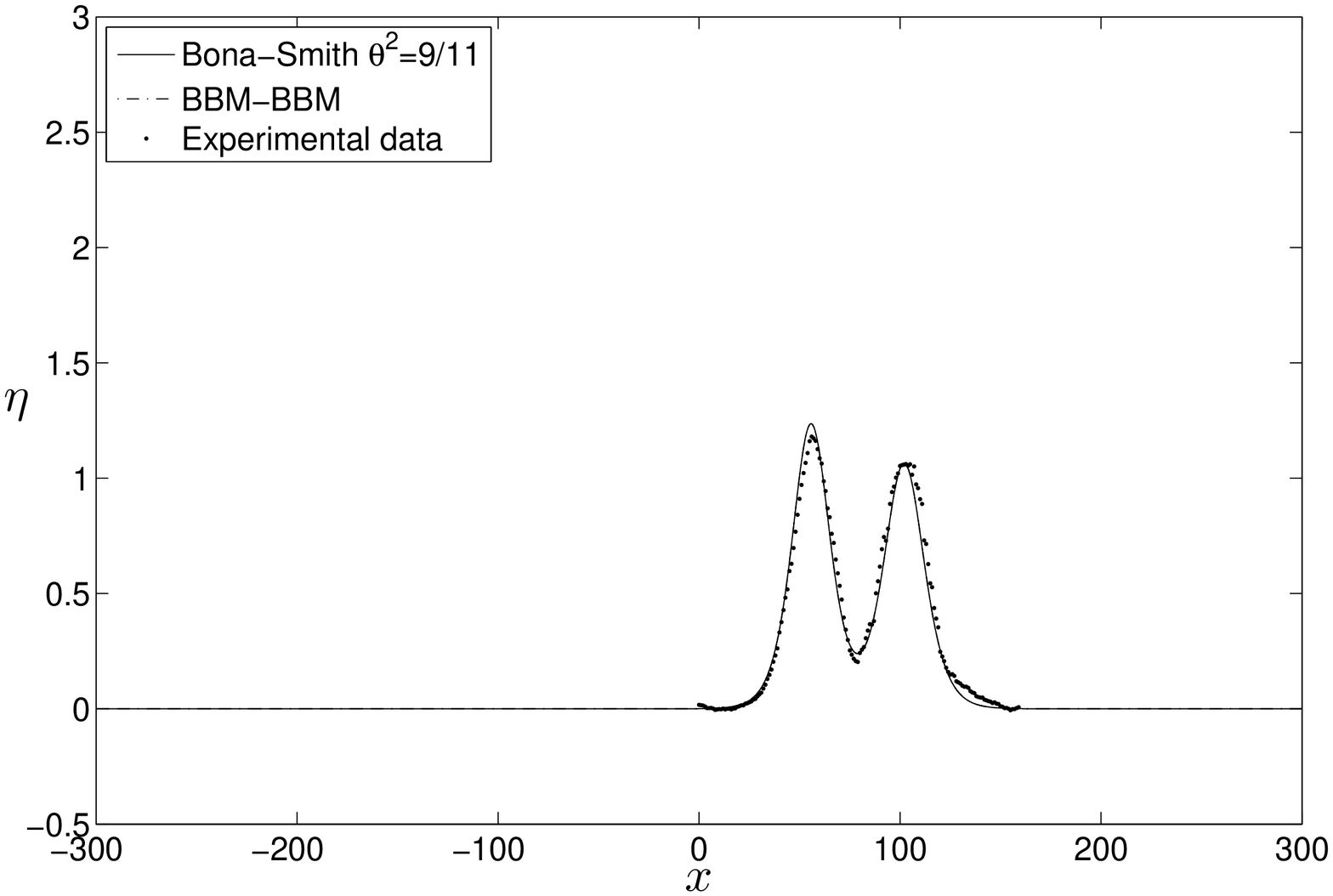}} 
\caption{Head-on collision of two solitary waves: ---: BBM-BBM, $--$: Bona-Smith ($\theta^2=9/11$), \textbullet: experimental data of \cite{CGHHS}}% 
\label{F15a}% 
\end{figure}

%%%%%%%%%%%%%%%%%%%%%%%%%%%%%%%%%%%%%%

\begin{figure}%
\ContinuedFloat
\centering
\subfigure[$t=19.84514 s$]{\includegraphics[scale=.33]{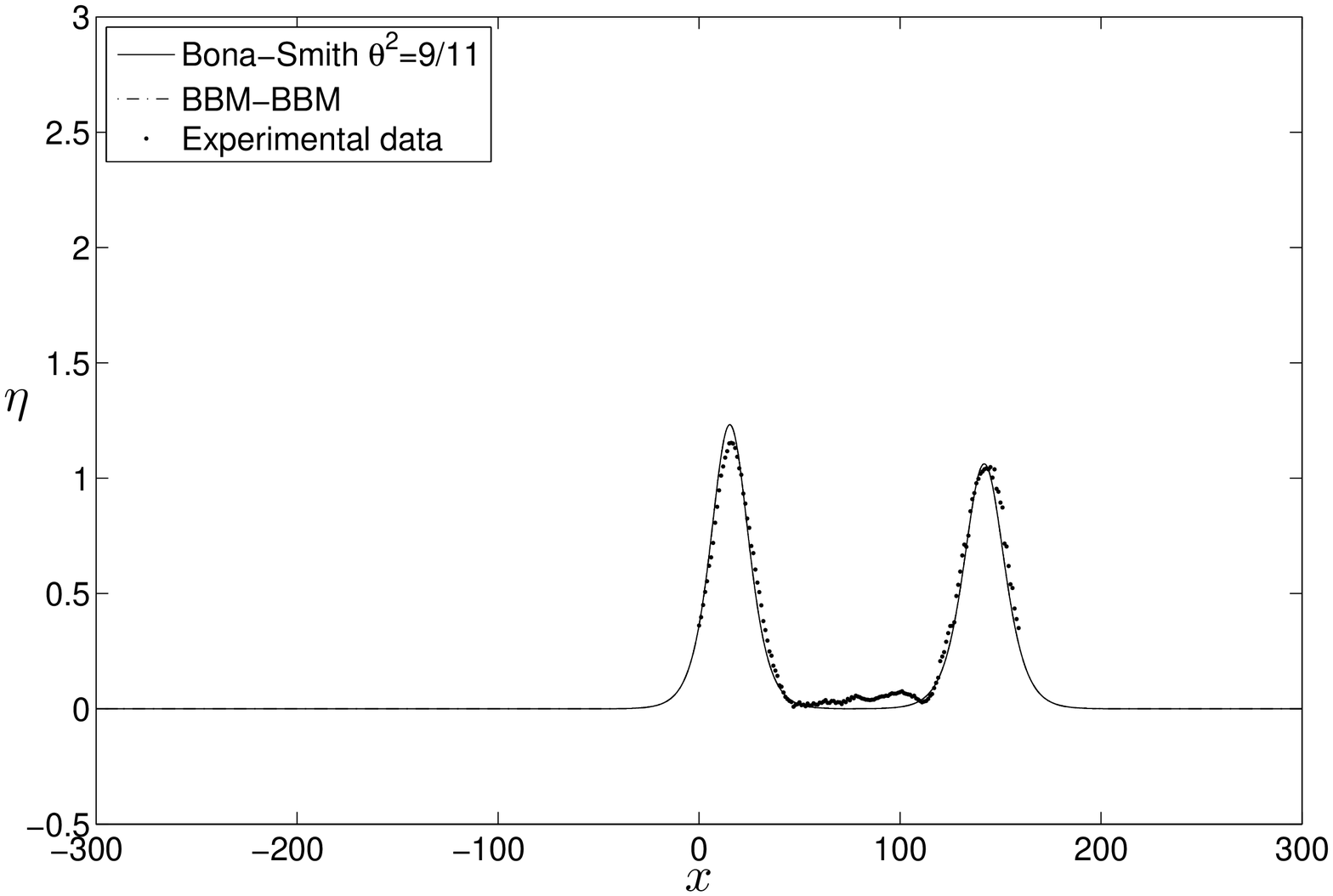}}
\subfigure[$t=20.49949 s$]{\includegraphics[scale=.33]{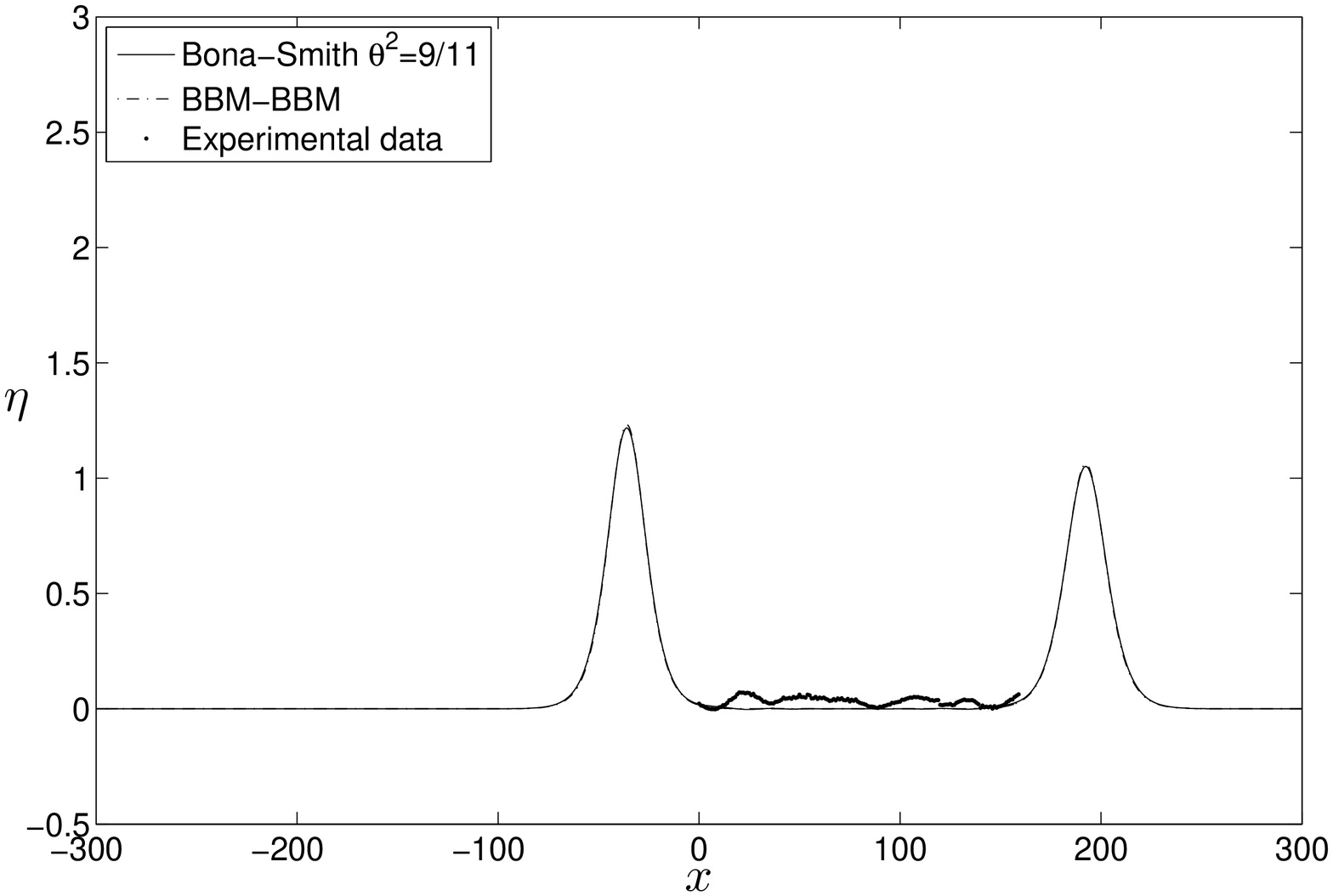}}
\caption{(Cont'd) Head-on collision of two solitary waves. ---: BBM-BBM, $--$: Bona-Smith ($\theta^2=9/11$), \textbullet: experimental data of \cite{CGHHS}}%
\label{F15b}%
\end{figure}

We observe that Boussinesq models converge to the same numerical solution with all numerical schemes we tested. A very good agreement with the experimental data is observed. The maximum height predicted by the numerical solution during the collision process is slightly higher in the case of the BBM-BBM system but the difference is negligible within the specific experimental scale. Furthermore, we observe similar underestimation of the maximum amplitude of the colliding waves compared to the experimental data, \cite{CGHHS}. This discrepancy might be explained by a possible "splash" phenomenon during the collision reported also earlier by T. Maxworthy, \cite{Maxworthy1976}. After the collision we observe that the phase shift of the solitary waves is the same in both numerical and experimental data, while the shape of the experimental solitary waves were not stabilized due to interactions with other small amplitude dispersive waves. We note that after the head-on collision of the waves small amplitude dispersive tails were developed, \cite{BC, ADM2, DDMM}.

The discrete mass for the Bona-Smith system is $I_0^h = 0.0059904310418$  and for the BBM-BMM system is $I_0^h = 0.0059199389479$ for all fluxes and reconstructions used. The variances in $I_1^h$ are mainly due to different types of reconstruction and not to the choice of numerical fluxes. In Table \ref{TINV} these values are reported.

\begin{table}%
\centering
\subtable[Bona-Smith]{
\begin{tabular}{|c|l|} 
\toprule%
           &   $I_1^h$ \\ \midrule
m-flux     &   0.000944236 \\ \hline
UNO2       &   0.00094423 \\ \hline
TVD2       &   0.00094 \\ \hline
WENO3      &   0.00094423\\ 
\bottomrule%
\end{tabular}}
\subtable[BBM-BBM]{
\begin{tabular}{|c|l|} 
\toprule%
           &   $I_1^h$ \\ \midrule
m-flux     &   0.00092793 \\ \hline
UNO2       &   0.00092793 \\ \hline
TVD2       &   0.00092 \\ \hline
WENO3      &   0.00092793\\ 
\bottomrule
\end{tabular}}
\caption{Preservation of the invariant $I_1^h$.}
\label{TINV}
\end{table}

\subsection{Overtaking collisions}

The overtaking collision of two solitary waves similarly to the head-on collision  incorporates nonlinear and dispersive effects. Overtaking collision has been studied recently in the case of bidirectional models in \cite{ADM2}. The interaction is similar to that of the unidirectional models but it was found that a new N-shape wavelet is generated during the interaction. This wavelet is of small amplitude and travels in the opposite direction to solitary waves and its shape depends on the Boussinesq system in use. Furthermore, as it was observed numerically and experimentally in \cite{CGHHS}, the interaction of two solitary waves during an overtaking collision is characterized by a mass exchange and not by a simple superposition of the solitary pulses. These pulses remain separate retaining two different maxima contrary to unidirectional models where they merge into a single pulse momentarily.

To study this interaction we solve numerically the Bona-Smith system (\ref{E1.8}) with $\theta^2 = 9/11$. Following the same process as before two solitary waves  were generated numerically  with speeds $c_{1,s}=1.2$ and $c_{2,s}=1.4$. We solved the system using all fluxes using UNO2 and WENO3 reconstructions with discretization parameters $\dx=0.01, \ \dt=0.005$ up to $T=600$.  During simulations we were able to observe the generation and propagation of a small N-shape wavelet. In all computations the invariants were $I_0^h=4.6098804880$, $I_1^h=5.116$ conserving the digits shown for all methods. 

%%%%%%%%%%%%%%%%%%%%%%%%%%%%%%%%%%%%%%

\begin{figure}%
\centering
\subfigure[$t=0$]{\includegraphics[scale=.33]{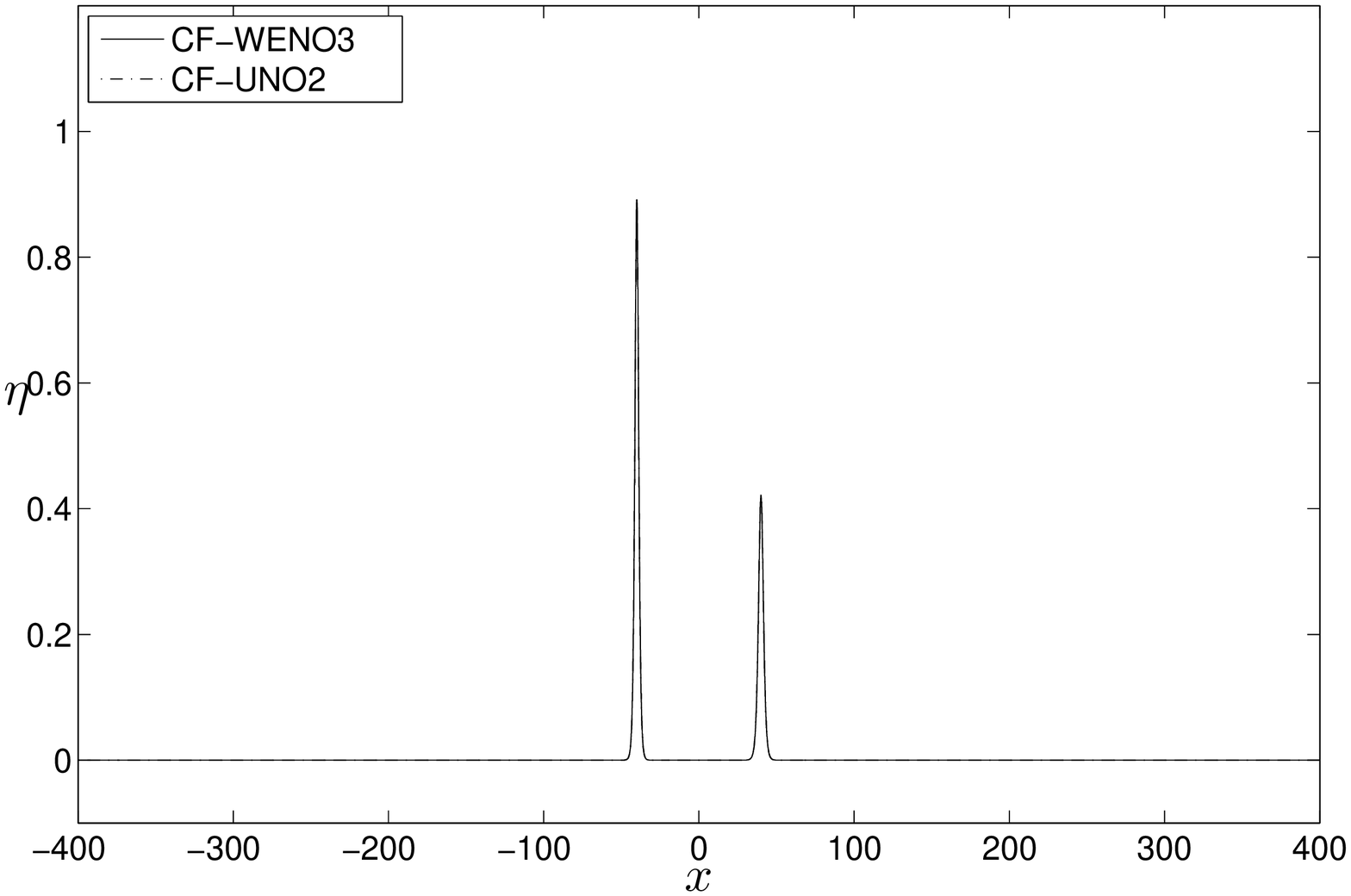}}
\subfigure[$t=350$]{\includegraphics[scale=.33]{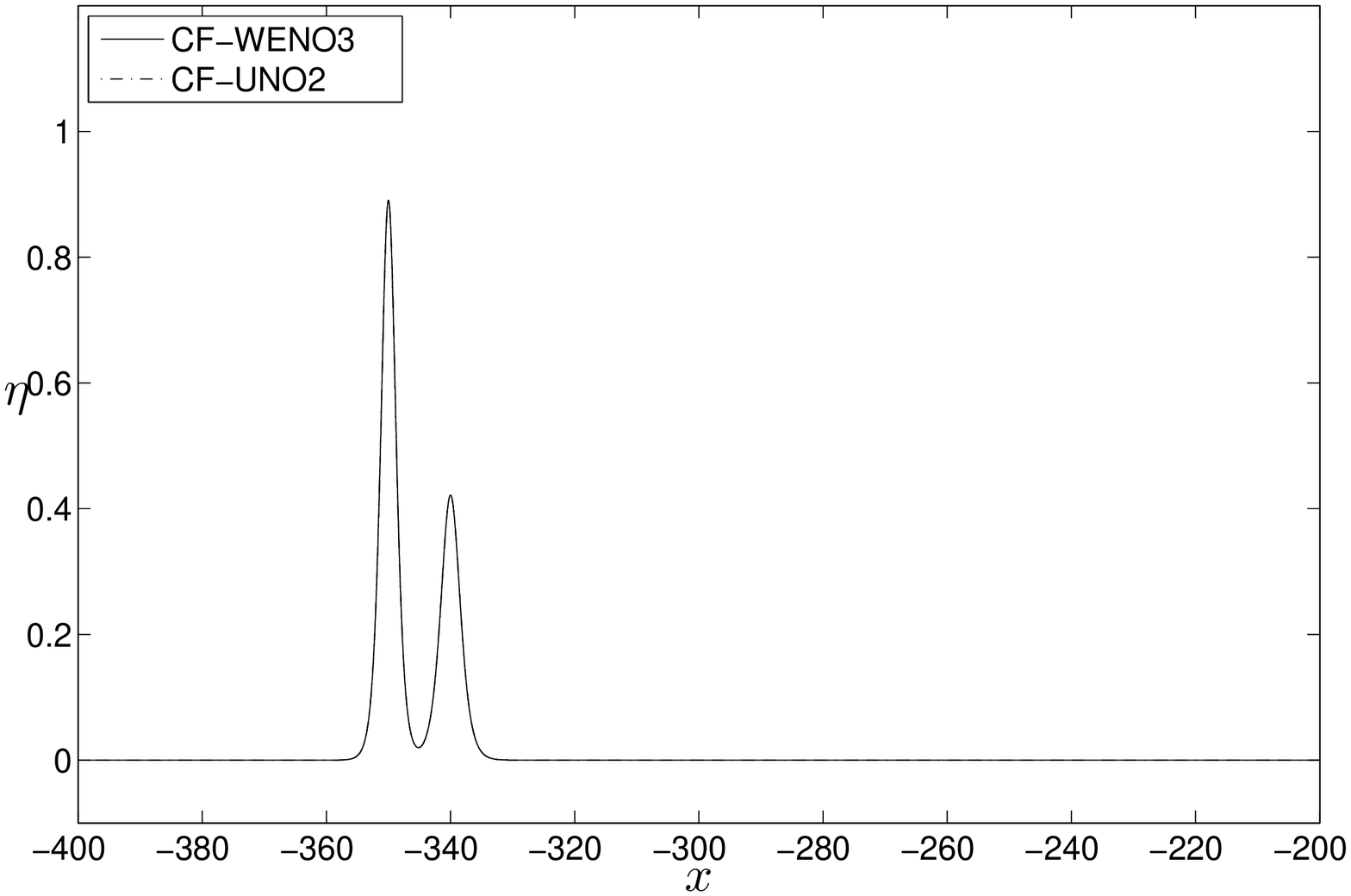}}
\subfigure[$t=400$]{\includegraphics[scale=.33]{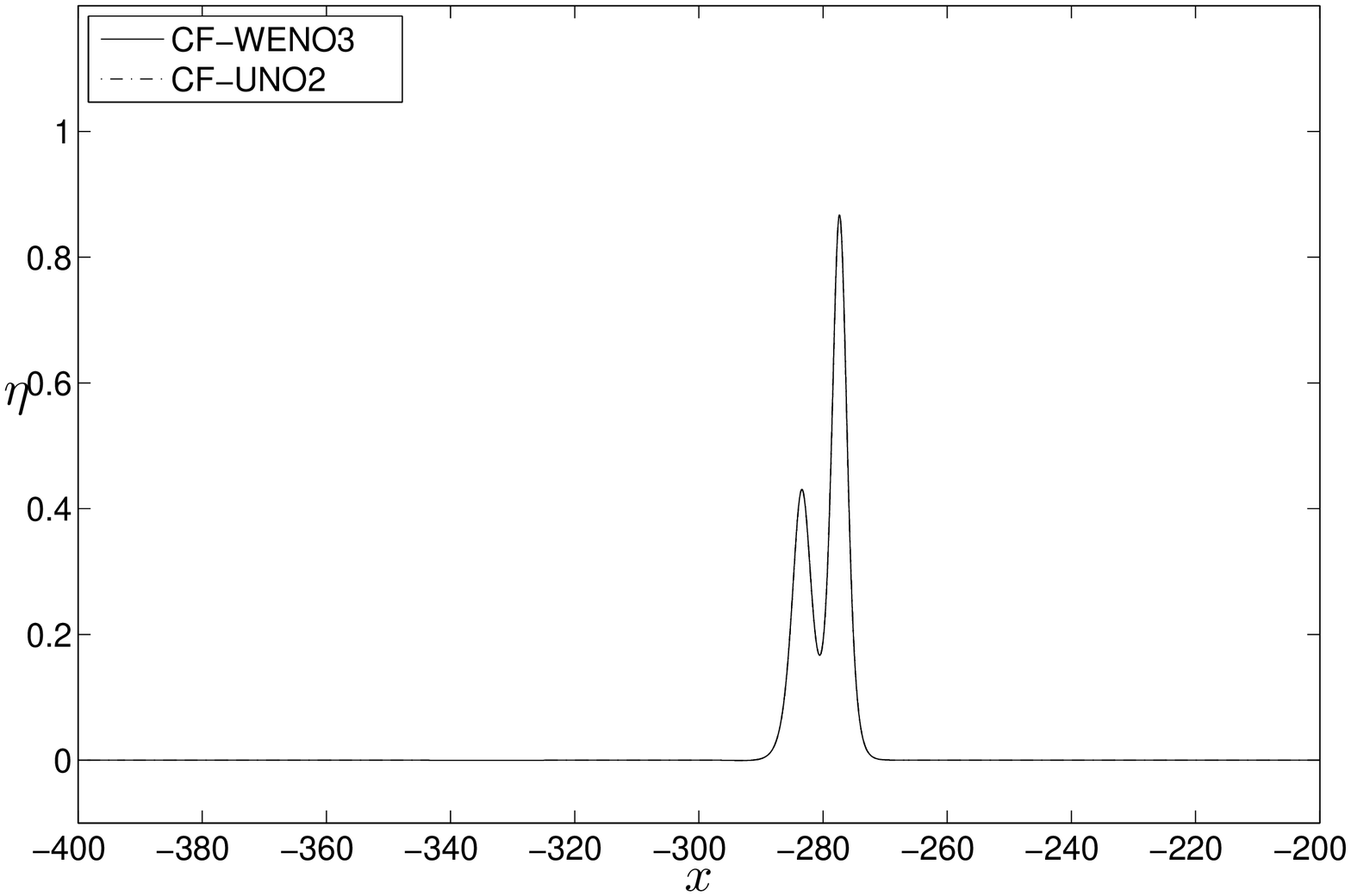}}
\subfigure[$t=600$]{\includegraphics[scale=.33]{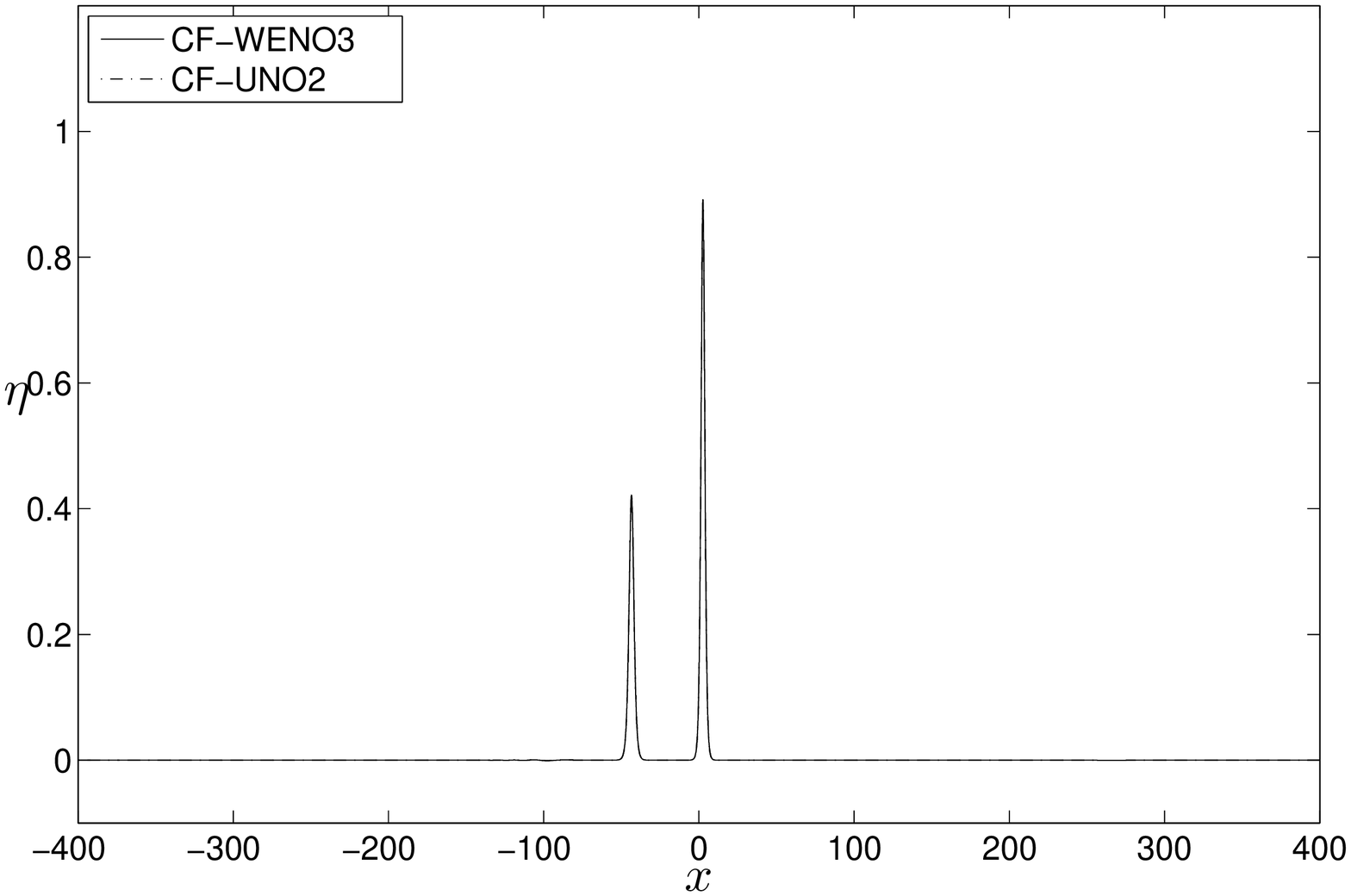}}
\caption{Overtaking collision of two solitary waves of the Bona-Smith system with $\theta^2=9/11$.}%
\label{F18}%
\end{figure}

%%%%%%%%%%%%%%%%%%%%%%%%%%%%%%%%%%%%%%
\begin{figure}%
\centering
\subfigure[$t=375$]{\includegraphics[scale=.33]{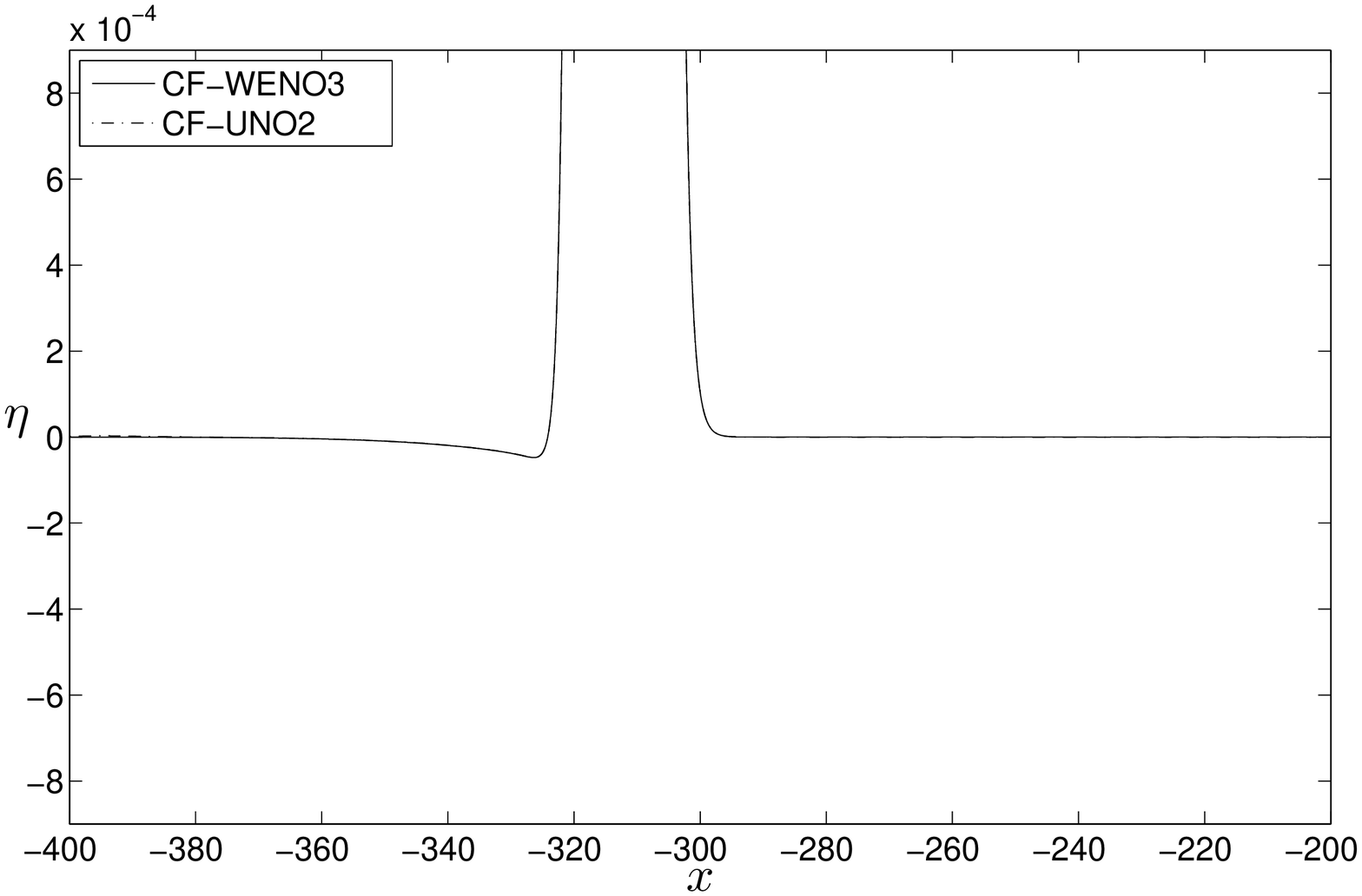}}
\subfigure[$t=400$]{\includegraphics[scale=.33]{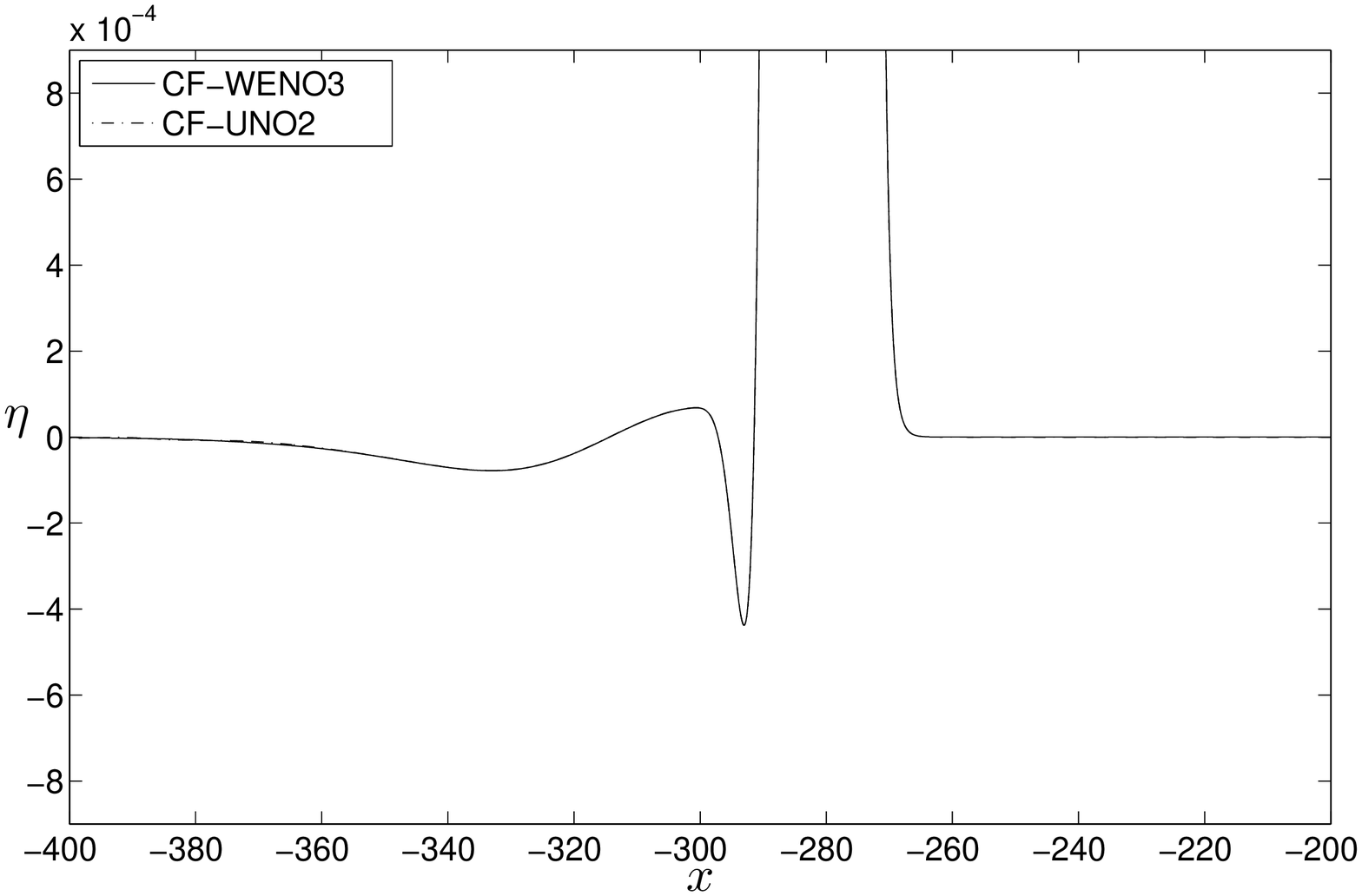}}
\subfigure[$t=425$]{\includegraphics[scale=.33]{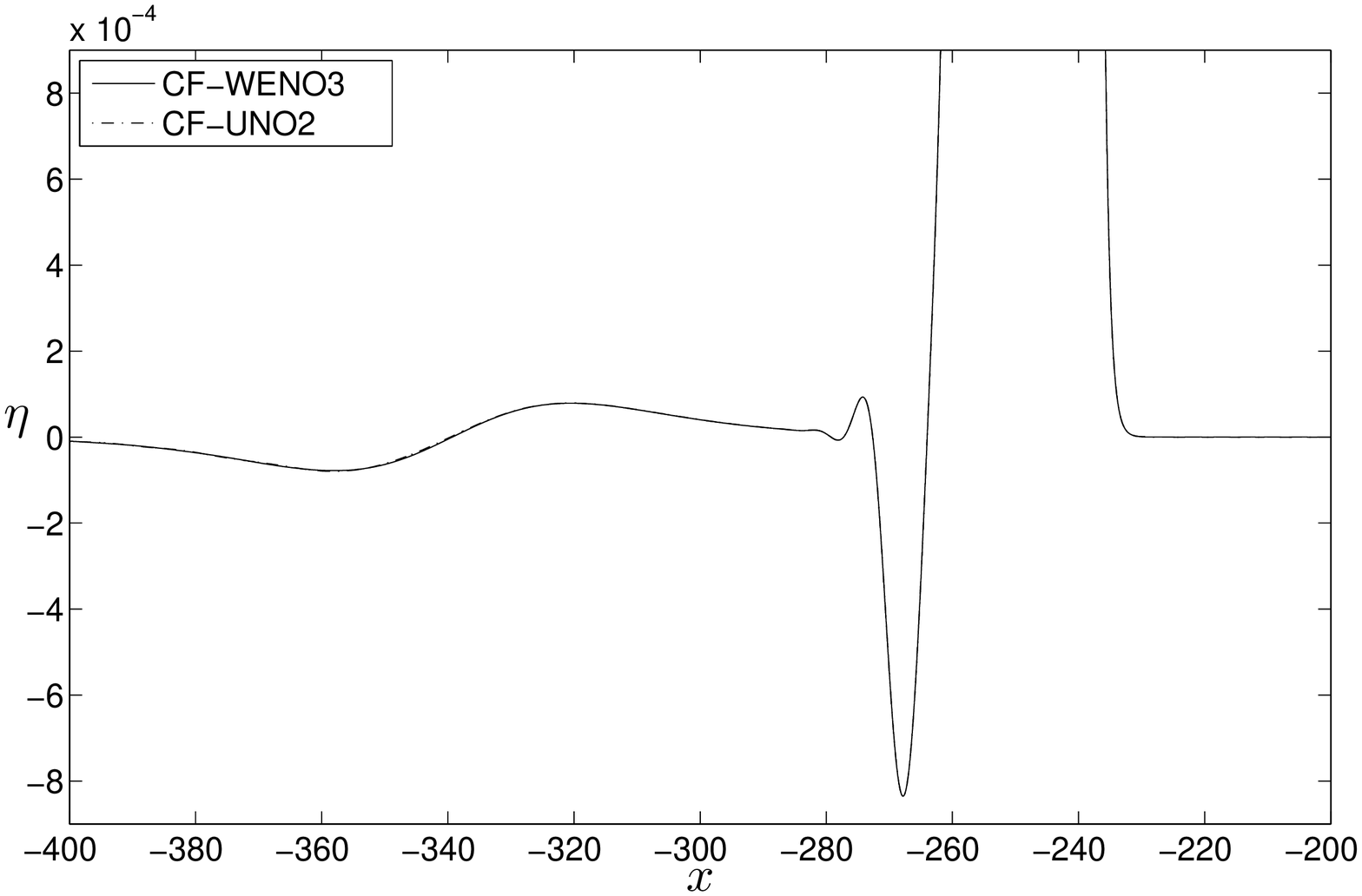}}
\subfigure[$t=600$]{\includegraphics[scale=.33]{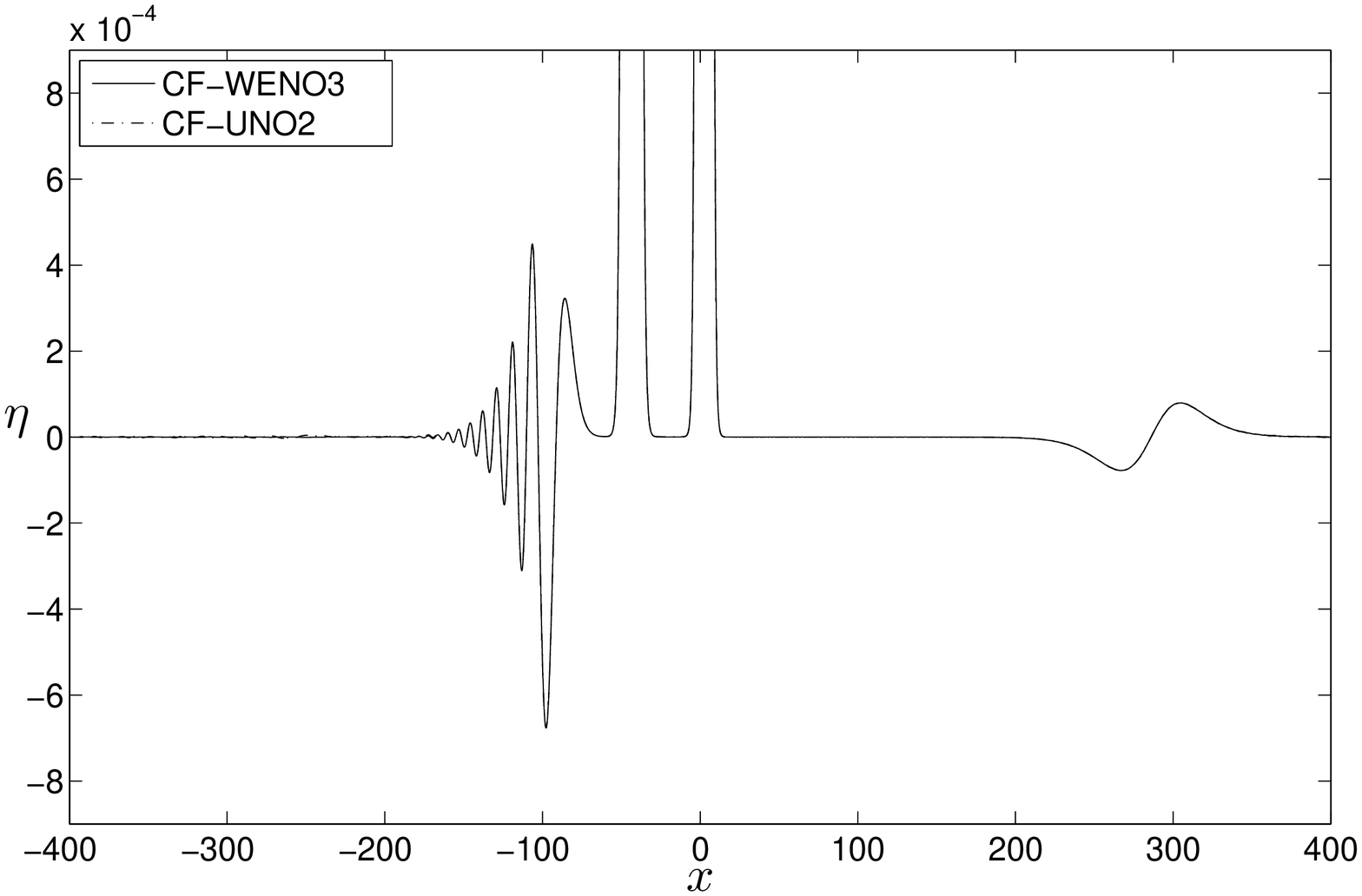}}
\caption{Generation of a wavelet during the overtaking collision of two solitary waves of the Bona-Smith system with $\theta^2=9/11$.}%
\label{F19}%
\end{figure}
%%%%%%%%%%%%%%%%%%%%%%%%%%%%%%%%%%%%%%
\begin{figure}%
\centering
\subfigure[$t=382.5$]{\includegraphics[scale=.333]{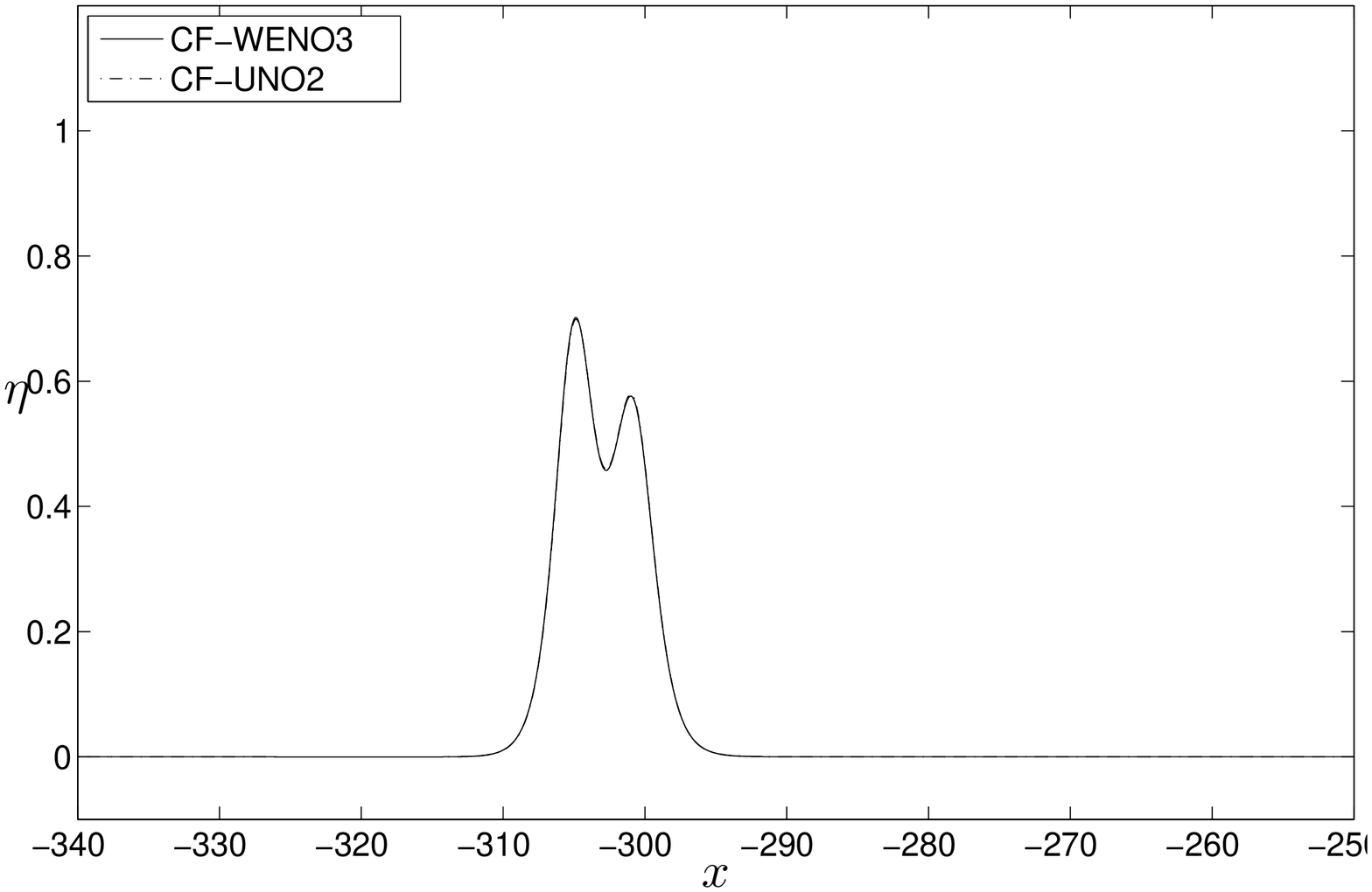}} 
\subfigure[$t=384$]{\includegraphics[scale=.333]{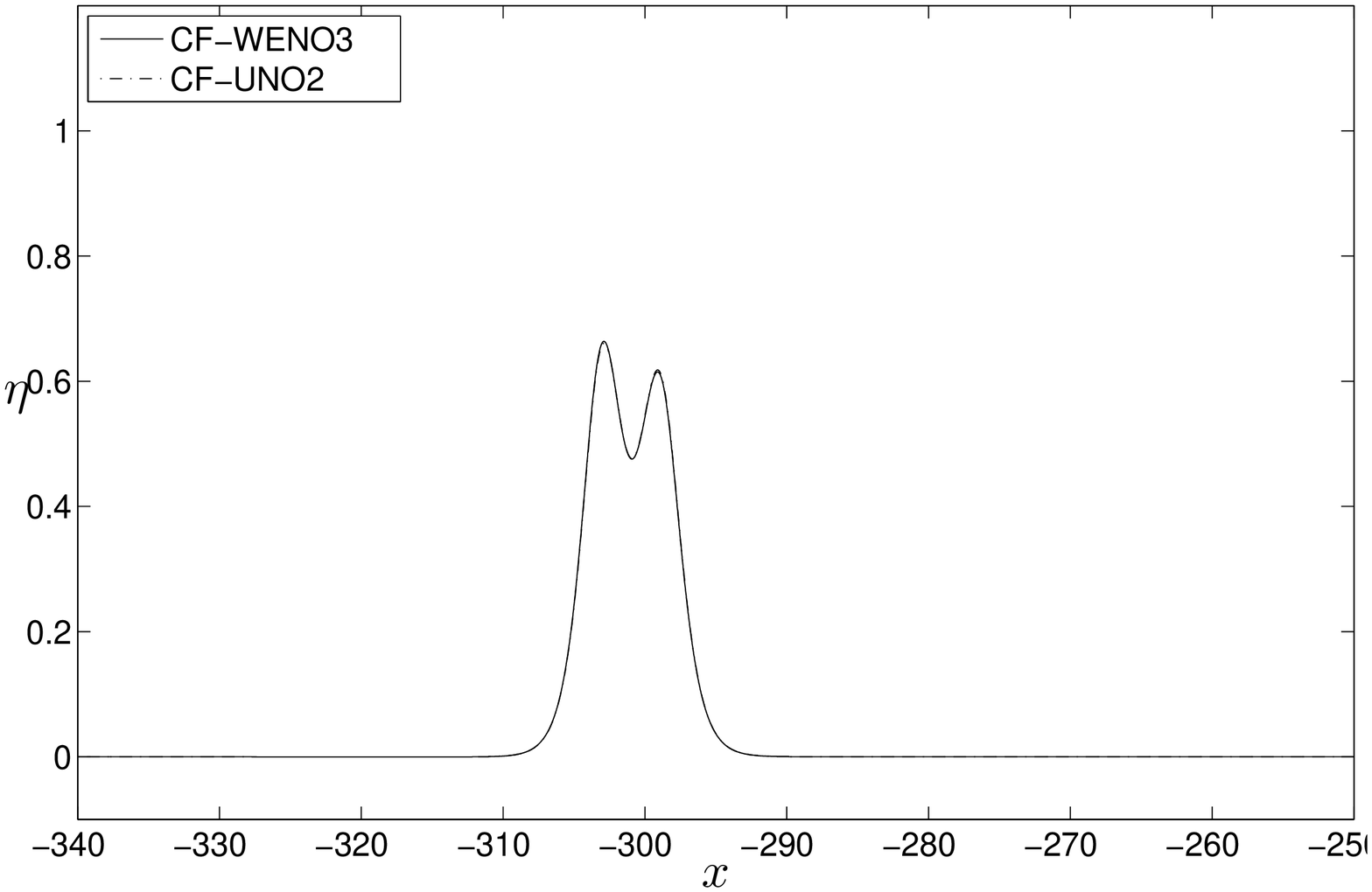}}
\subfigure[$t=385$]{\includegraphics[scale=.333]{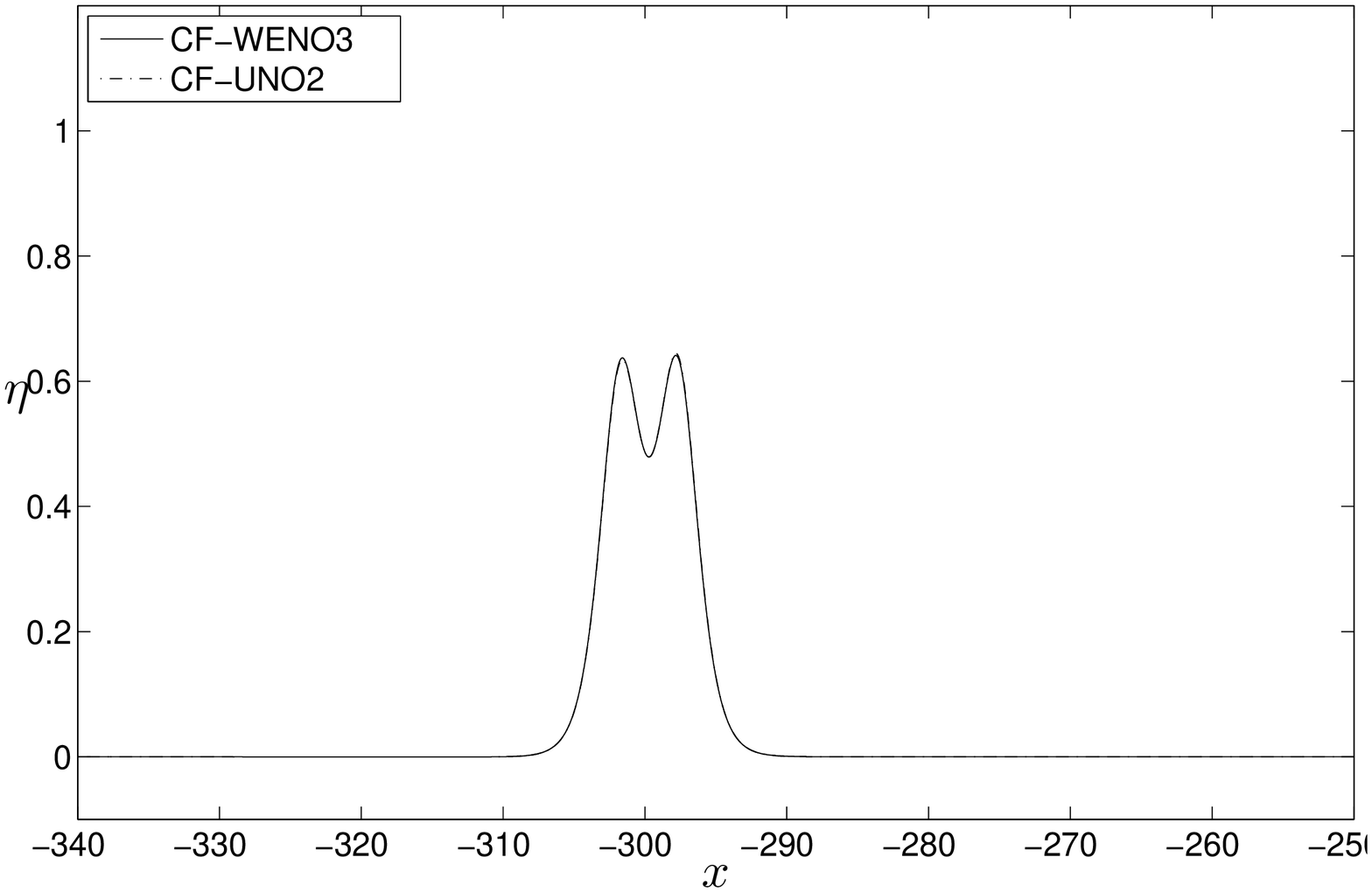}}
\subfigure[$t=386$]{\includegraphics[scale=.333]{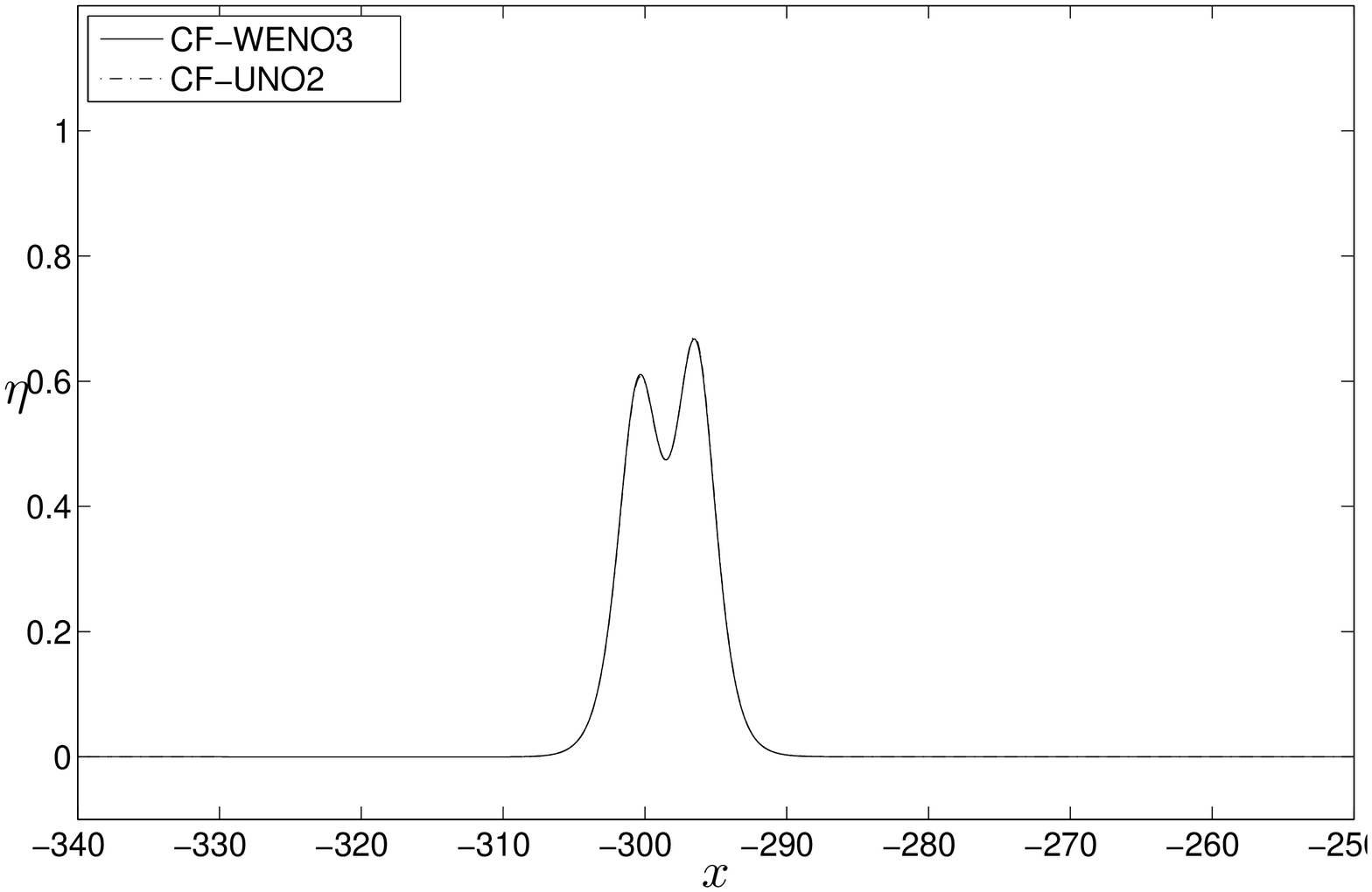}}
\caption{Overtaking collision of two solitary waves of the Bona-Smith system with $\theta^2=9/11$: mass exchange process.}%
\label{F19.1}%
\end{figure}
%%%%%%%%%%%%%%%%%%%%%%%%%%%%%%%%%%%%%%

In Figure \ref{F18} we present the interaction of two solitary waves. Figure \ref{F19} shows a magnification on the generation of a small wavelet along with the generation of dispersive tails as an effect of the inelastic interaction of two waves. In Figure \ref{F19.1} we observe that the overtaking collision is accompanied by an exchange of mass between pulses while both peaks are permanently present. The situation is different for unidirectional models where two pulses merge during a few time-steps to travel as a single pulse. Up to the graphic resolution we could not observe any difference in numerical solutions between UNO2 and WENO3 reconstructions.

%%%%%%%%%%%%%%%%%%%%%%%%%%%%%%%%%%%%%%

\subsection{Small dispersion effect}

\begin{figure}%
\centering
\subfigure[CF-UNO2]{\includegraphics[scale=.333]{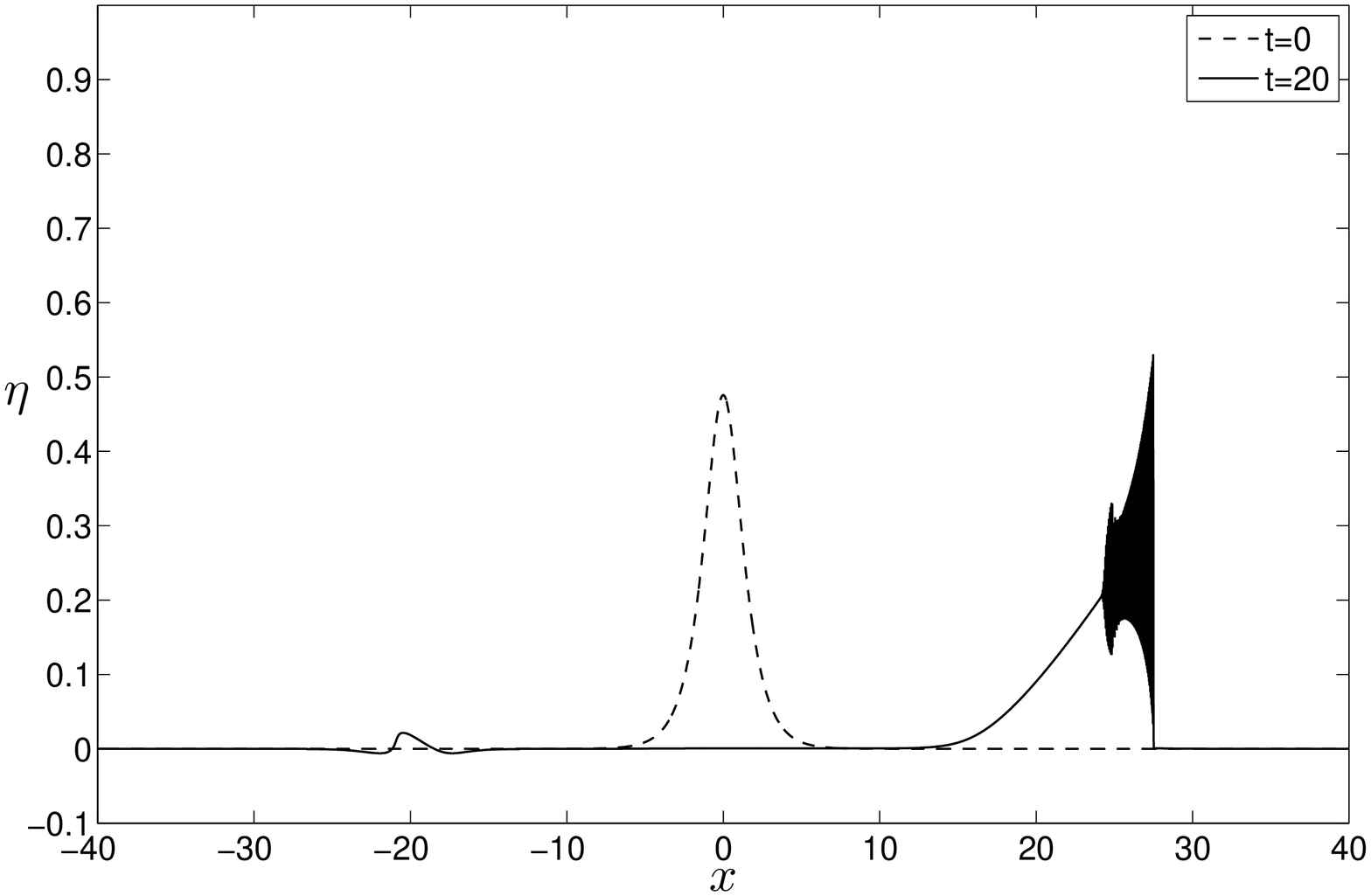}}
\subfigure[CF-WENO3]{\includegraphics[scale=.333]{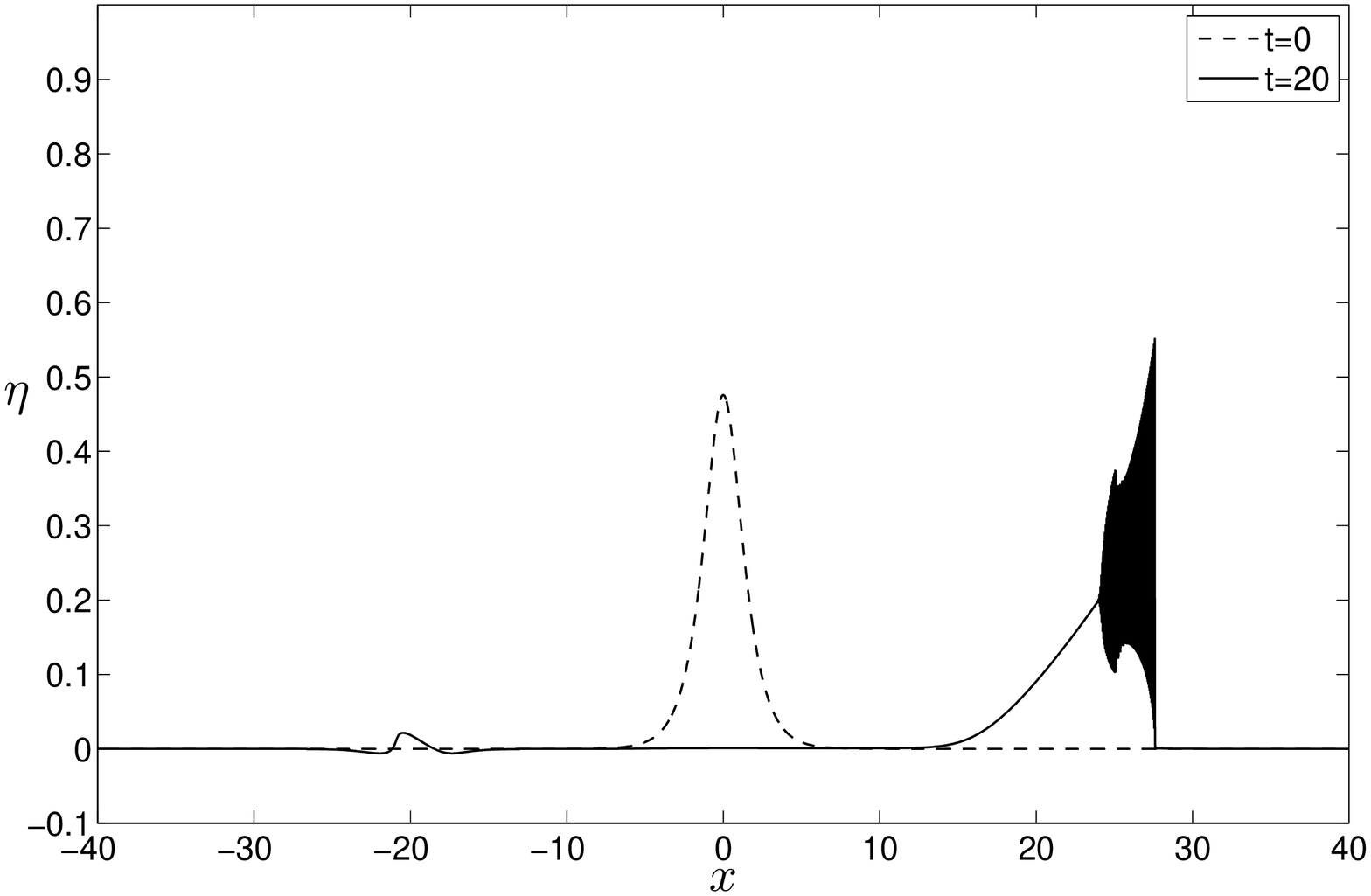}}
\subfigure[CF-UNO2]{\includegraphics[scale=.333]{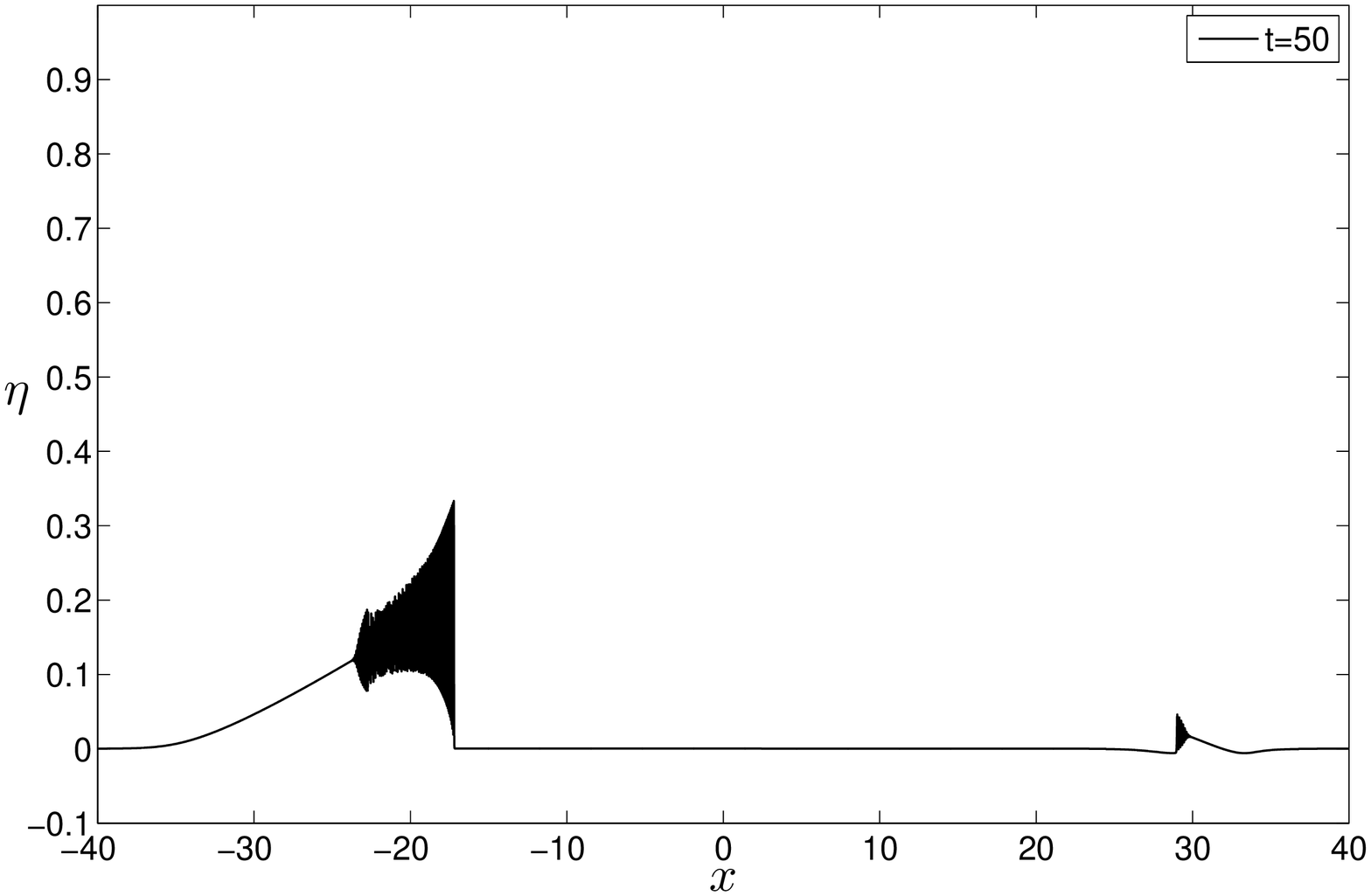}}
\subfigure[CF-WENO3]{\includegraphics[scale=.333]{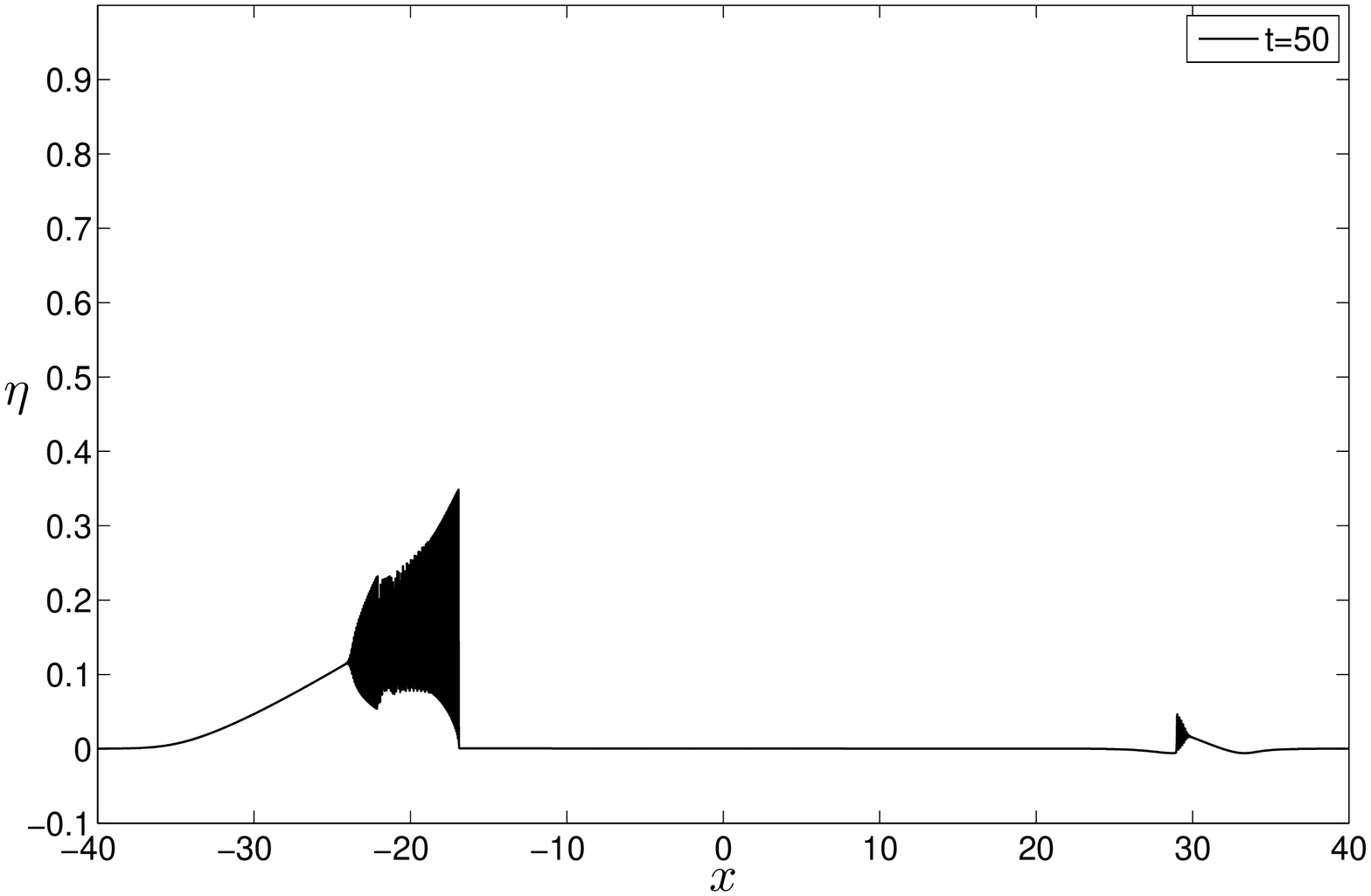}}
\caption{The small dispersion effect onto classical Boussinesq equations solutions.}%
\label{F20}%
\end{figure}

In this section we study the small dispersion effects on solitary waves of the classical Boussinesq system. The motivation for this study is the lack of theory supporting the breaking phenomena in Boussinesq systems contrary to the KdV equation. For this reason we employ the Boussinesq system with $a=b=c=0$, $d=10^{-5}$ and we take the solitary wave of the Boussinesq systems (\ref{E1.6}) as an initial condition. In Figure \ref{F20} we present numerical results obtained with CF-UNO2 and CF-WENO3 schemes. In these experiments we take $\dx=0.001$ and $\dt=\dx/2$. The invariant $I_0^h$ is $1.629096452537$ preserving the digits shown during all simulations. The invariant $I_1^h$ is not preserved by this model since the coefficient $b$ is not equal to $d$. The oscillations generated in the case of the WENO3 reconstruction were larger compared to those generated by the UNO2 reconstruction. Moreover, a new W-shaped wavelet is generated traveling to the left. This small wavelet finishes by producing a secondary breaking very similar to that of the initial solitary wave.

%%%%%%%%%%%%%%%%%%%%%%%%%%%%%

%%%%%%%%%%%%%%%%%%%%%%%%%%%%%%%%%%%%%%%%%%%%%%%
%%%%% SECTION
%%%%%%%%%%%%%%%%%%%%%%%%%%%%%%%%%%%%%%%%%%%%%%%

\section{Boussinesq system with variable bottom: runup of long waves}\label{sec:bottom}

\begin{figure}%
% Generated with LaTeXDraw 2.0.2
% Thu Jan 14 13:16:19 CET 2010
% \usepackage[usenames,dvipsnames]{pstricks}
% \usepackage{epsfig}
% \usepackage{pst-grad} % For gradients
% \usepackage{pst-plot} % For axes
\scalebox{0.8} % Change this value to rescale the drawing.
{
\begin{pspicture}(0,-3.03)(16.64,3.03)
\psline[linewidth=0.04](0.2,3.01)(5.4,-2.99)(16.6,-2.99)(16.6,-2.99)(16.62,-3.01)
\psline[linewidth=0.04cm,linestyle=dashed,dash=0.16cm 0.16cm](5.4,-2.99)(0.2,-2.99)
\pscustom[linewidth=0.04]
{
\newpath
\moveto(2.4,0.41)
\lineto(2.9,0.41)
\curveto(3.15,0.41)(3.65,0.41)(3.9,0.41)
\curveto(4.15,0.41)(4.65,0.41)(4.9,0.41)
\curveto(5.15,0.41)(5.6,0.46)(5.8,0.51)
\curveto(6.0,0.56)(6.45,0.61)(6.7,0.61)
\curveto(6.95,0.61)(7.4,0.66)(7.6,0.71)
\curveto(7.8,0.76)(8.2,0.86)(8.4,0.91)
\curveto(8.6,0.96)(9.05,1.06)(9.3,1.11)
\curveto(9.55,1.16)(10.0,1.21)(10.2,1.21)
\curveto(10.4,1.21)(10.85,1.16)(11.1,1.11)
\curveto(11.35,1.06)(11.8,0.91)(12.0,0.81)
\curveto(12.2,0.71)(12.6,0.51)(12.8,0.41)
\curveto(13.0,0.31)(13.45,0.16)(13.7,0.11)
\curveto(13.95,0.06)(14.4,0.06)(14.6,0.11)
\curveto(14.8,0.16)(15.25,0.21)(15.5,0.21)
\curveto(15.75,0.21)(16.2,0.26)(16.4,0.31)
}
\psline[linewidth=0.04cm,linestyle=dotted,dotsep=0.16cm](0.0,0.41)(16.6,0.41)
\usefont{T1}{ptm}{m}{n}
\rput(4.711406,-2.68){$\beta$}
\psline[linewidth=0.04cm,linestyle=dashed,dash=0.16cm 0.16cm,arrowsize=0.05291667cm 2.0,arrowlength=1.4,arrowinset=0.4]{<->}(9.0,0.41)(9.0,-2.99)
\psline[linewidth=0.04cm,linestyle=dashed,dash=0.16cm 0.16cm,arrowsize=0.05291667cm 2.0,arrowlength=1.4,arrowinset=0.4]{<->}(10.4,1.21)(10.4,-2.99)
\usefont{T1}{ptm}{m}{n}
\rput(8.291407,-0.68){$D$}
\usefont{T1}{ptm}{m}{n}
\rput(11.091406,-0.48){$H$}
\usefont{T1}{ptm}{m}{n}
\rput(2.8514063,0.72){$0$}
\end{pspicture} 
}
\caption{Sketch of the problem setup.}\label{F6.1}
\end{figure}
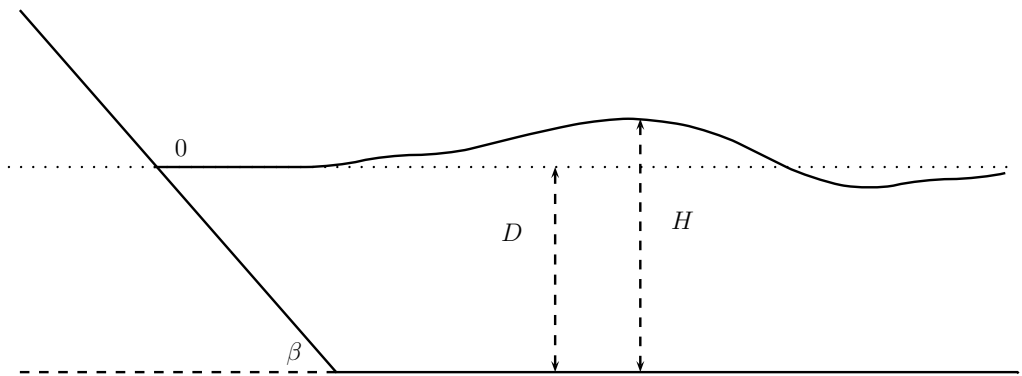

The shallow water equations are routinely used to predict a tsunami wave runup and, subsequently, constitute inundation maps for tsunami hazard areas. One of the main questions we address in this study is whether the inclusion of dispersive effects is beneficial for the description of the wave/shore interaction. In this section we perform a comparison of numerical solutions to Boussinesq equations \eqref{E.Per5}, shallow water equations  \eqref{eq:E1.1} (solved by the same numerical method) and experimental measurements made by C.E.~Synolakis \cite{Synolakis1987} and J.A.~Zelt \cite{Zelt1991}. In these experiments we consider a bottom of the form, Figure \ref{F6.1}, 
$$
-D(x)=\left\{\begin{array}{ll}-x\, \tan\beta,& x\leq \cot\beta, \\ -1, & x>\cot\beta,  \end{array} \right.
$$  

%%%%%%%%%%%%%%%%%%%%%%%%%%%%%%%%%%%%%%%%%%%%%%%

In all experiments  over a flat bottom, $D(x)=D_0$, we use an approximate solitary wave solution of the following form:
\begin{align}
\eta_0(x)&=A_s{\sech}^2\left(\lambda (x-X_0) \right), \quad u_0(x)=-c_s\frac{\eta_0(x)}{D_0+\eta_0(x)},\label{SW1} \\
\lambda&=\sqrt{\frac{3A_s}{4(1+A_s)}}, \quad c_s=\sqrt{g}\frac{\sqrt{6}(1+A_s)}{\sqrt{3+2A_s}}\cdot \frac{\sqrt{(1+A_s)\log (1+A_s)-A_s}}{A_s},  \label{SW2}
\end{align}
where $A_s$ denotes the amplitude, $c_s$ is the correct speed of the solitary wave propagation for classical Boussinesq equations.

The first three experiments we tested are described in \cite{Synolakis1987} and deal with the runup of solitary waves on a shore with a mild slope of $1:19.85$. The first is a non-breaking solitary wave with dimensionless and scaled amplitude $A_s/D_0=0.0185$, the second one is a nearly breaking solitary wave with $A_s/D_0=0.04$, while the third experimental setup is a breaking solitary wave with $A_s/D_0=0.28$.

System (\ref{E.Per5}) has some advantages over other asymptotically equivalent models with variable bottom. Namely, it shows excellent stability properties even for nearly breaking waves on the shore. However, for the simulation of strong breaking events, it is beneficial to include friction or dissipative terms taking into account turbulence generation.

We also considered two experiments from \cite{Zelt1991} concerning the runup of solitary waves on a shore with steep slope $1:2.74$. These experiments shed some light on the differences between dispersive and non-dispersive models. 

Finally we consider a non-uniform sloping shore  that contains a small pond demonstrating the capability of the modified Peregrine's system to handle simultaneously and correctly dispersive effects in two basins with different mean sea levels.

In the sequel $t$ denotes the dimensionless time scaled by the quantity $\sqrt{g/D_0}$. Furthermore, we denote by $R$ the height of the last dry cell at a specific time instance. In our computations a cell is considered as dry if the total water depth $H_i$ inside is less than $5\cdot 10^{-14}$. The quantity $R$ will also be referred to as \emph{runup}. The maximum runup will be denoted by $R_\infty$. In all experiments the discretization parameters were taken to be equal $\dx=0.05, \ \dt=\dx/10$, unless otherwise mentioned. Further, we compute in all cases the discrete mass $I_0^h$ and show the preserved digits. We use the KT and CF schemes combined with the TVD2 and UNO2 reconstructions. The CF-scheme appeared to be less dissipative and we emphasize the results of this method.

\subsection{Runup of a solitary wave on a gradual slope $\beta=2.88^\circ$ with $A_s/D_0=0.0185$}

\begin{figure}%
\centering
\subfigure[$t=30$]{\includegraphics[scale=0.333]{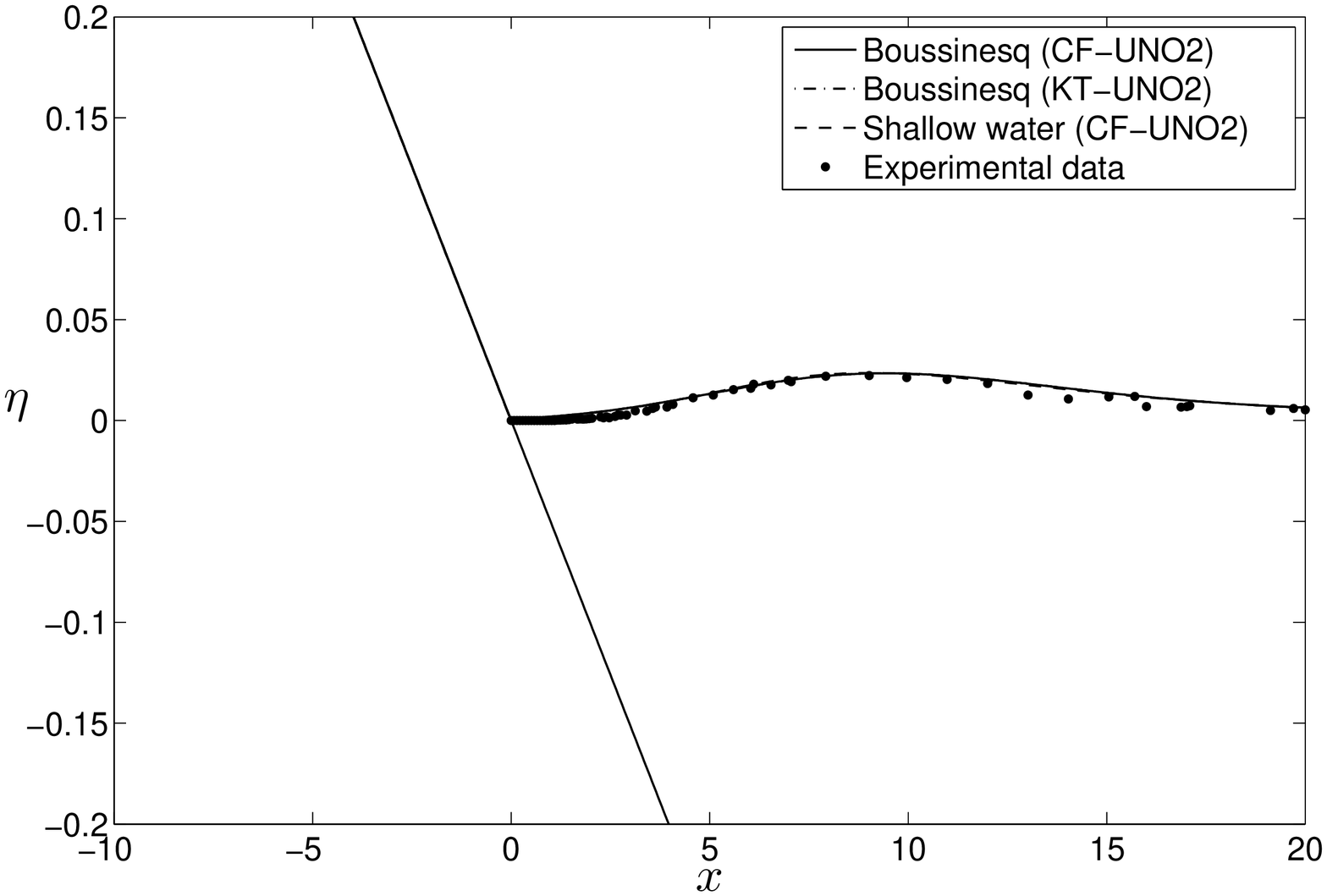}}
\subfigure[$t=40$]{\includegraphics[scale=0.333]{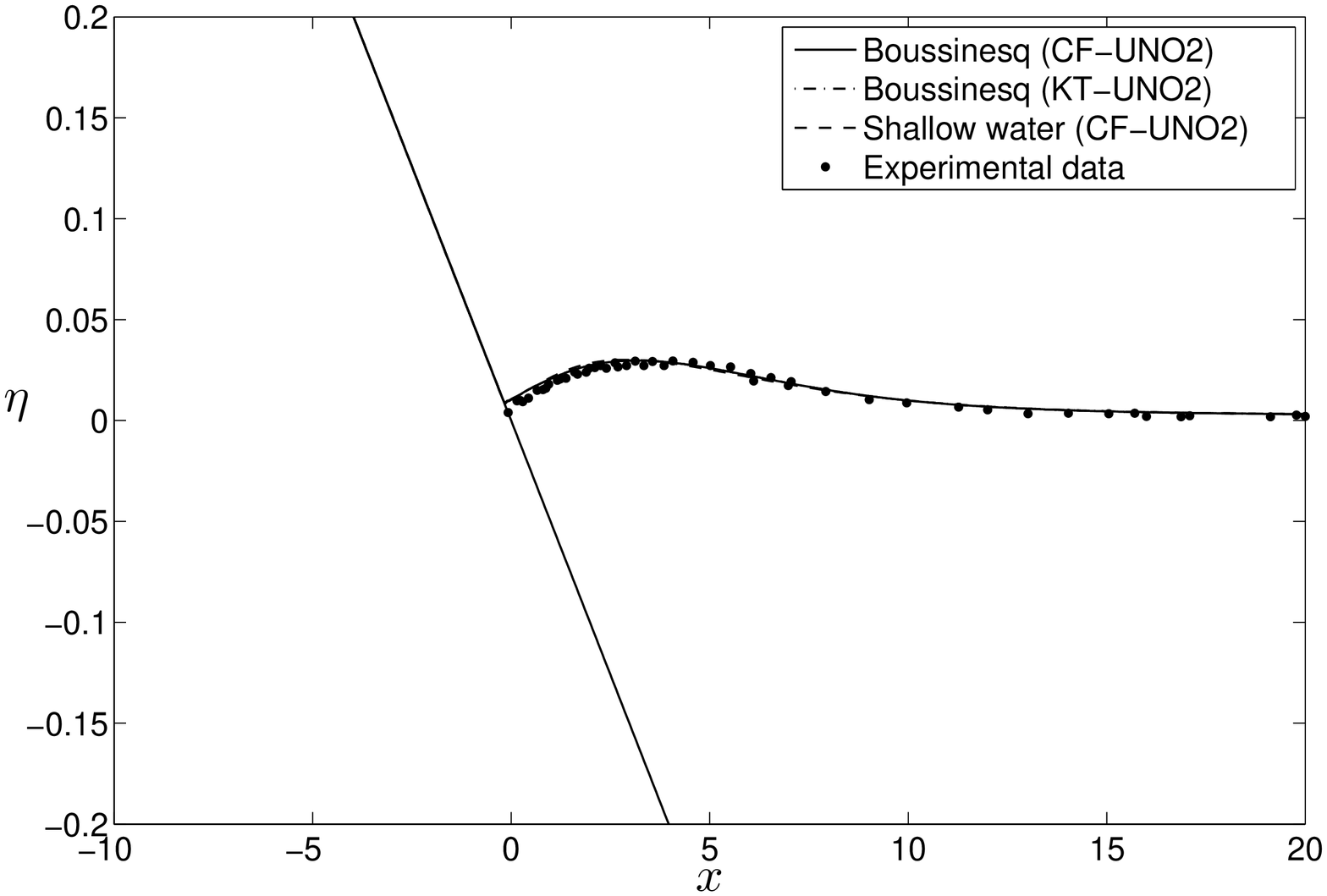}}
\subfigure[$t=50$]{\includegraphics[scale=0.333]{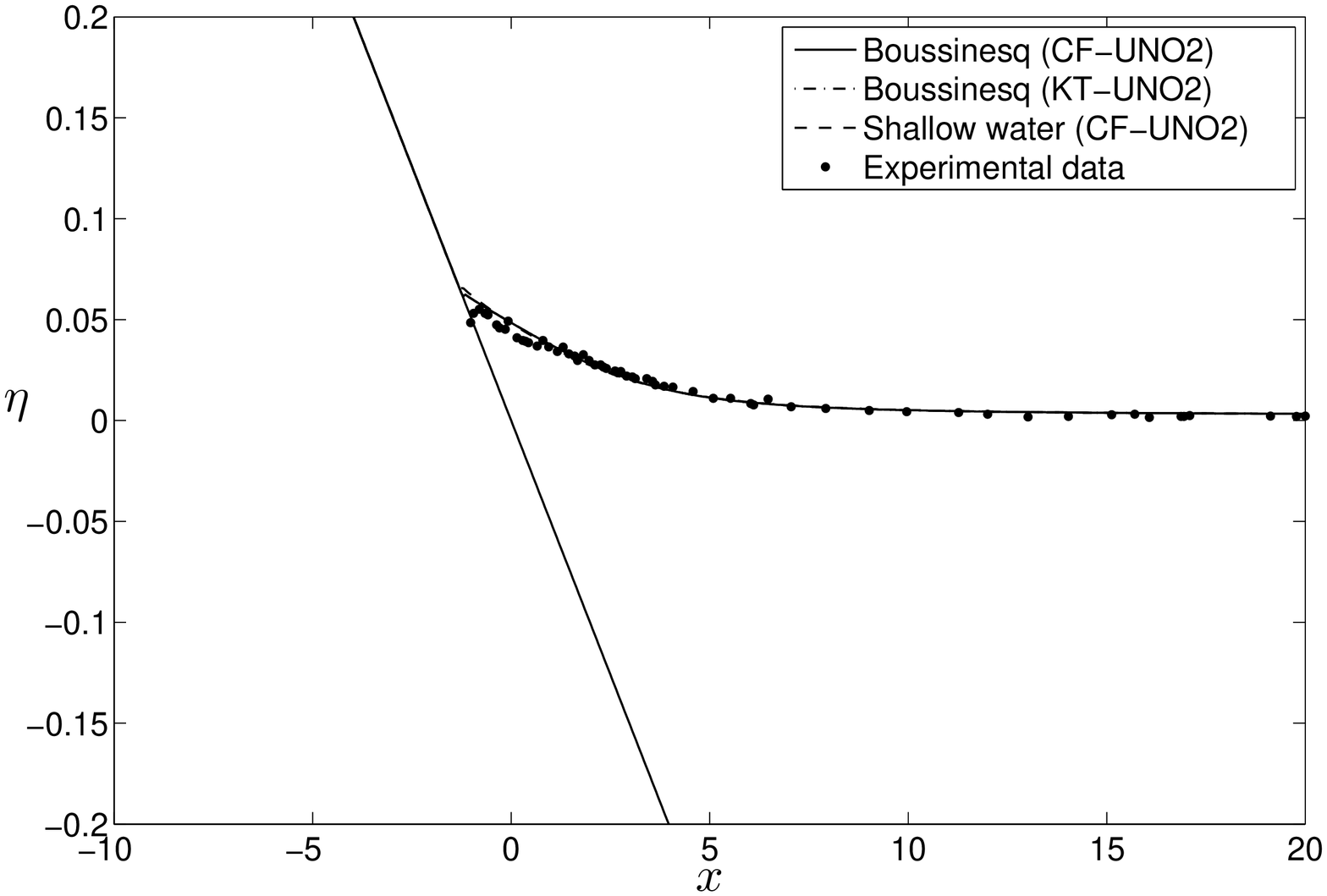}}
\subfigure[$t=70$]{\includegraphics[scale=0.333]{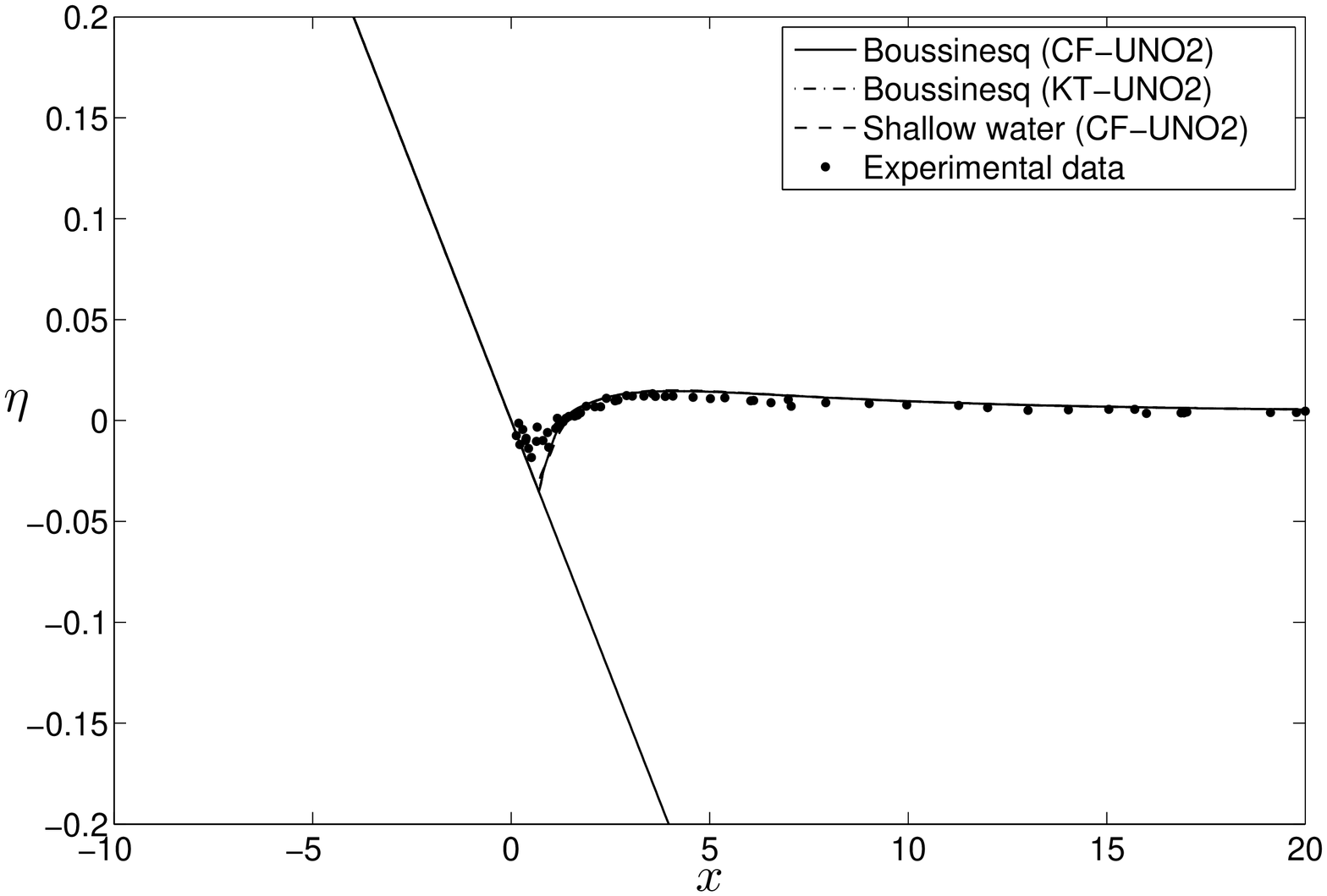}}
\caption{Solitary wave runup on a sloping shore: $A_s = 0.0185$ case.}%
\label{F6i.1}%
\end{figure}

\begin{figure}%
\centering
\includegraphics[scale=0.65]{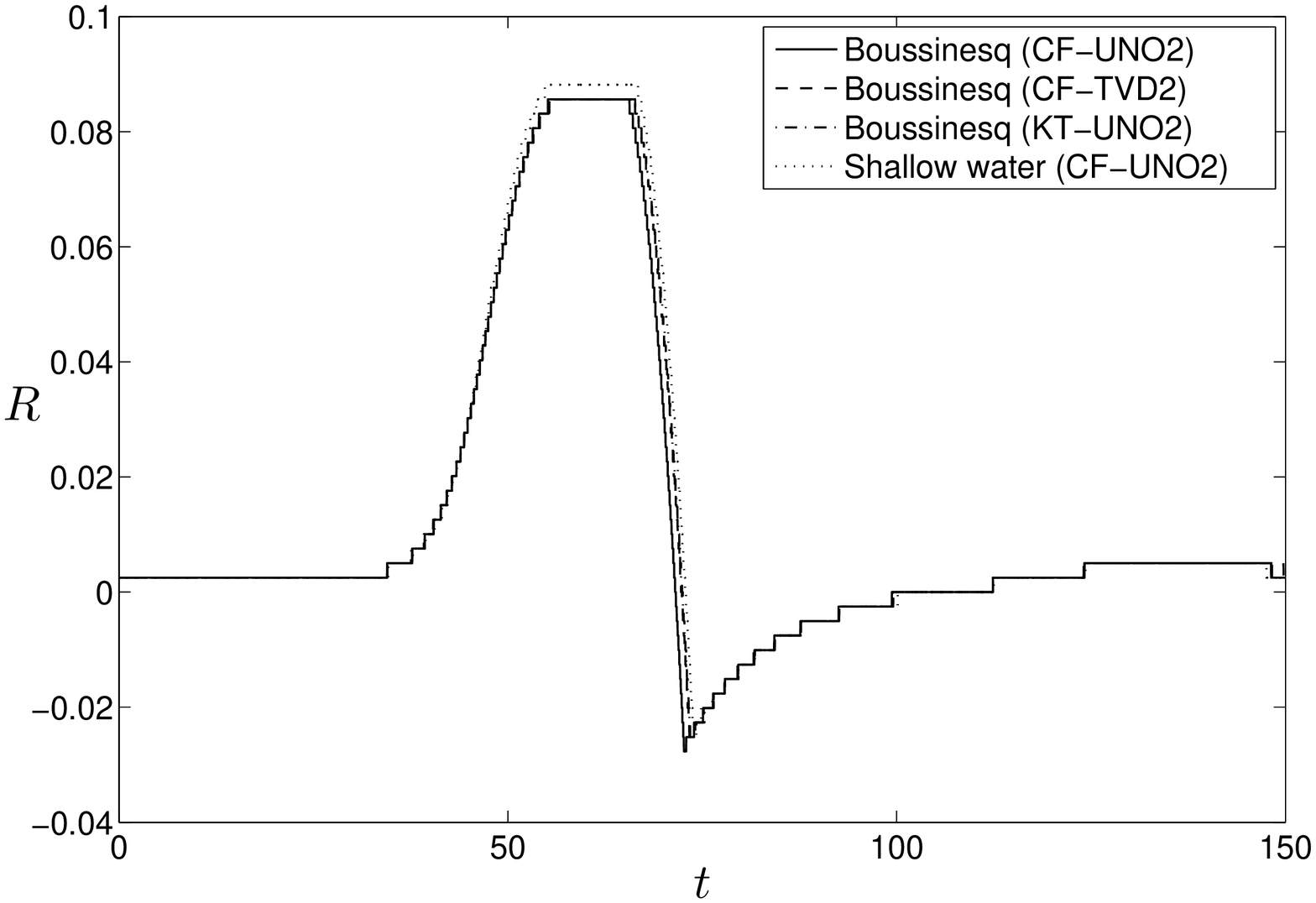} 
\caption{Runup value $R$ as a function of time:  $A_s = 0.0185$ case.}%
\label{F6i.2}%
\end{figure}

We consider first the simplest case --- the runup of a non-breaking solitary wave. In this experiment we take an initial solitary wave with the amplitude $A_s = 0.0185$, $D_0 = 1$ and $X_0 = 19.85$ in $I=[-10,70]$ and a mildly sloping shore $1:19.85$. This specific solitary wave does not break \cite{Synolakis1987} and the solution remains smooth during the runup and the rundown processes. In Figure \ref{F6i.1} we show several profiles of numerical solutions to Boussinesq and  shallow water equations along with the experimental data of \cite{Synolakis1987}. We observe that both models converge to the same solution.  The runup as well as  the rundown in this experiment is predicted very well. The runup value $R$ for both models is almost the same. The maximum runup is  $R_{\infty}\approx 0.085$ for the Boussinesq system, while for NSWE is $R_\infty\approx 0.088$. The experimental value reported in \cite{Synolakis1987} is equal to $R_{\infty}\approx 0.078$. In Figure \ref{F6i.2}  the runup $R$ as a function of time is represented. The discrete mass is preserved $I_0^h = 60.3667671231$ conserving the digits shown for both models.

%%%%%%%%%%%%%%%%%%%%%%%%%%%%
%%%%%%%%%%%%%%%%%%%%%%%%%%%%

%%%%%%%%%%%%%%%%%%%%%%%%%%%%%%%%%

\subsection{Runup of a solitary wave on a gradual slope $\beta=2.88^{\circ}$ with $A_s/D_0=0.04$} \label{nolake}

\begin{figure}%
\centering
\includegraphics[scale=.45]{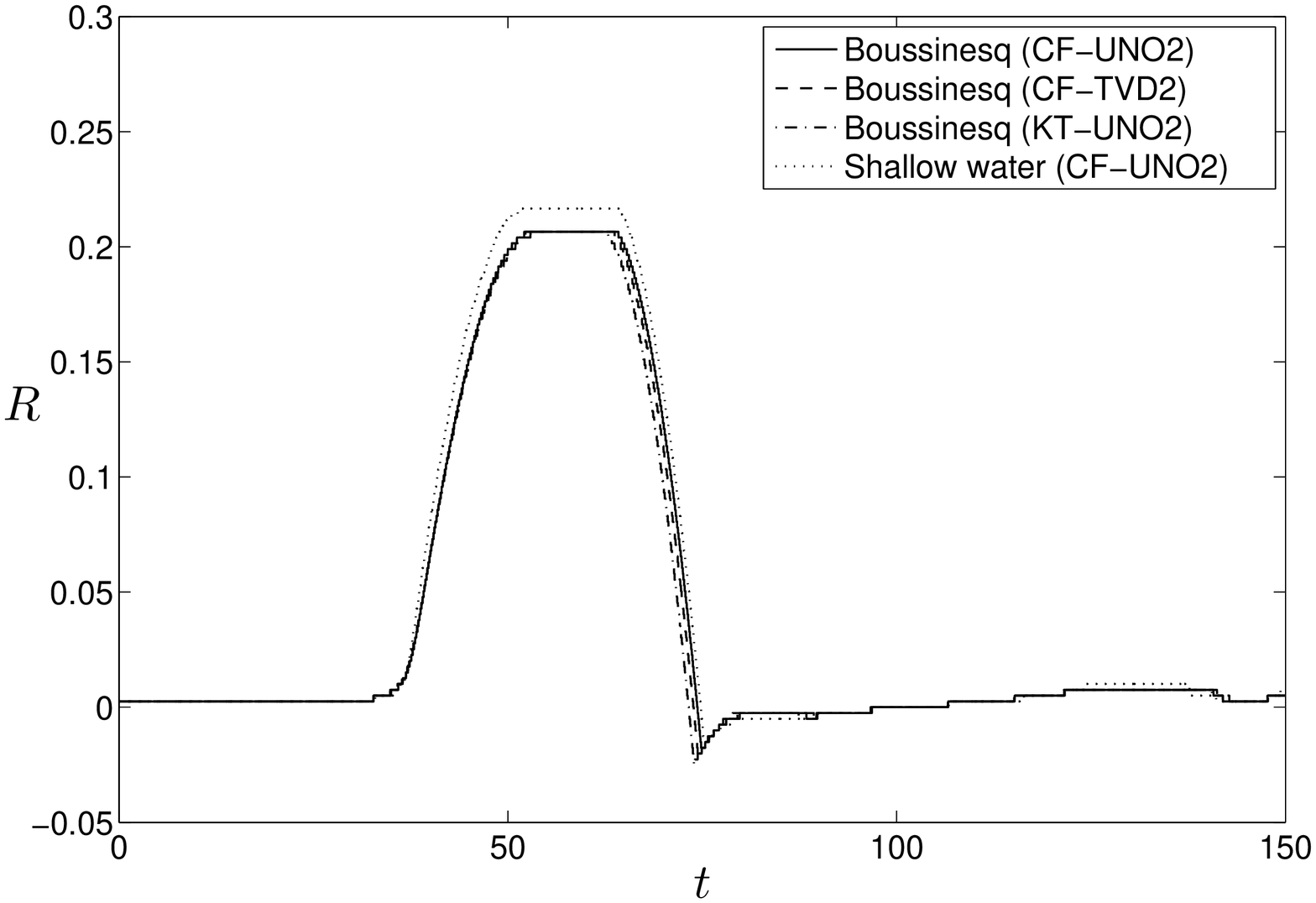}
\caption{Runup value $R$ as a function of time: $A_s=0.04$ case.}%
\label{F6ii.2}%
\end{figure}

We consider the same sloping shore as before. We study the runup of a solitary wave with amplitude $A_s = 0.04$, placed initially at $X_0 = 19.85$ in $I = [-10,70]$. The solitary wave does not break during the runup phase. Breaking occurs during the rundown process as in experimental observations \cite{Synolakis1987}. Results of the numerical simulations are presented in Figure \ref{F6ii.1}. 
In Figure \ref{F6ii.2} the evolution of the runup value is shown. The maximum runup for the Boussinesq system is $R_\infty\approx 0.20$ and $R_\infty\approx 0.21$ for shallow water system. The experimental value reported in \cite{Synolakis1987} is $R_\infty\approx 0.156$.

\begin{figure}%
\centering
\includegraphics[scale=0.45]{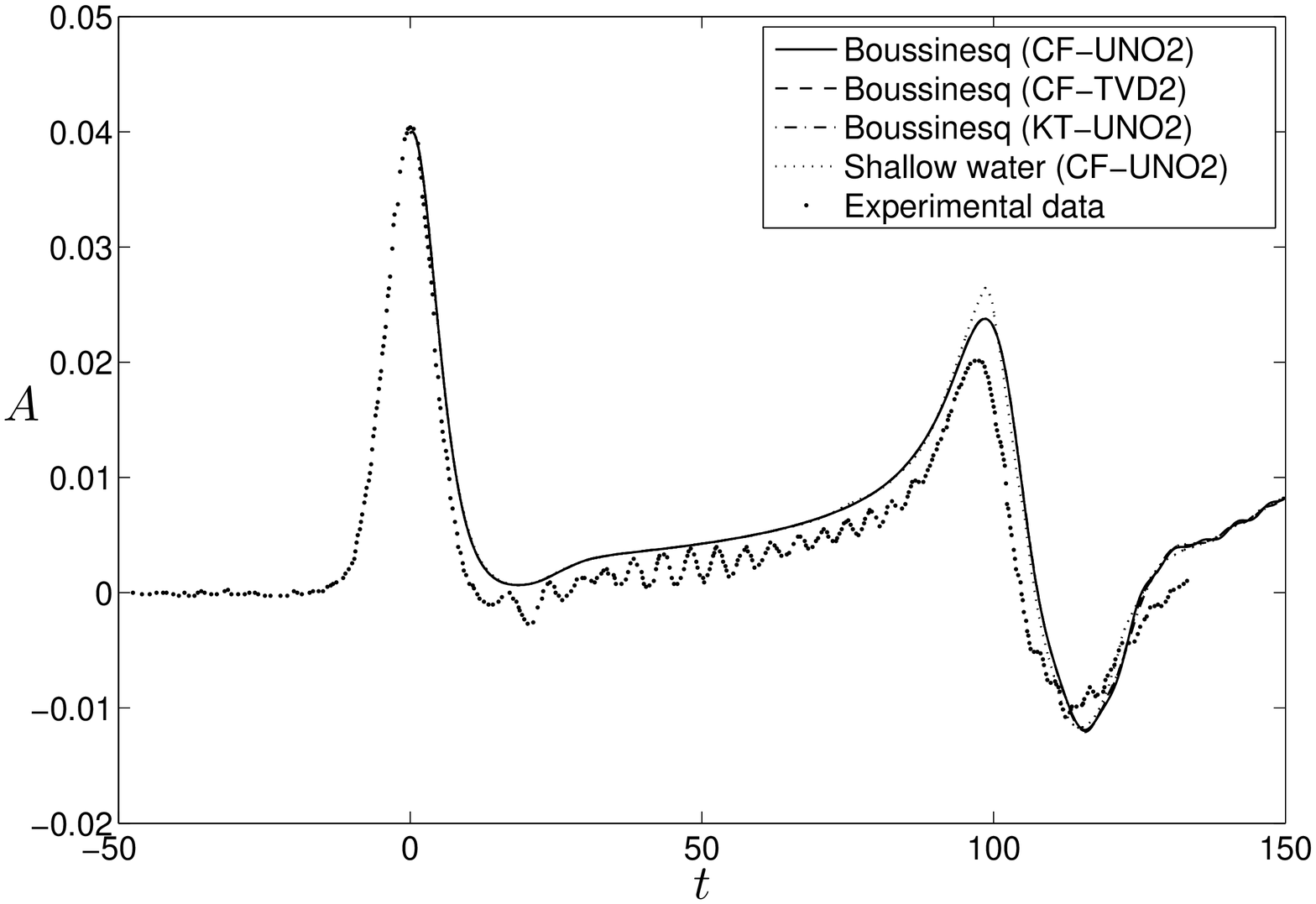} 
\caption{Free surface elevation at $x = 32.1$ m.}%
\label{F6ii.3}%
\end{figure}

In Figure \ref{F6ii.3} we perform a comparison with tide gauge data (free surface elevation measured in \cite{Synolakis1987}) collected at 32.1 m from the still shoreline position. We observe again a good agreement between the dispersive and nondispersive models. The discrete mass is preserved, $I_0^h= 60.5210181987$ conserving the digits shown.

%%%%%%%%%%%%%%%%%%%%%%%%%
\begin{figure}%
\centering
\subfigure[$t=20$]{\includegraphics[scale=0.333]{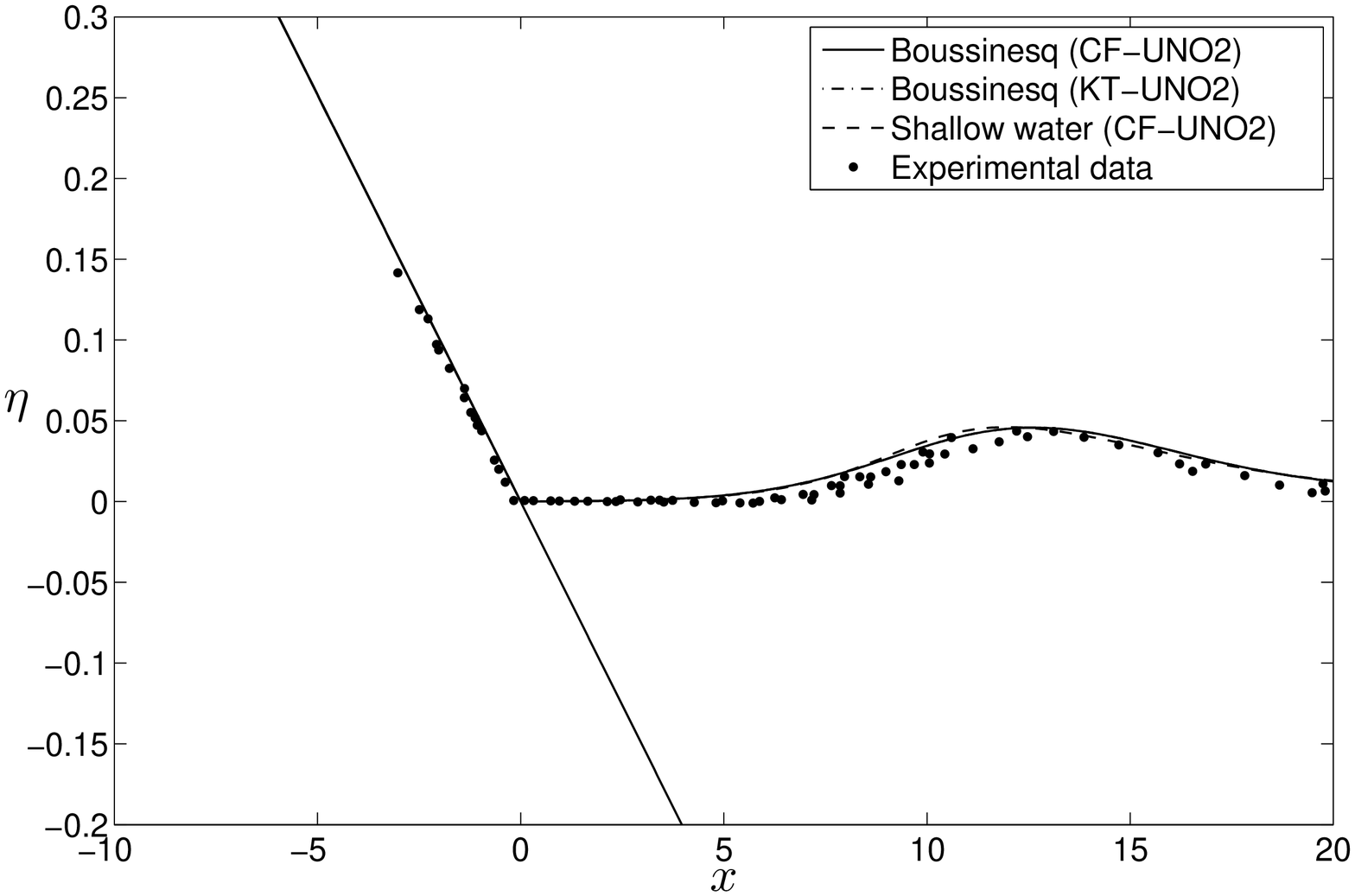}}
\subfigure[$t=26$]{\includegraphics[scale=0.333]{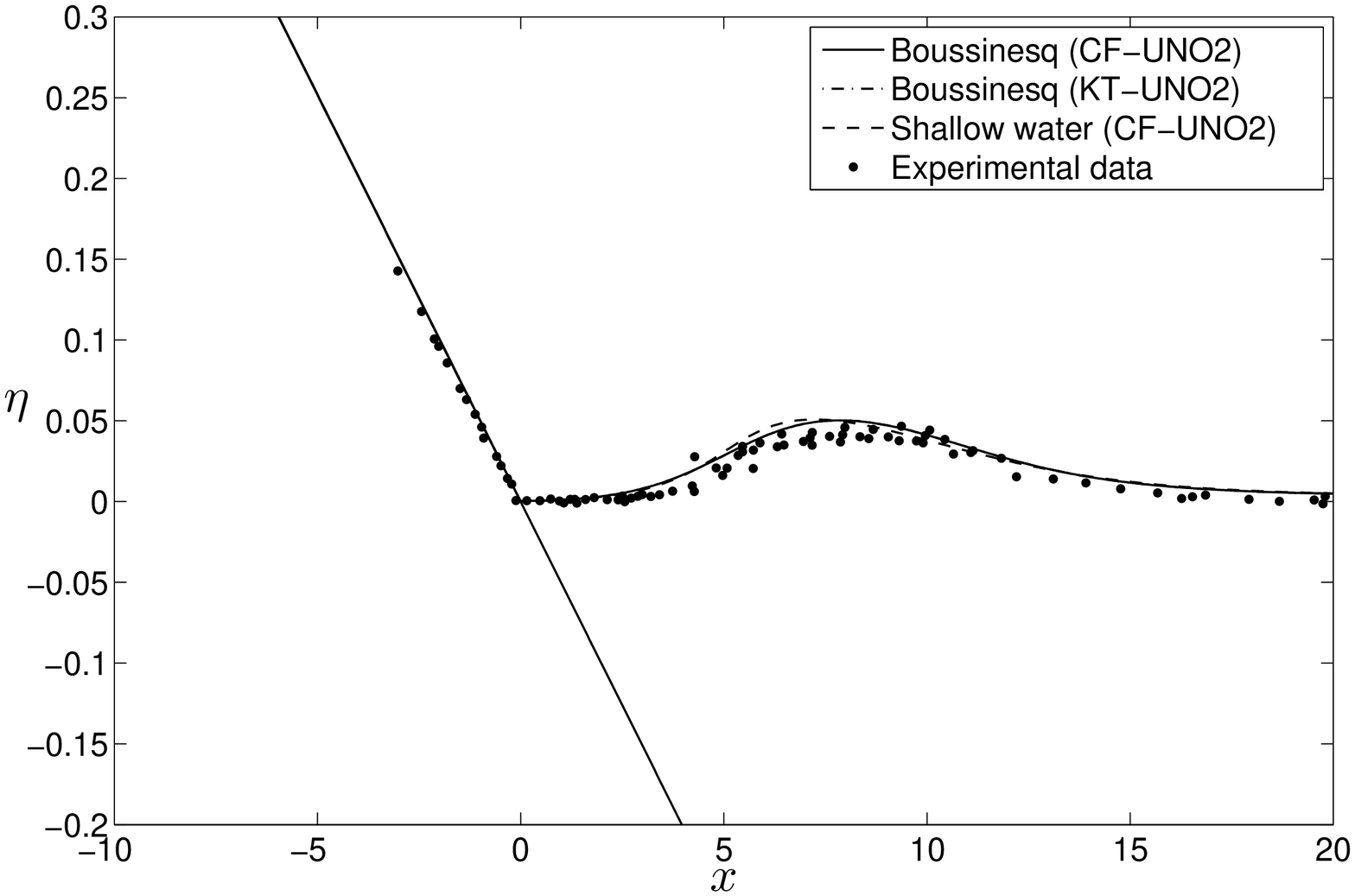}}
\subfigure[$t=32$]{\includegraphics[scale=0.333]{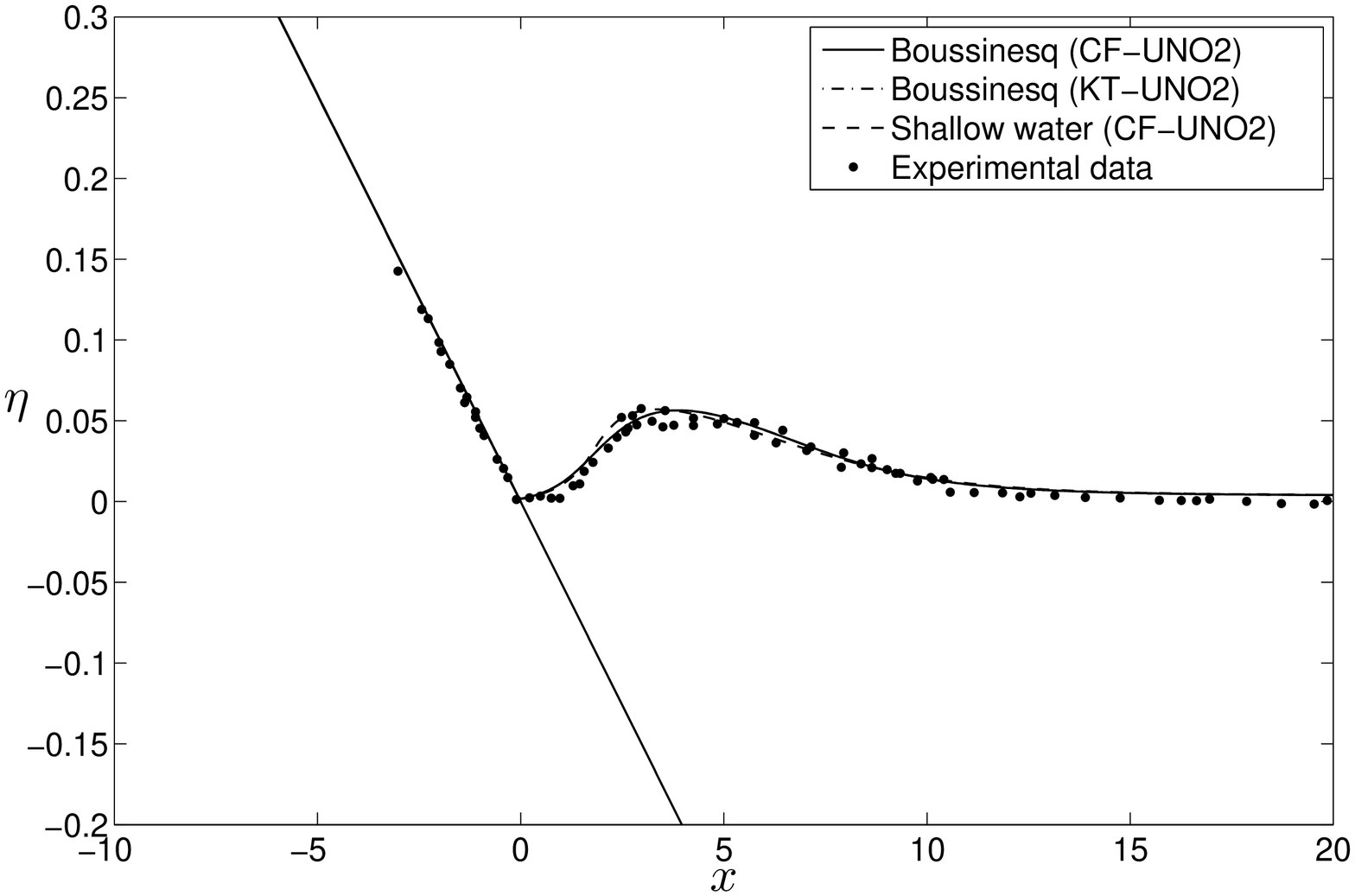}}
\subfigure[$t=38$]{\includegraphics[scale=0.333]{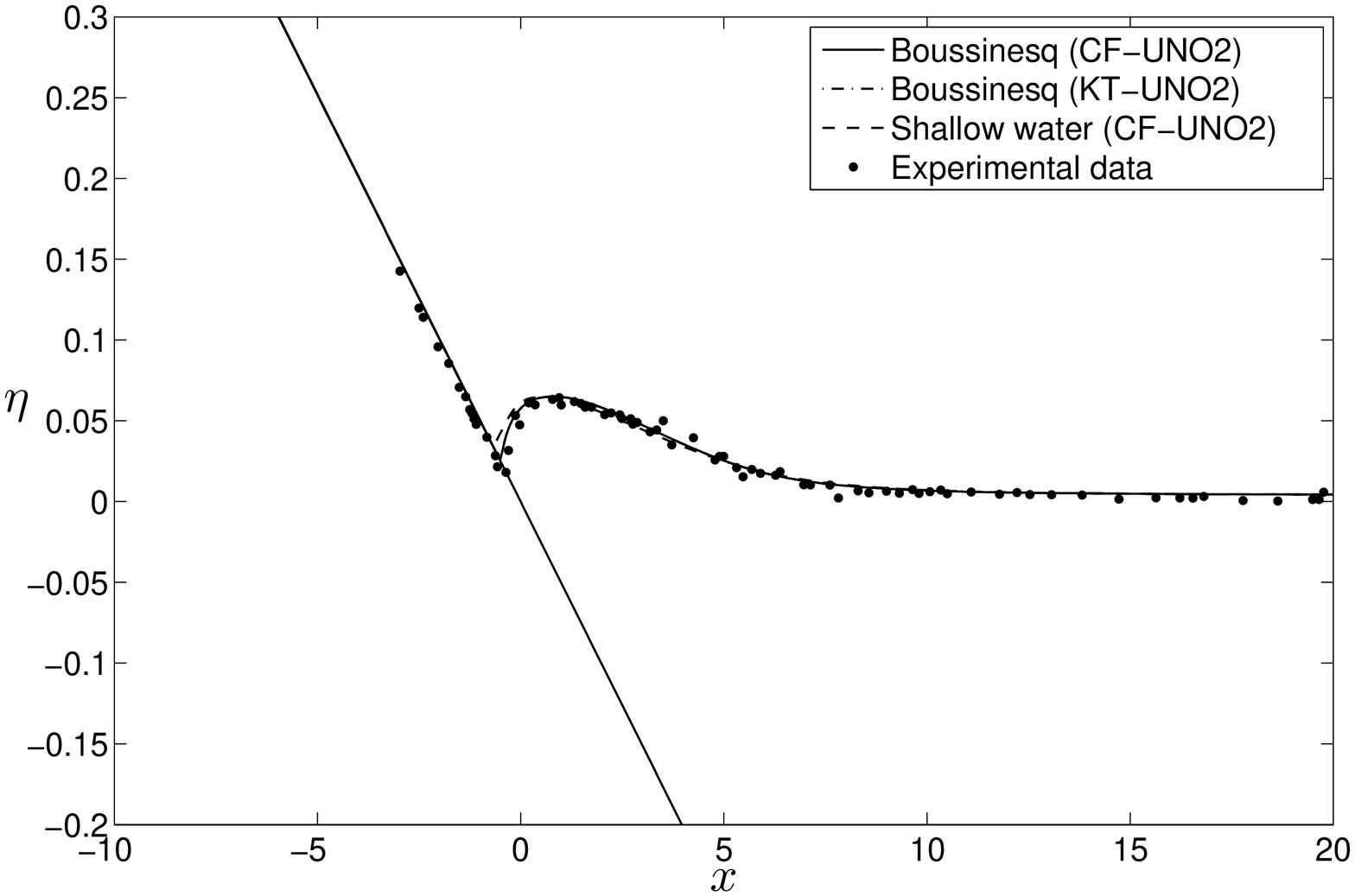}}
\subfigure[$t=44$]{\includegraphics[scale=0.333]{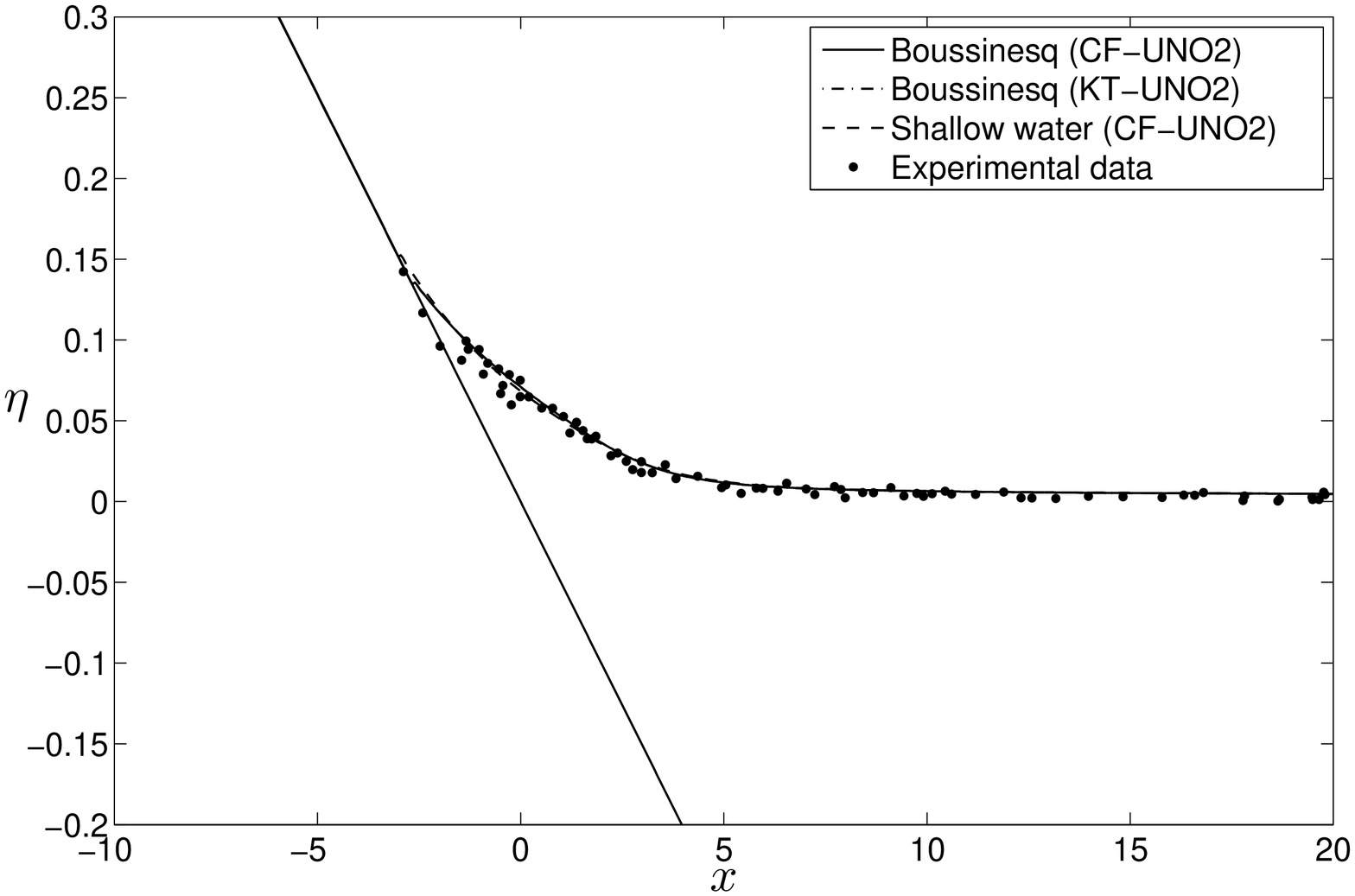}}
\subfigure[$t=50$]{\includegraphics[scale=0.333]{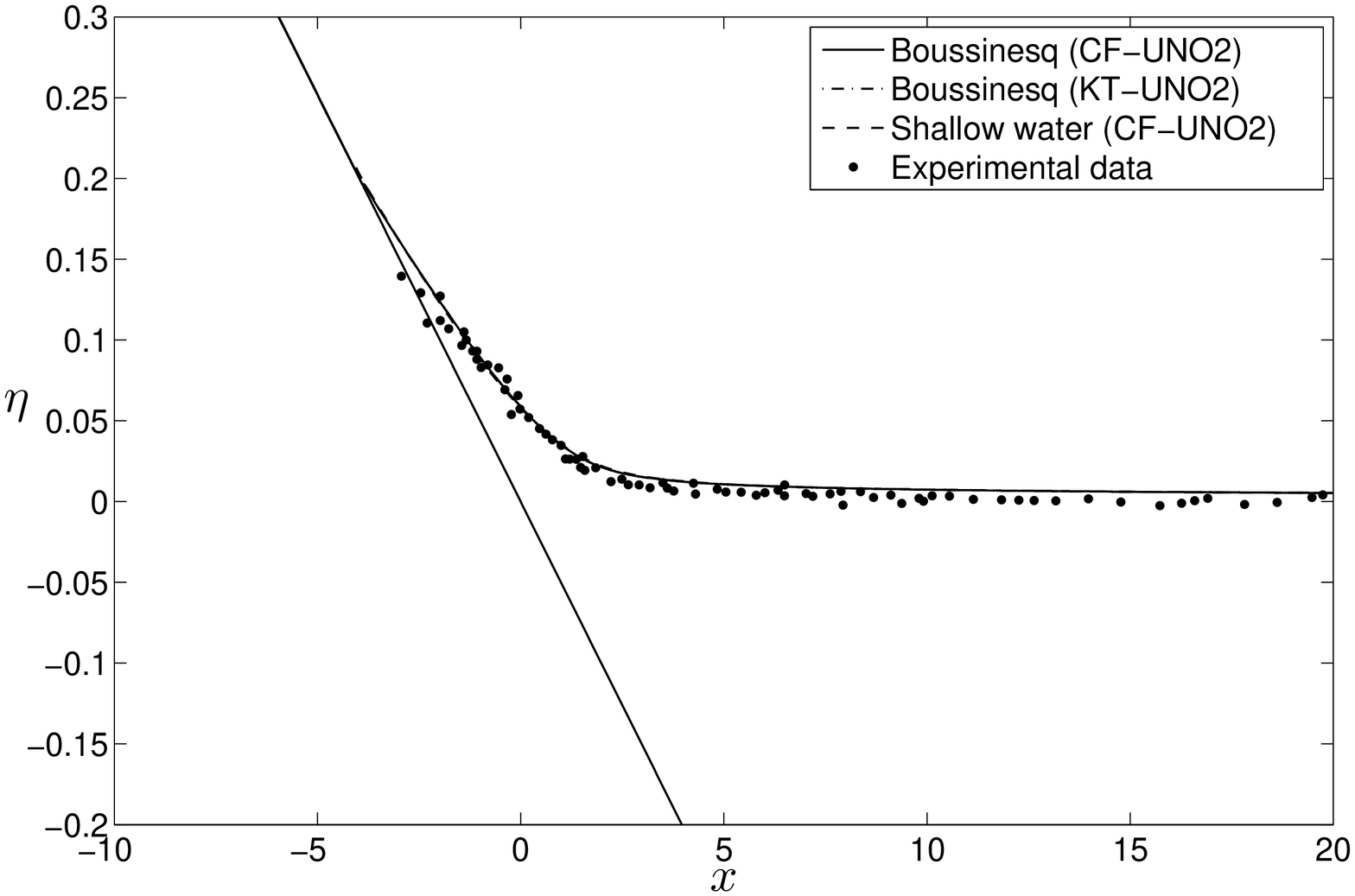}}
\caption{Solitary wave runup on a sloping shore: $A_s = 0.04$ case.}%
\label{F6ii.1}%
\end{figure}
%%%%%%%%%%%%%%%%%%%%%%%%%
\begin{figure}%
\ContinuedFloat
\centering
\subfigure[$t=56$]{\includegraphics[scale=0.333]{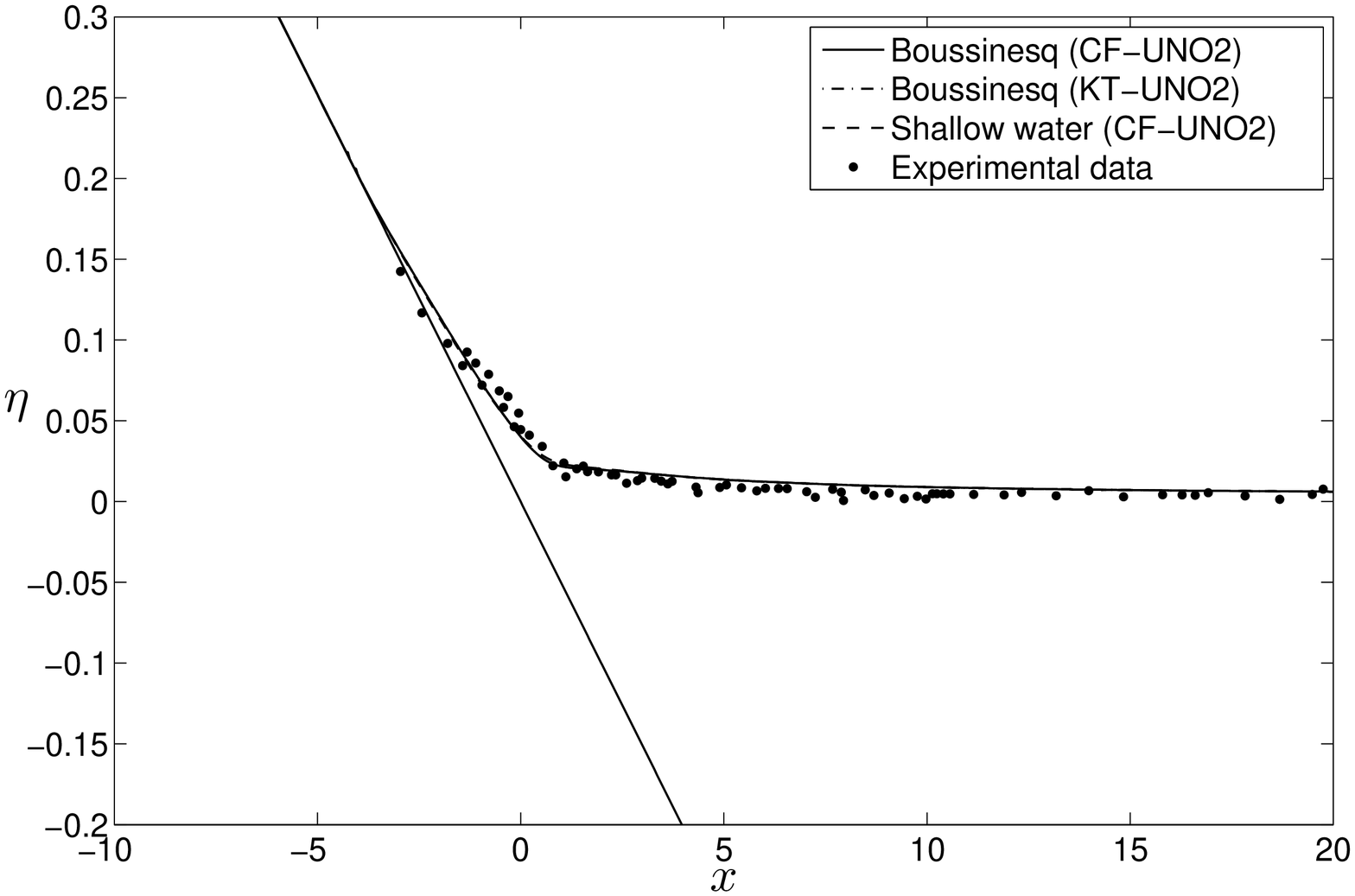}}
\subfigure[$t=62$]{\includegraphics[scale=0.333]{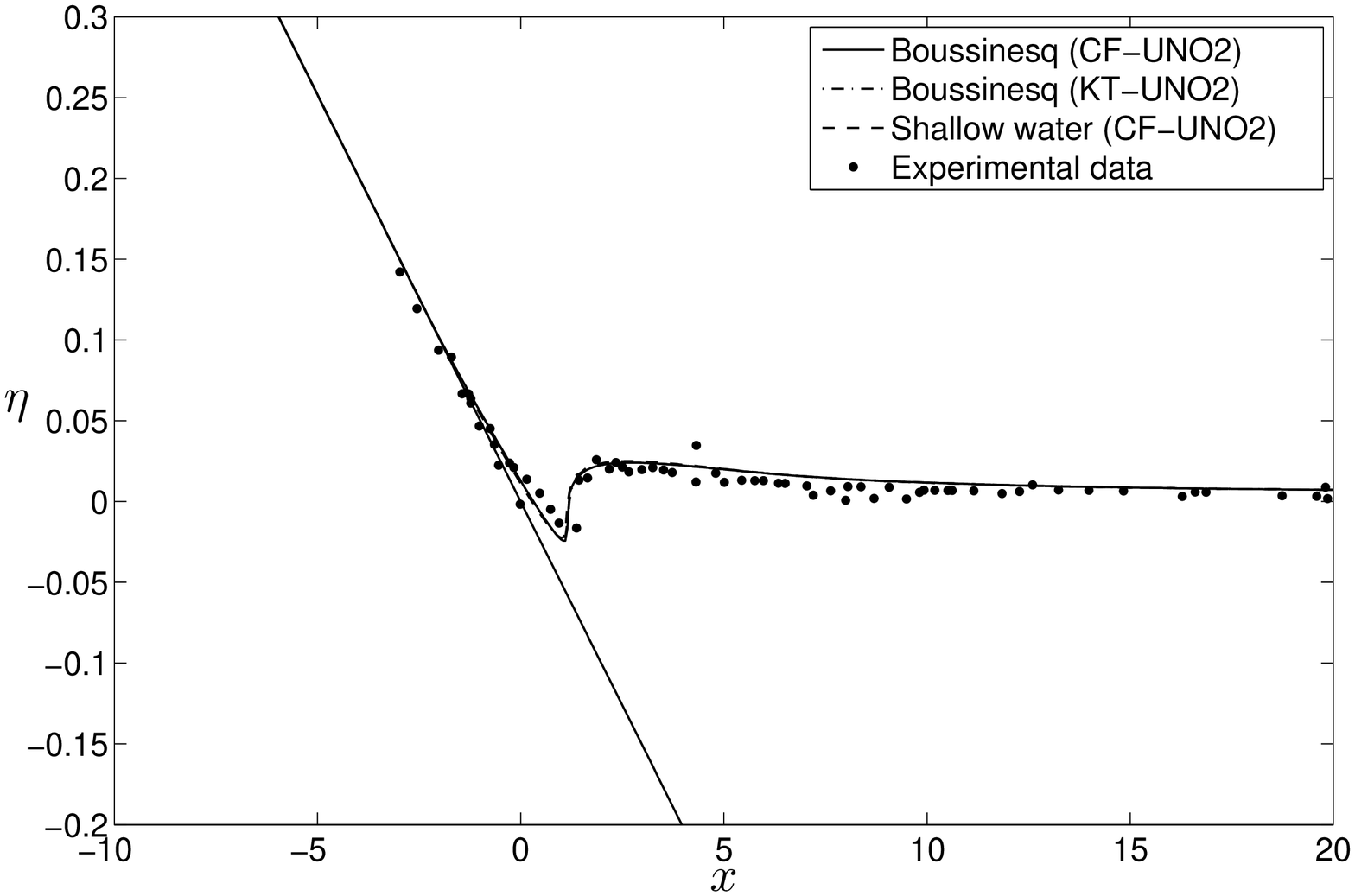}}
\caption{(Cont'd) Solitary wave runup on a sloping shore: $A_s = 0.04$ case.}%
\label{F6ii.1a}%
\end{figure}
%%%%%%%%%%%%%%%%%%%%%%%%%%%%%%%%%
%%%%%%%%%%%%%%%%%%%%%%%%%%%%%%%%%
%%%%%%%%%%%%%%%%%%%%%%%%%%%%%%%%%

\subsection{Runup of a solitary wave on a gradual slope $(\beta=2.88^{\circ})$ with $A_s=0.28$}

Finally we present the stiffest case of a solitary wave with amplitude $A_s = 0.28$, placed initially at $X_0 = 19.85$ in $I=[-10,60]$. This specific initial condition is characterized by the wave breaking phenomenon before even reaching the shoreline. Strictly speaking, in this case Boussinesq model is not valid unless a wave breaking mechanism is considered, cf. \cite{Zelt1991}.  In this case the approximate solitary wave given by formulas (\ref{SW1})-(\ref{SW2}) with $A_s=0.28$ it is outside the range of validity of the specific system. We proceed by constructing numerically a more accurate solitary wave following the \emph{cleaning procedure} of \cite{BC}. (Note that since the linearized system \eqref{E.Per5} coinsides with the linearized classical Boussinesq system we conclude that there exist classical solitary wave solutions of at least small amplitude, cf. \cite{DMII}). We constructed a solitary wave with $A_s\cong 0.28$ with $\dx=0.1$ and $\dx=0.05$ while $\dt=\dx/10$. 

\begin{figure}%
\centering
\subfigure[Before cleaning]{\includegraphics[scale=0.33]{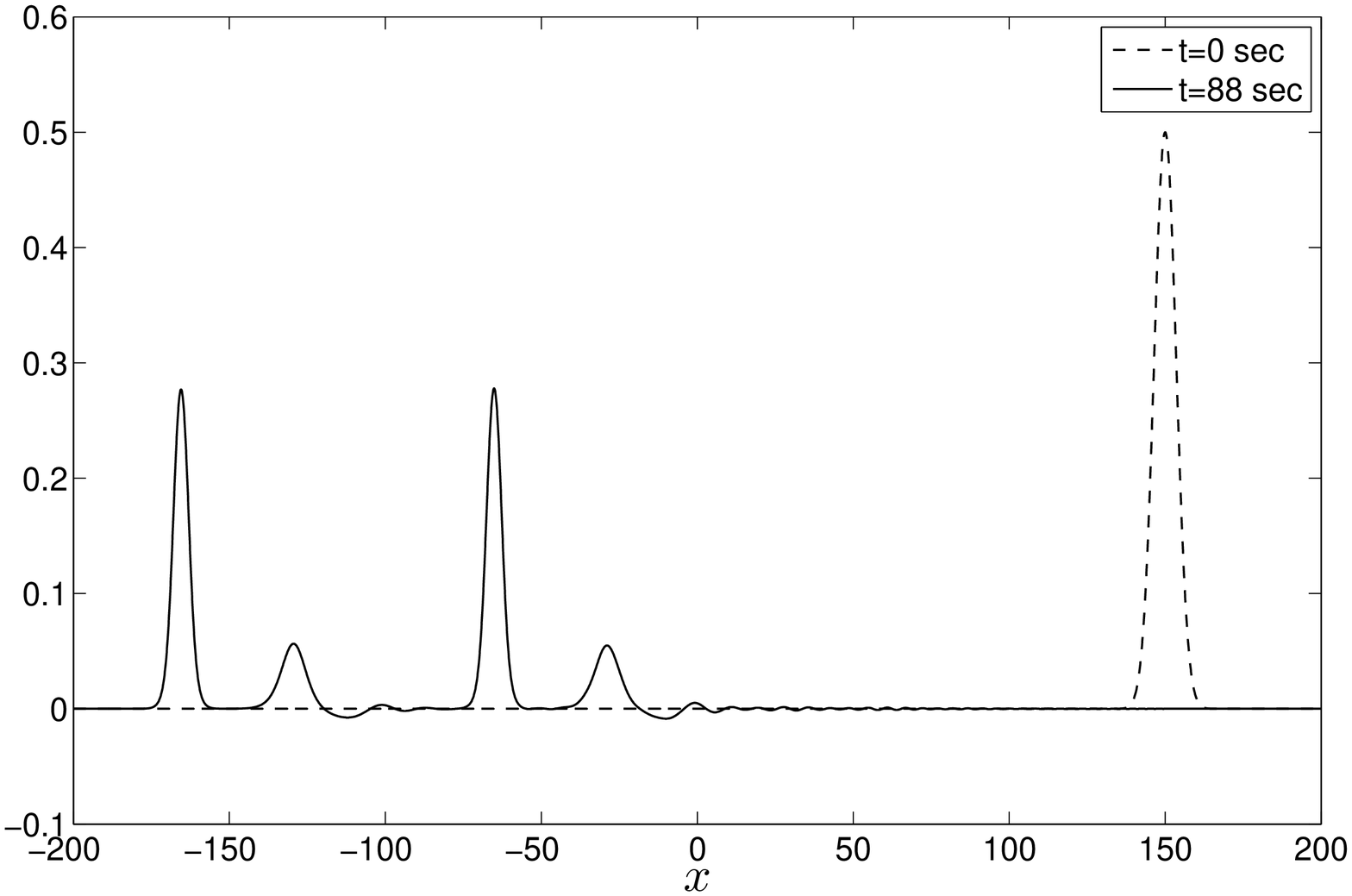}}
\subfigure[$t=0$ after cleaning and translation]{\includegraphics[scale=0.33]{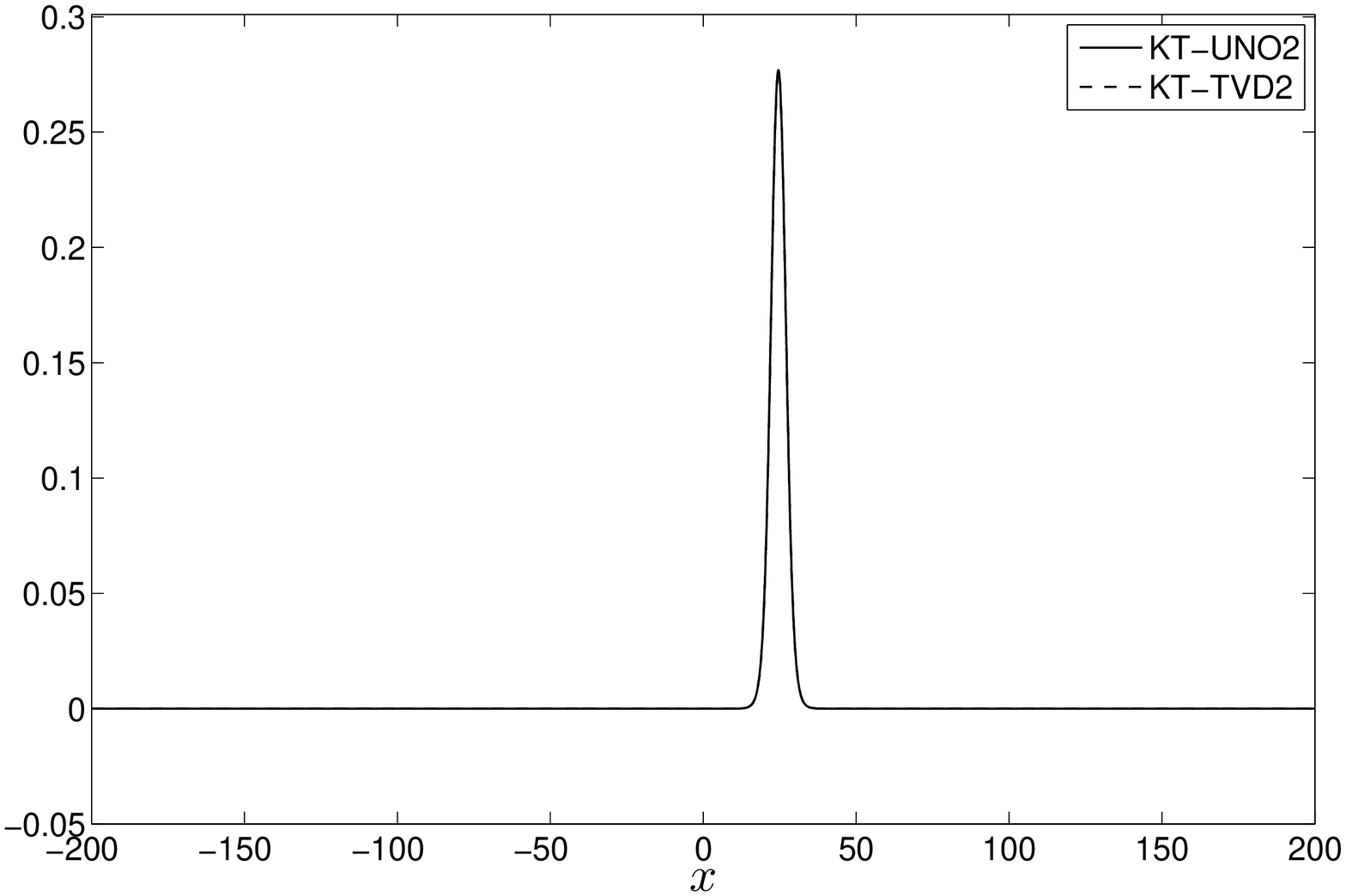}}
\subfigure[$t=50$ after cleaning]{\includegraphics[scale=0.33]{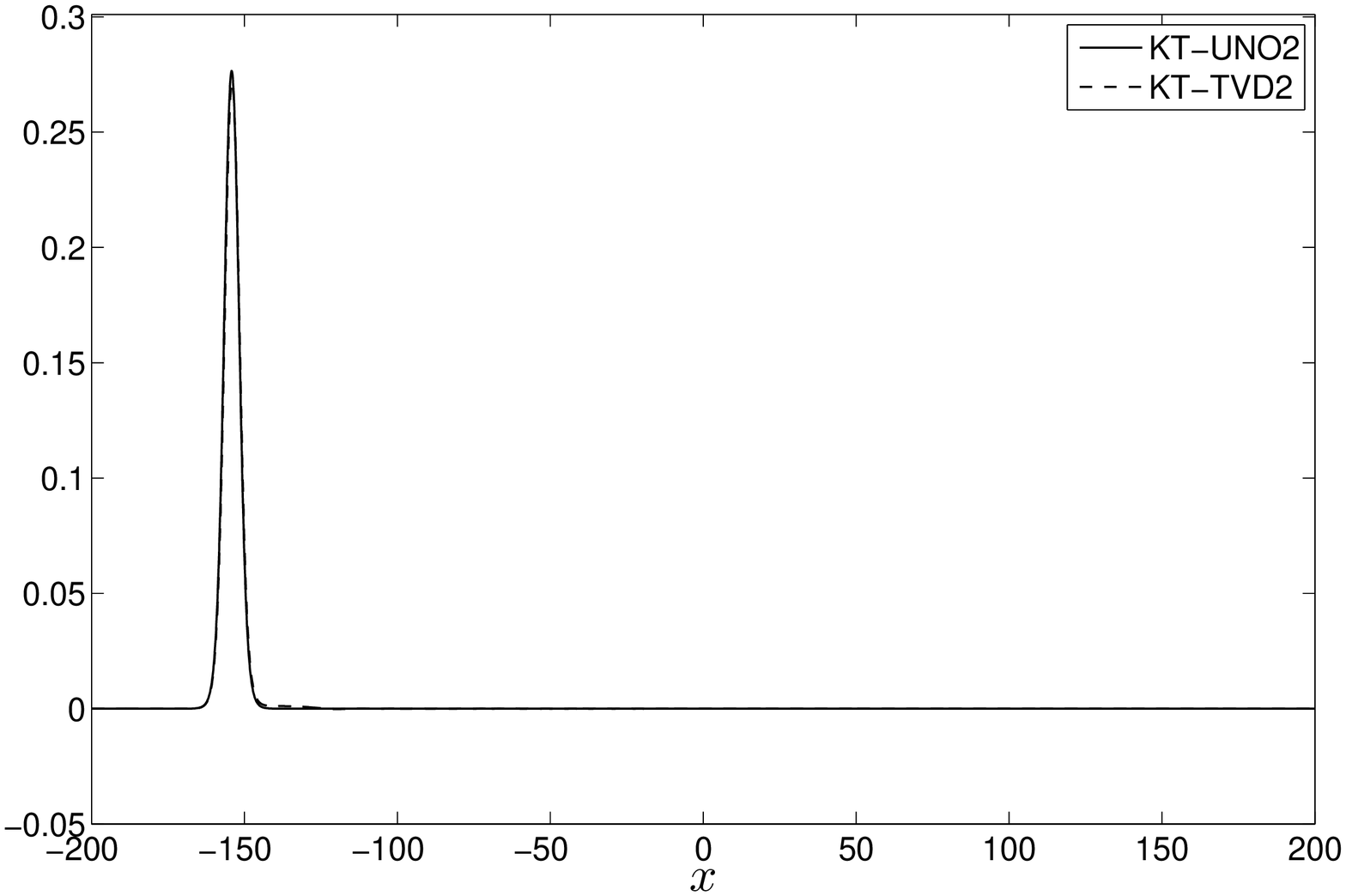}}
\subfigure[magnification of (c)]{\includegraphics[scale=0.33]{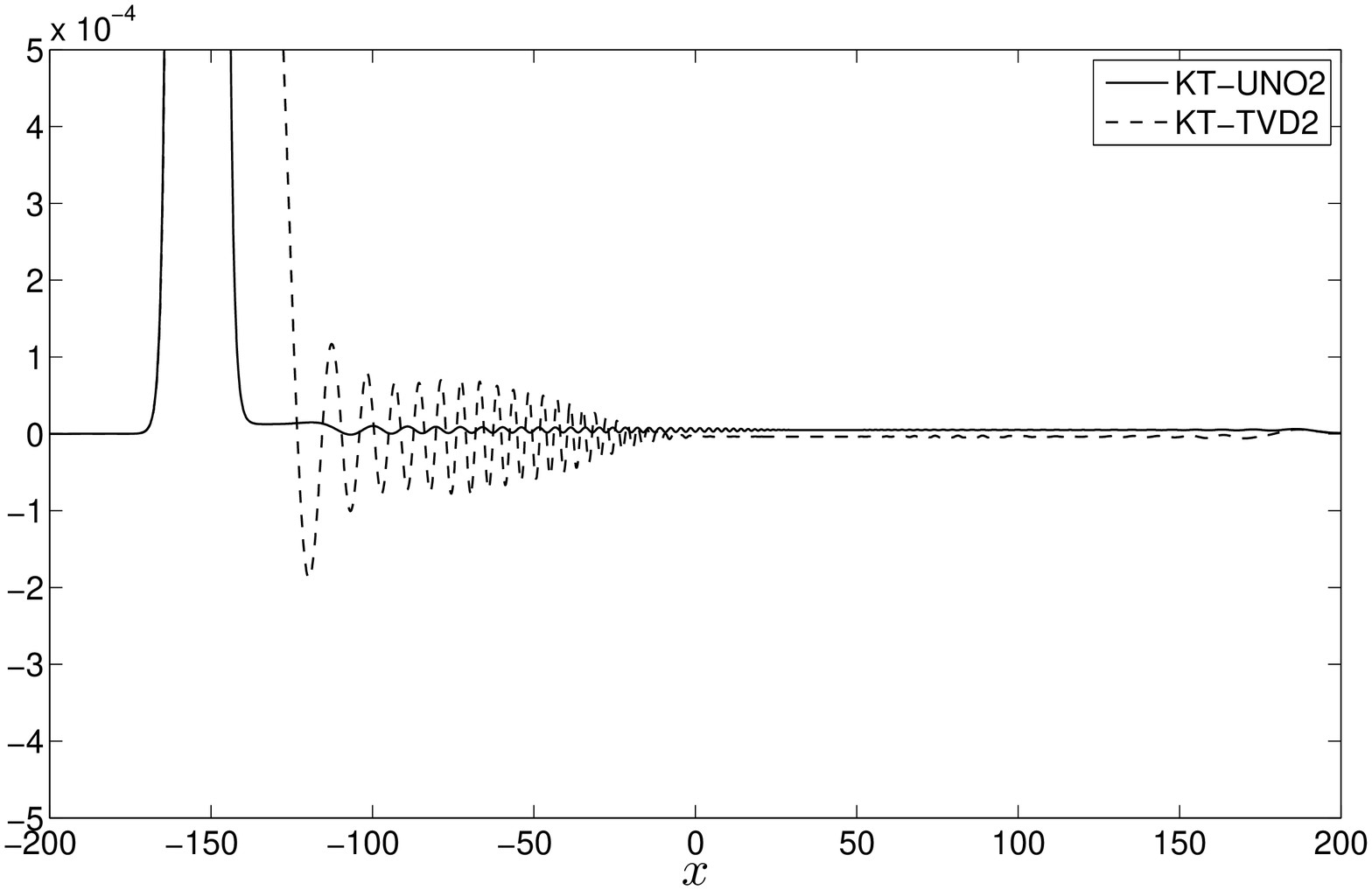}}
\subfigure[evolution of the amplitude]{\includegraphics[scale=0.5]{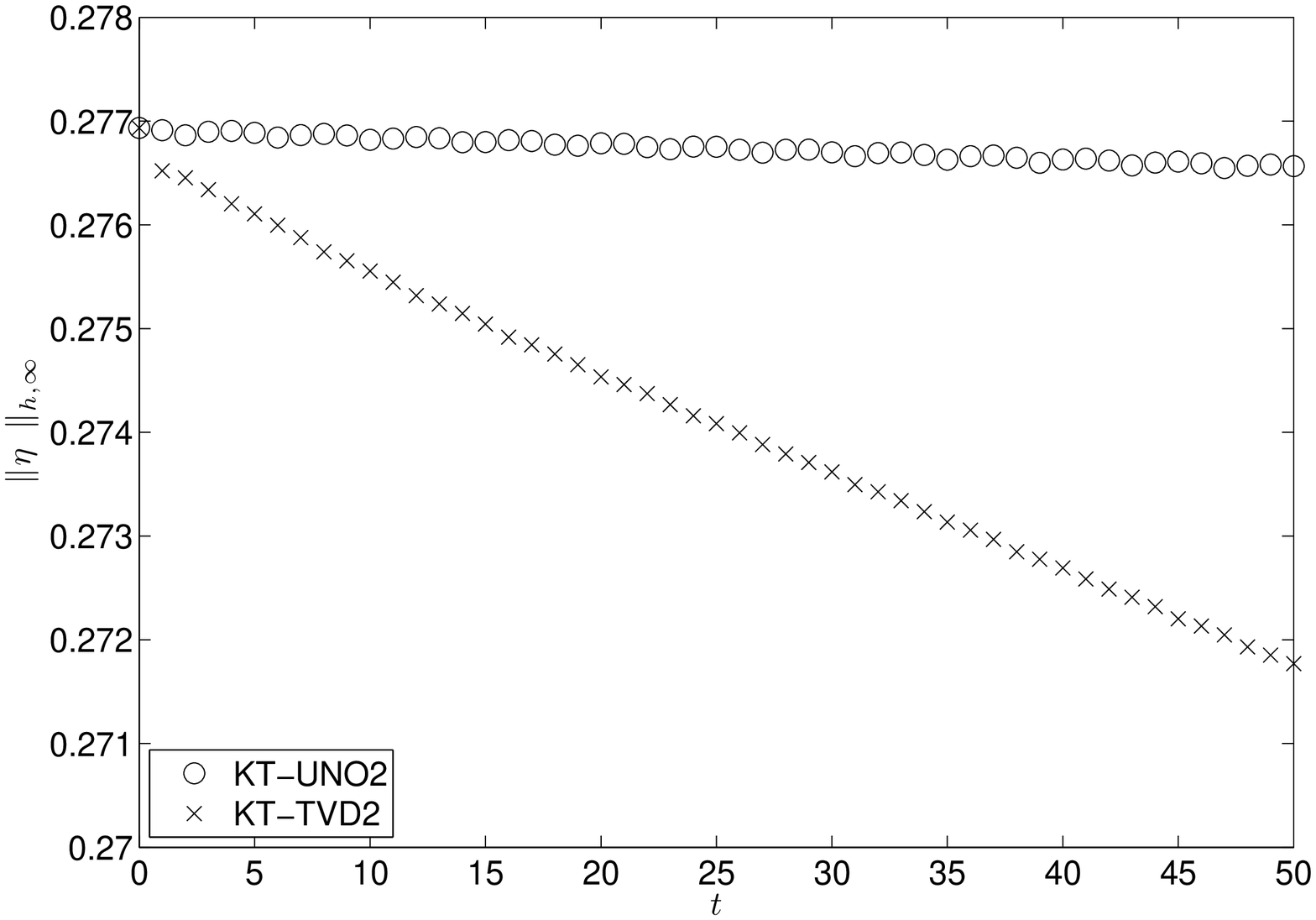}}
\caption{Generation of a solitary wave with $A_s\cong 0.28$ applying cleaning}%
\label{FClean}%
\end{figure}

Specifically, we consider the interval $[-200,200]$,  and the initial condition $\eta_0(x)=0.5\exp(-(x-150)^2/25)$, $u_0=0$ (with reflective boundary conditions). The initial condition is resolved into two solitary waves traveling in opposite directions. We observed that the left-traveling solitary wave at $t=88$ was separated enough from the rest of the solution. This solitary wave is isolated by  keeping the numerical solution in the interval $[-200,-151]$ and setting it equal to zero to the rest of the interval. The solution is then translated to the right at $X_0=19.85$ and we let it propagate. We observe that the clean solitary wave propagates like the analytical solution of the Bona-Smith system with analogous behavior between the TVD2 and UNO2 reconstruction. In Figure \ref{FClean} we present the results for the KT scheme since the results of the CF scheme are completely analogous and no difference can be observed. In Figure \ref{FClean} (a) the rightward traveling waves were reflected by the right boundary.

%%%%%%%%%%%%%%%%%%%%%%%%%
%%%%%%%%%%%%%%%%%%%%%%%%%%%%%%%%%  

In order to ensure the stability of the simulation and to study the runup, instead of smoothing, filtering or adding extra dissipative terms, we simply exclude the contribution of the term $Q_{xxt}$ in the vicinity of the shoreline (where $D_i<0.3$). Wave transition between these two regions appeared to be smooth as one may witness on Figure \ref{F6iii.1}. After this slight modification, the algorithm became more robust for large amplitude breaking waves without creating any unphysical oscillations.
 
In this experiment  friction appeared to play a significant role during the runup process, contrary to previous cases. The maximum runup computed without taking into account the friction of the bottom was far away from the experimentally measured values. For this reason, and only in this specific test case we included the empirical friction term \eqref{SFri} into the momentum conservation equation \eqref{E.Per5}, with coefficient $c_m=2\cdot 10^{-4}$. The friction term is discretized according to  \eqref{FVS}. This discretization preserves the positivity of all numerical schemes we tested. Mass conservation in this experiment was perfect $I_0^h=51.7504637472$ preserving the digits shown.

In Figure \ref{F6iii.1} we show the propagation of a breaking wave including its runup and rundown. We observe a significant difference between shallow water system and the dispersive model during the wave propagation. Discrepancies are present in the amplitude and in the phase speed simultaneously. However the dispersive model solution approximates better the measurements of J.A.~Zelt \cite{Zelt1991}. Nevertheless, we have to underline that the runup and rundown are fairly well described by both models. The maximum runup value according to the dispersive and nondispersive models is $R_\infty \approx 0.47$ which is in the range of $[0.42, 0.53]$ of the theoretical prediction of C.E.~Synolakis, \cite{Synolakis1987}. There is no single experimental value reported for the maximum runup in \cite{Synolakis1987} due to practical difficulties in generating a solitary wave of such large amplitude. Finally, we mention that the specific technique for handling the breaking wave leads to more accurate results for the rundown process than one presented in \cite{Zelt1991}.

%%%%%%%%%%%%%%%%%%%%%%%%%%%%%%%%%%%%%%
\begin{figure}%
\centering
\subfigure[$t=10$]{\includegraphics[scale=0.333]{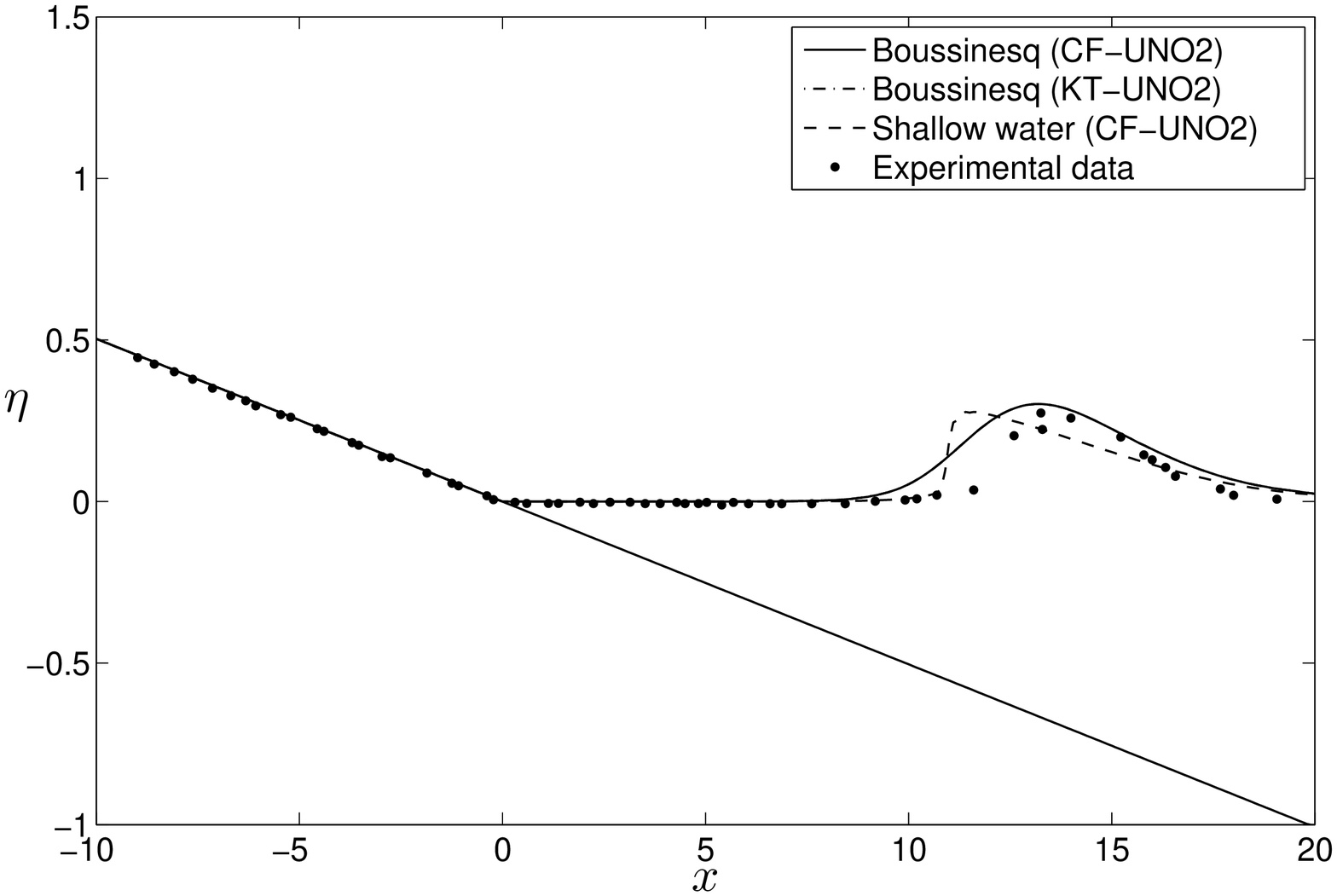}}
\subfigure[$t=15$]{\includegraphics[scale=0.333]{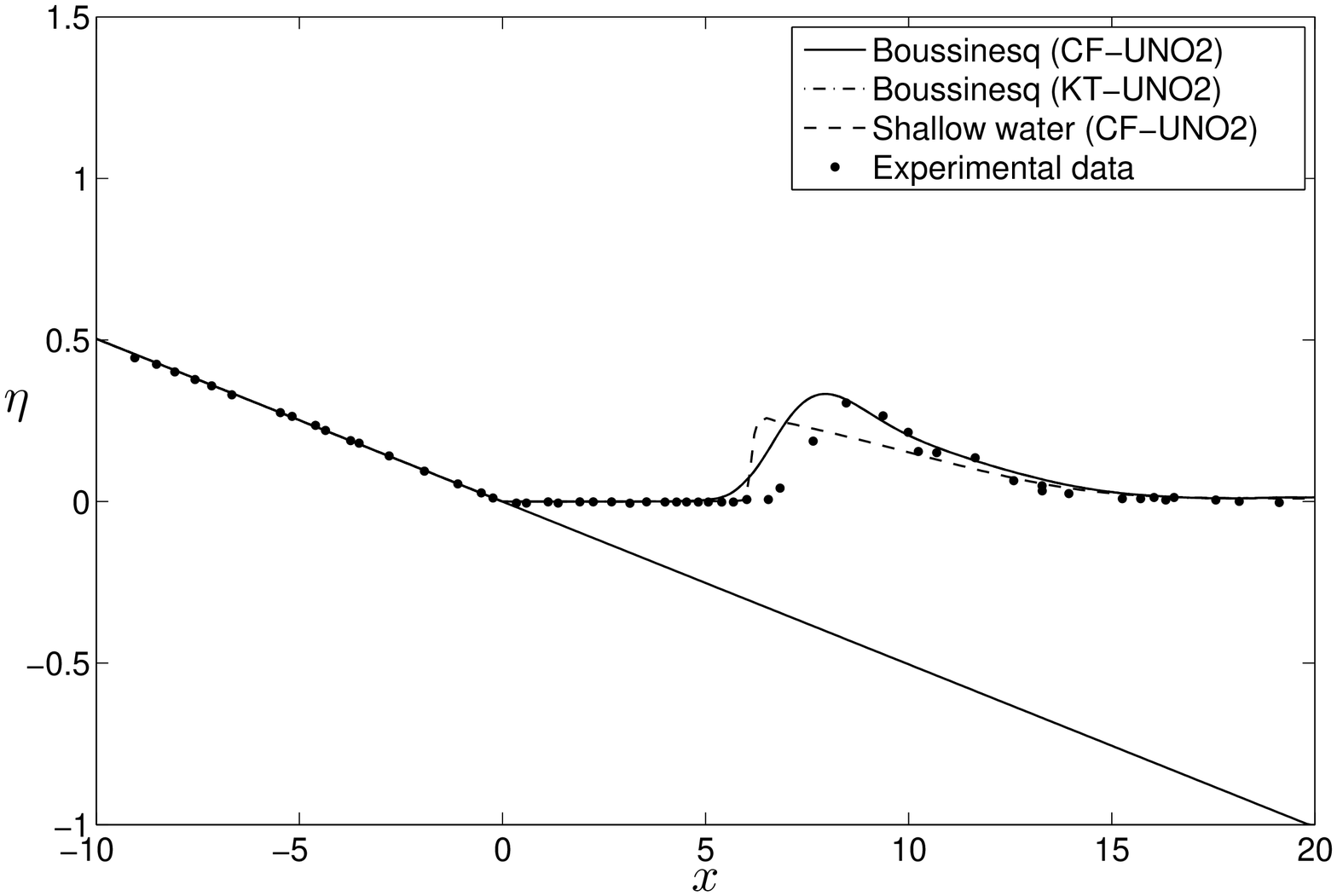}}
\subfigure[$t=20$]{\includegraphics[scale=0.333]{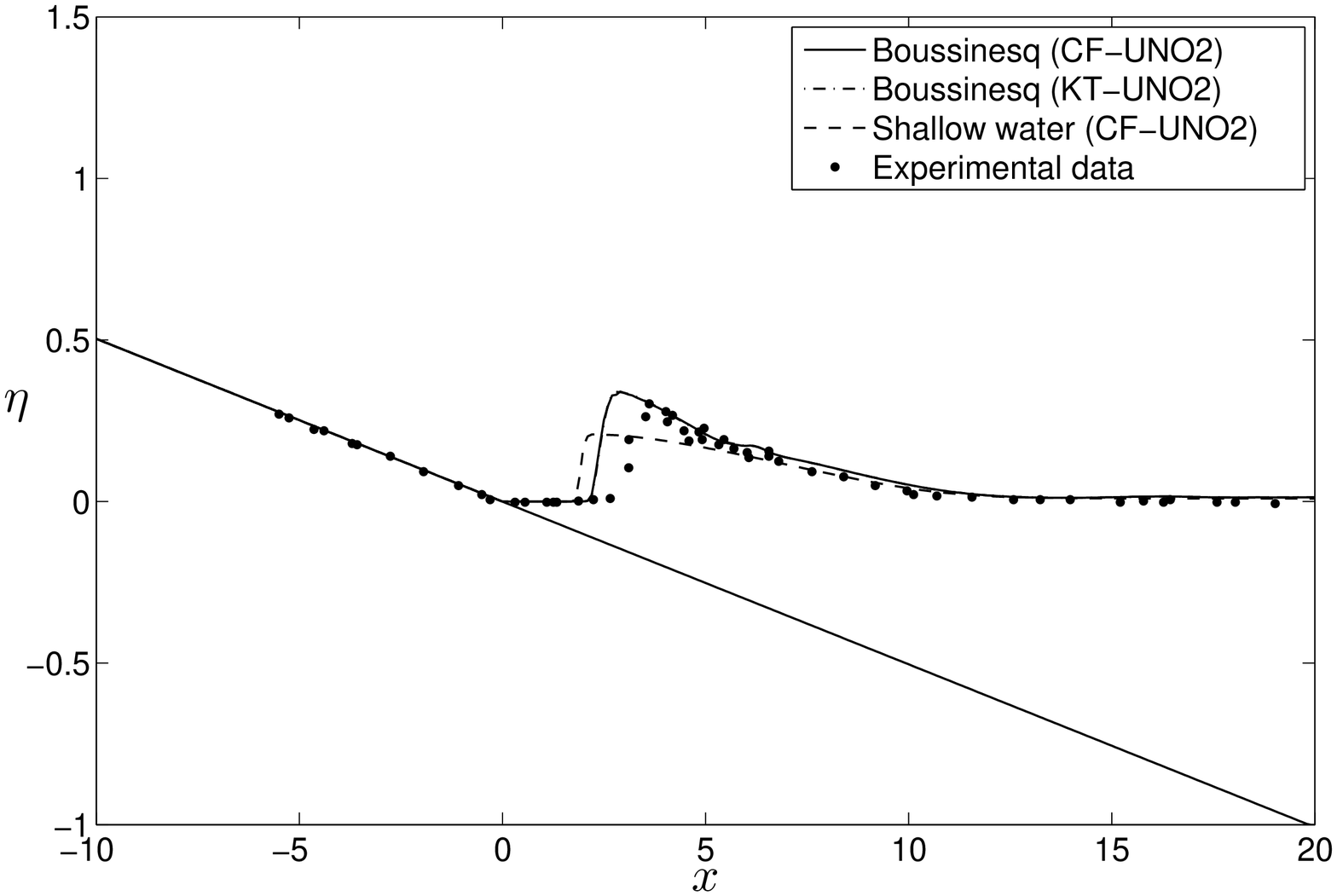}}
\subfigure[$t=25$]{\includegraphics[scale=0.333]{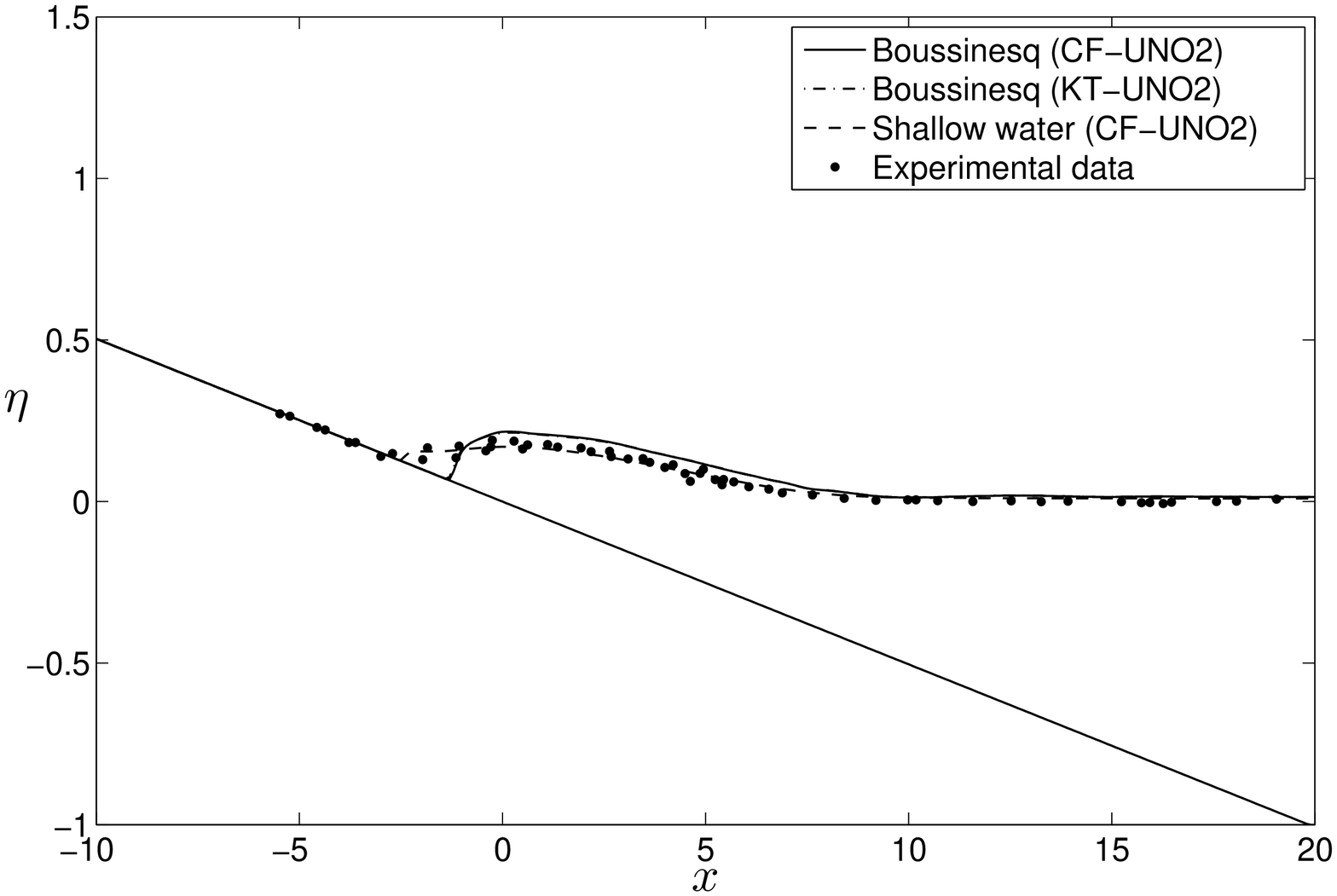}}
\caption{Solitary wave runup on a sloping shore: $A_s=0.28$ case.}%
\label{F6iii.1}%
\end{figure}
%%%%%%%%%%%%%%%%%%%%%%%%%
\begin{figure}%
\ContinuedFloat
\centering
\subfigure[$t=30$]{\includegraphics[scale=0.333]{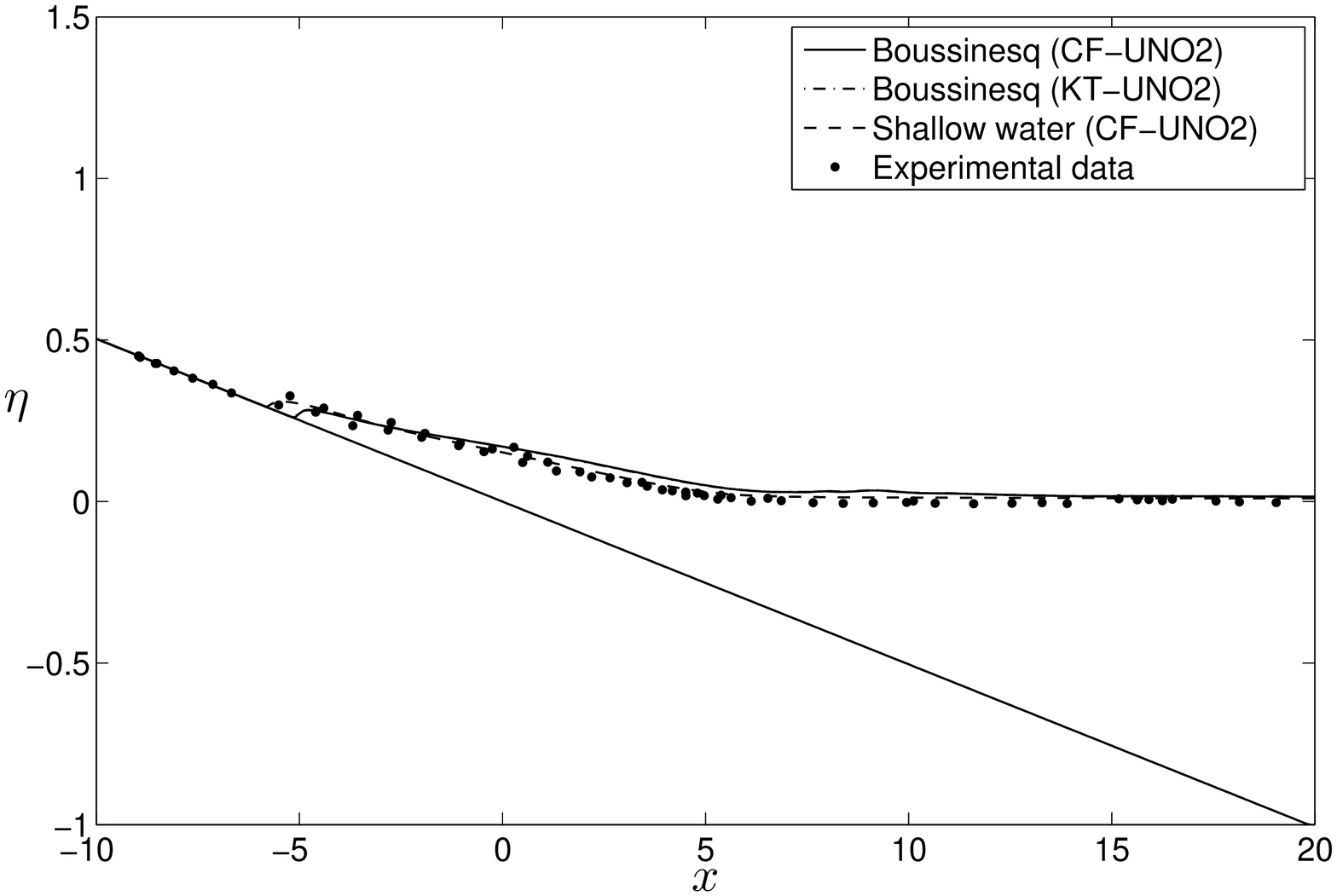}}
\subfigure[$t=45$]{\includegraphics[scale=0.333]{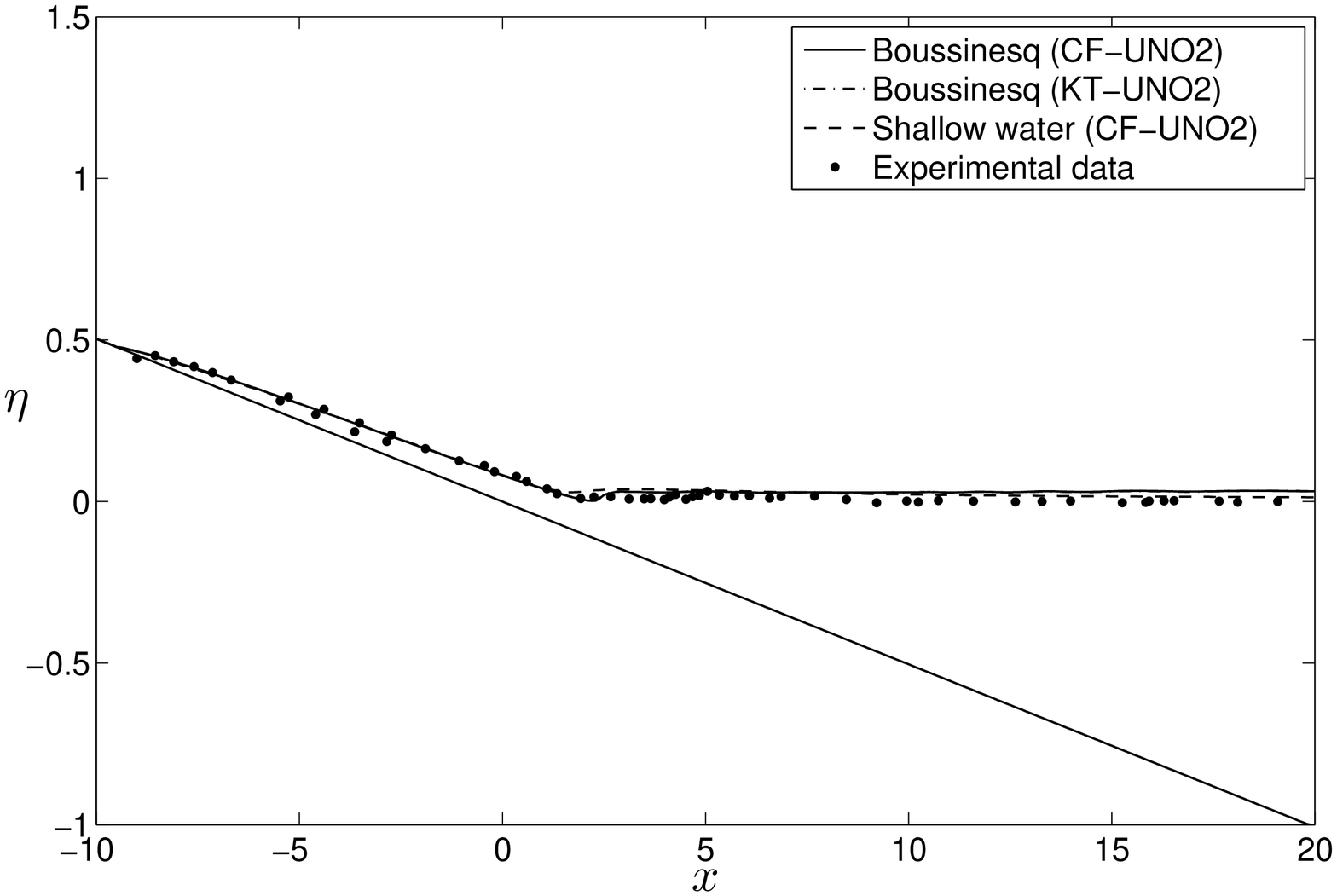}}
\subfigure[$t=55$]{\includegraphics[scale=0.333]{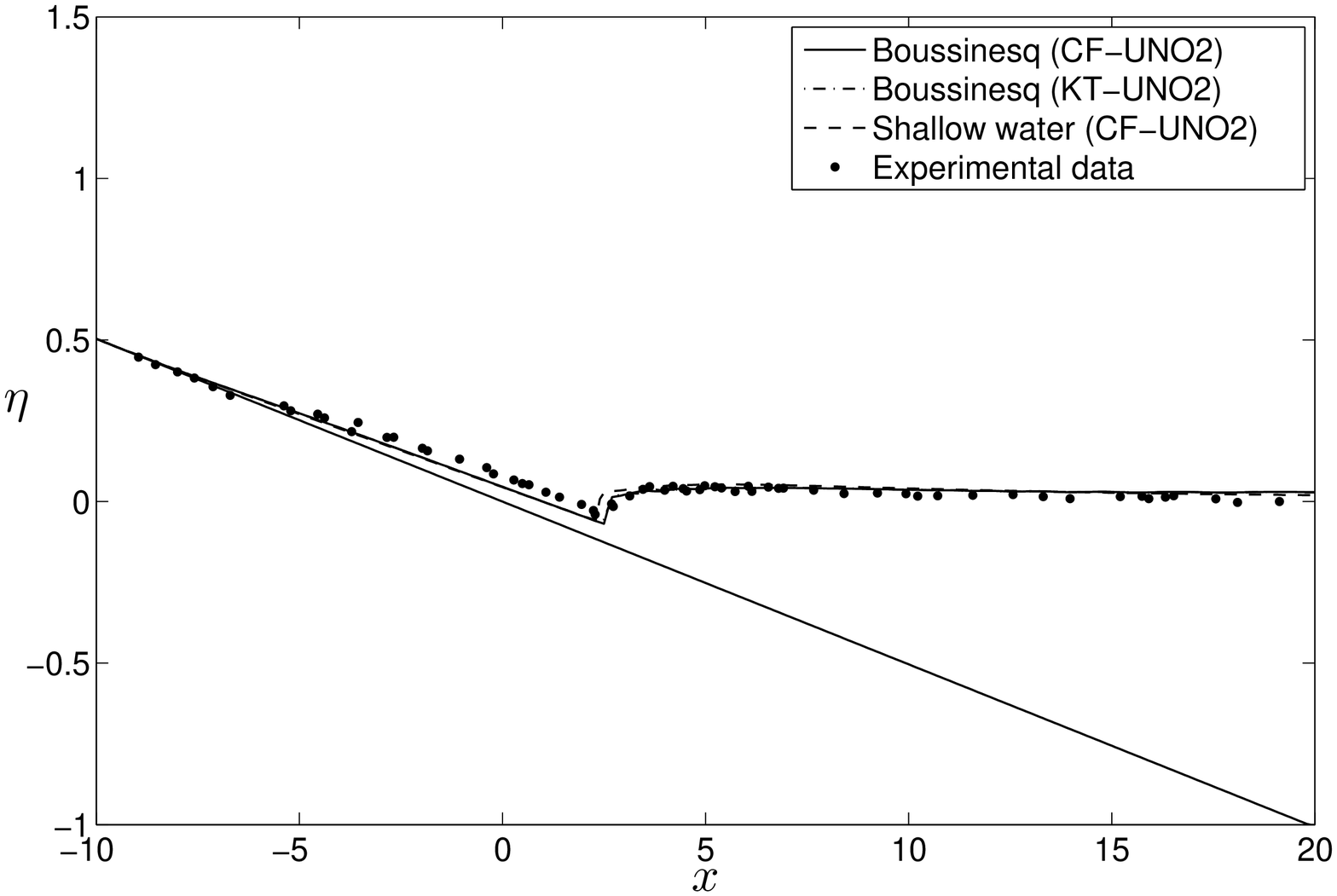}}
\subfigure[$t=60$]{\includegraphics[scale=0.333]{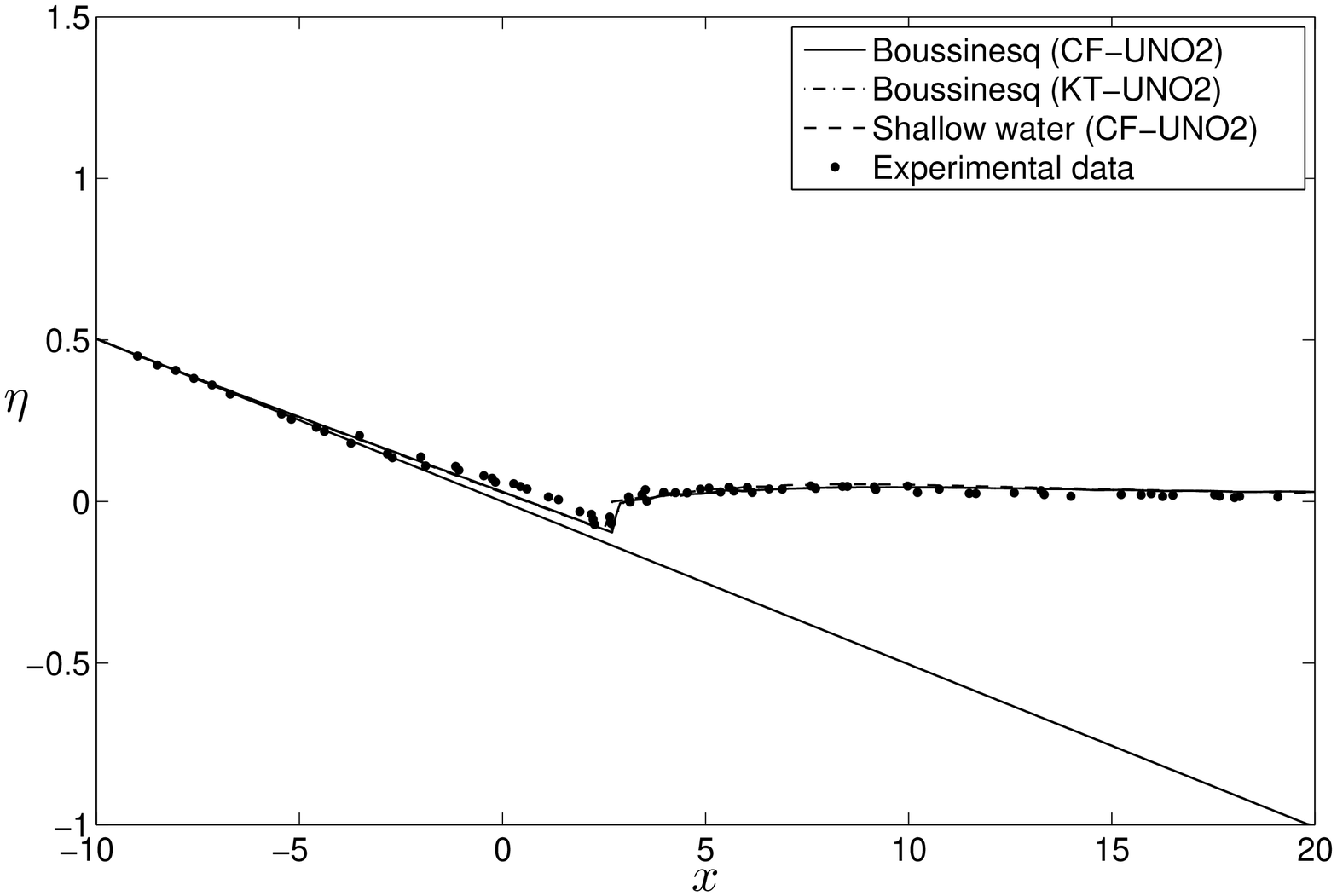}}
\subfigure[$t=70$]{\includegraphics[scale=0.333]{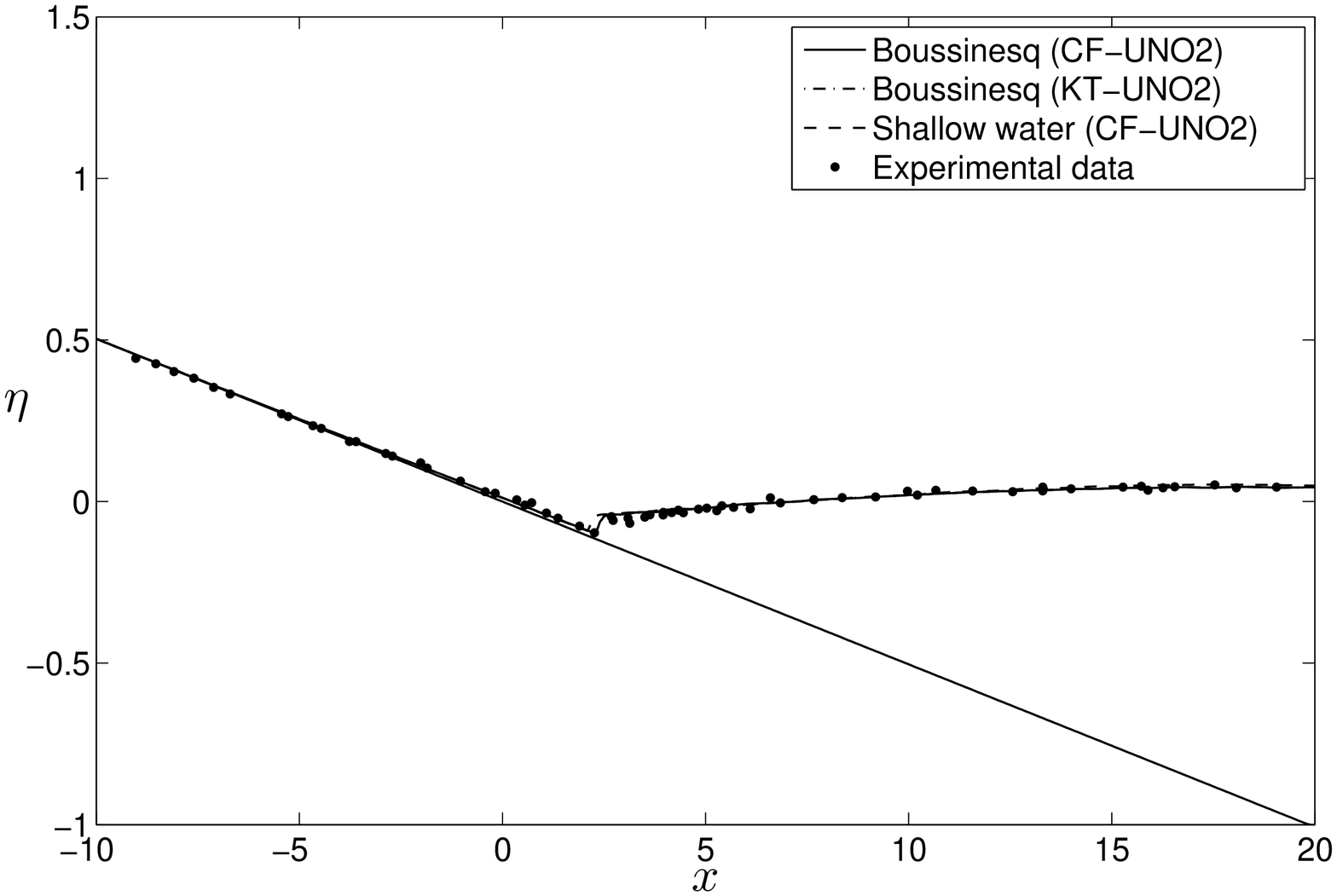}}
\subfigure[$t=80$]{\includegraphics[scale=0.333]{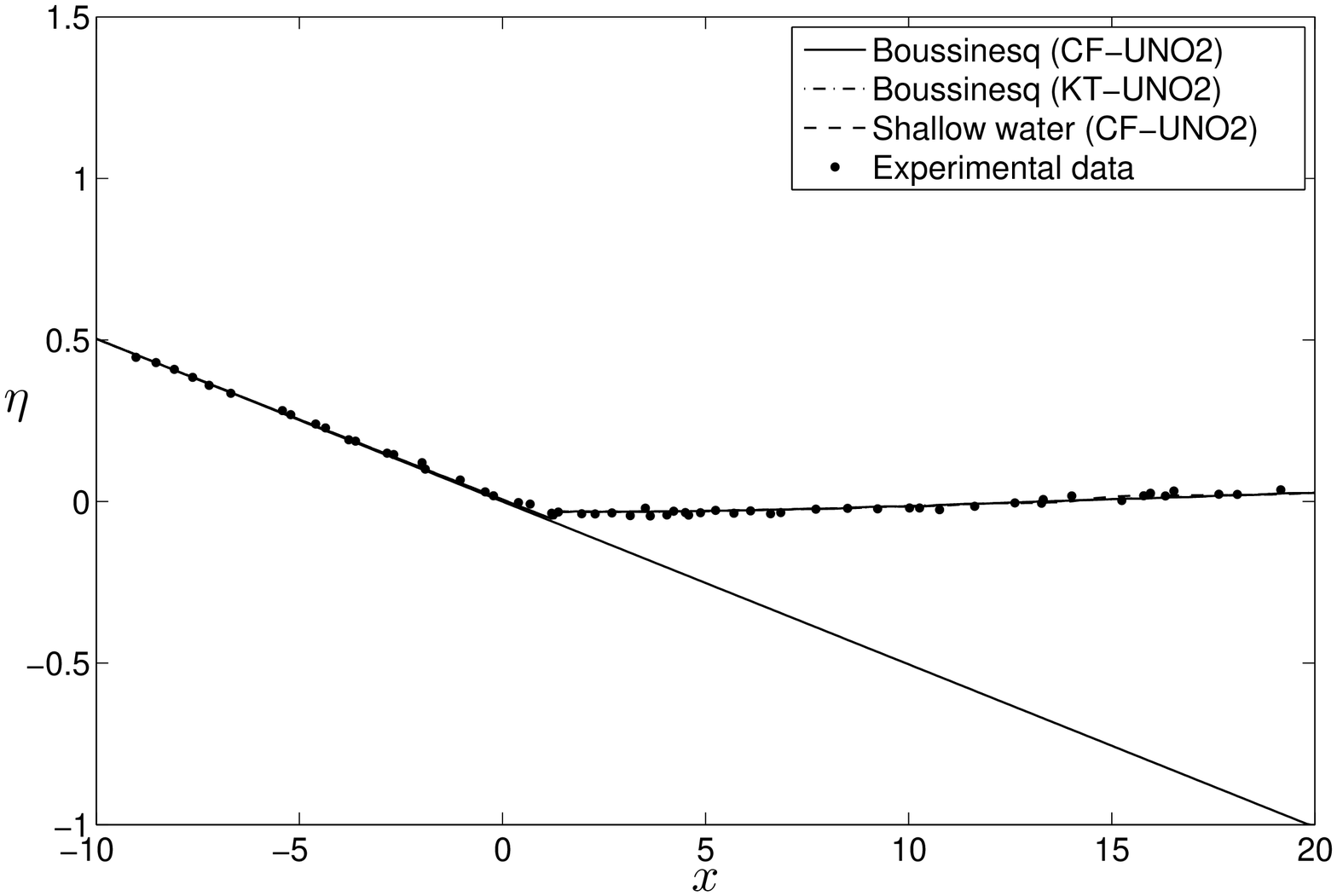}}
\caption{(Cont'd) Solitary wave runup on a sloping shore: $A_s=0.28$ case.}%
\label{F6iii.1a}%
\end{figure}
%%%%%%%%%%%%%%%%%%%%%%%%%

\subsection{Solitary wave runup on a steep slope $(\beta=20^{\circ})$}

Now we present two experiments pointing out some further differences in solutions to dispersive and nondispersive models. These experiments were performed by J.A.~Zelt, \cite{Zelt1991}. We consider  two waves in $I=[-10,30]$ with amplitudes $A_s=0.12$ and $A_s=0.2$ initially located at $X_0=8.85$ and $X_0=10.62$ respectively. These waves propagate onto a steep sloping shore $1:2.74$. A very fine grid of $\dx=0.01, \ \dt=\dx/100$ is used to guarantee the accuracy and stability of simulations.

As it was observed in \cite{Zelt1991} both waves do not break during the runup but the second one generates a strong breaking event during the rundown. Friction does not play an important role in these experiments. Consequently, no friction term is included into the models. 

\begin{figure}%
\centering
\subfigure[$A_s=0.12$]{\includegraphics[scale=.333]{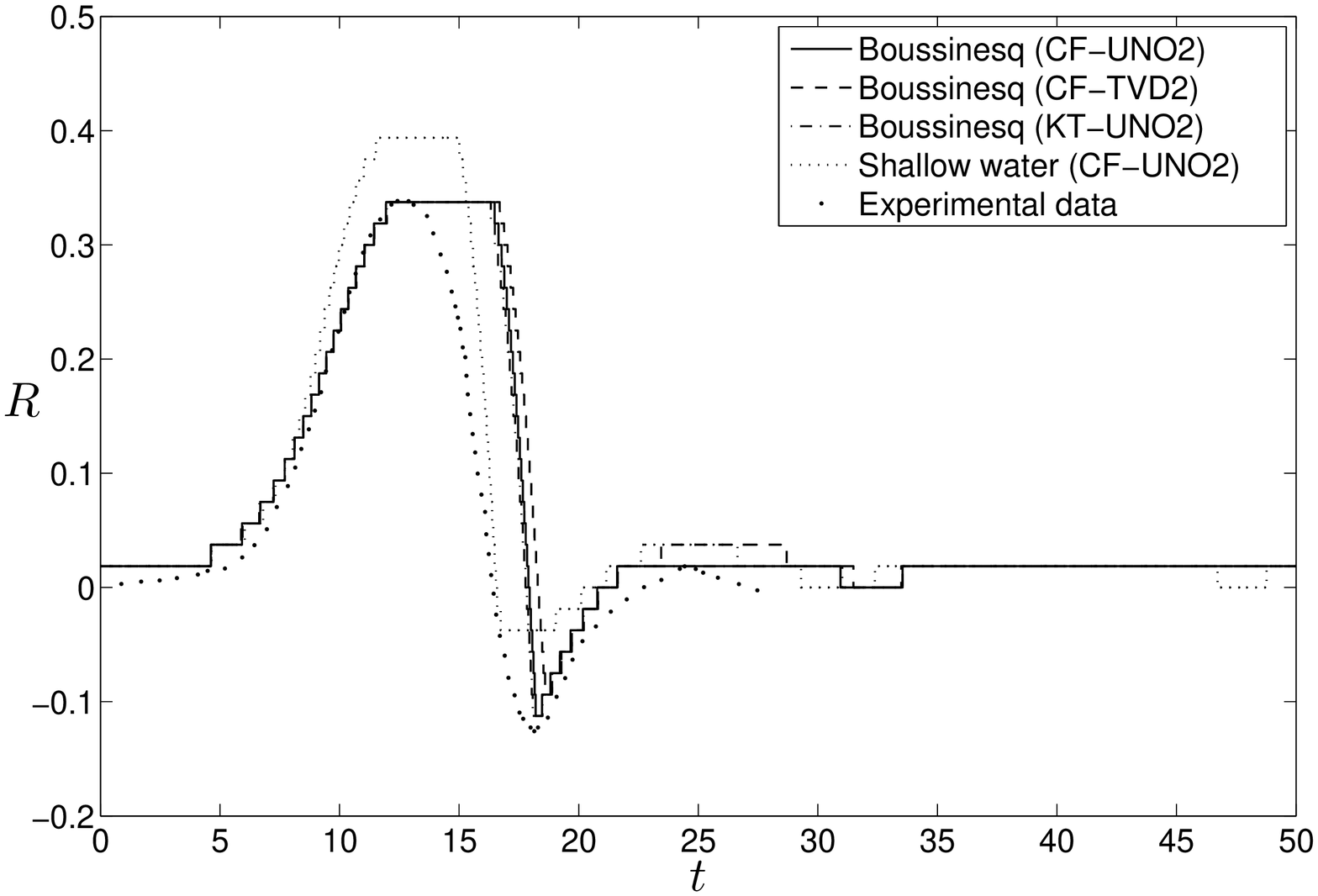}}
\subfigure[$A_s=0.2$]{\includegraphics[scale=.333]{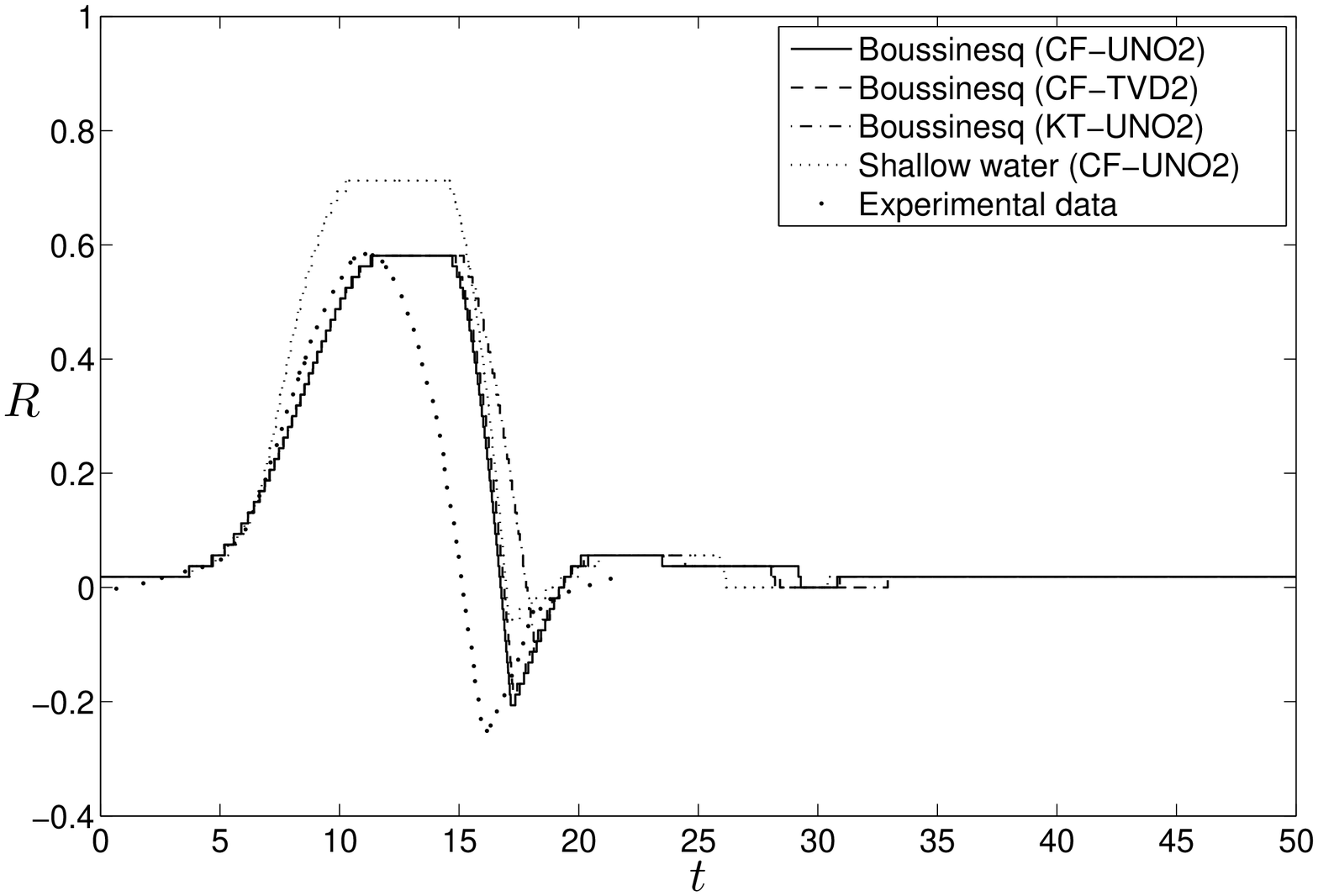}}
\caption{Runup value $R$ as a function of time.}%
\label{F6iiii.1}%
\end{figure}

Figure \ref{F6iiii.1} shows the runup value $R$ as a function of time. We observe  for both models that there is a phase lag compared to the experimental data. We believe that this discrepancy can be removed by changing the definition of the last dry cell. We also observe that shallow water equations over-predict the maximum runup and the minimum rundown while the Boussinesq model predicts correctly the extrema  in both cases.

\begin{figure}%
\centering
\subfigure[$t=15.71$]{\includegraphics[scale=.333]{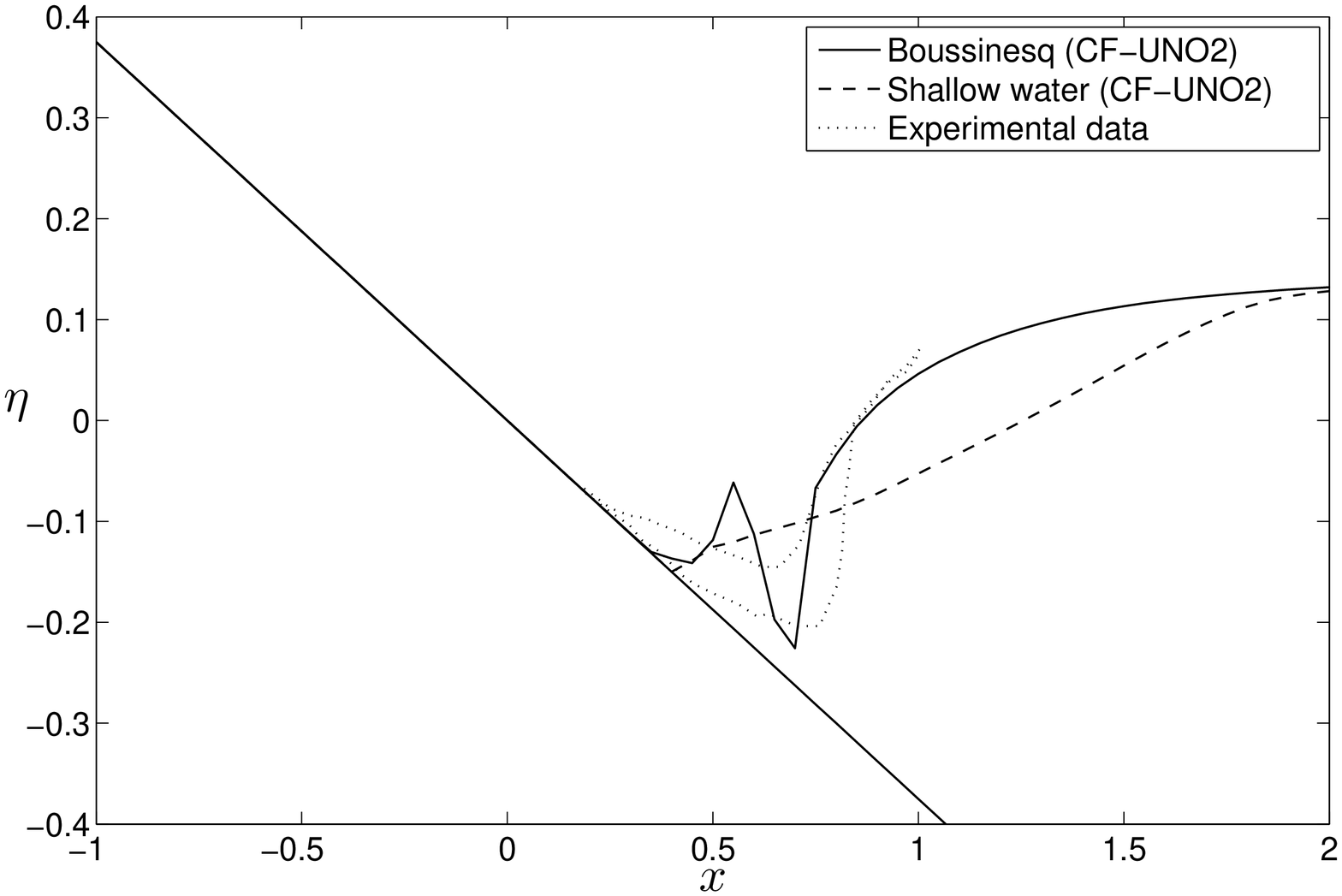}}
\subfigure[$t=16.62$]{\includegraphics[scale=.333]{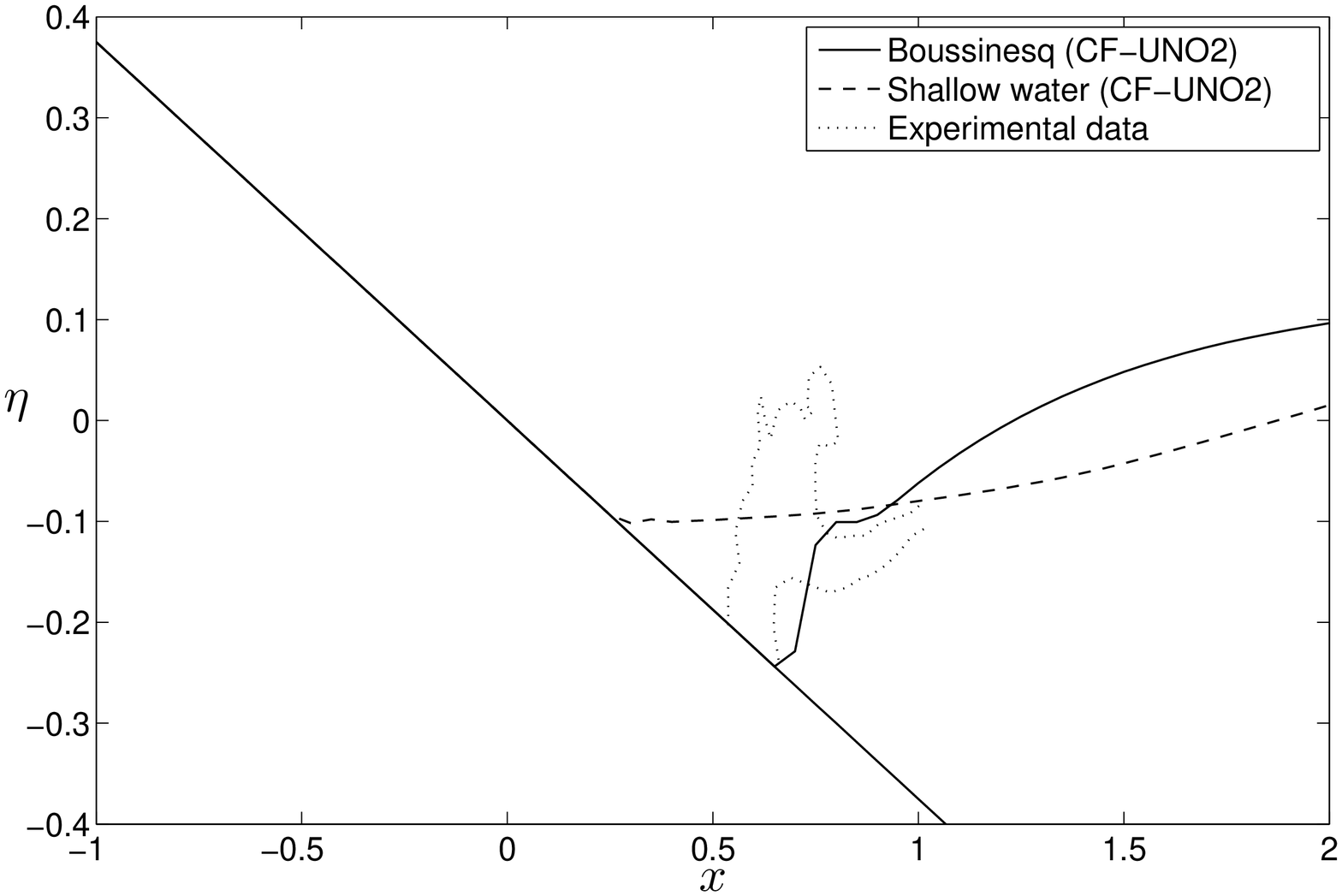}}
\subfigure[$t=17.32$]{\includegraphics[scale=.333]{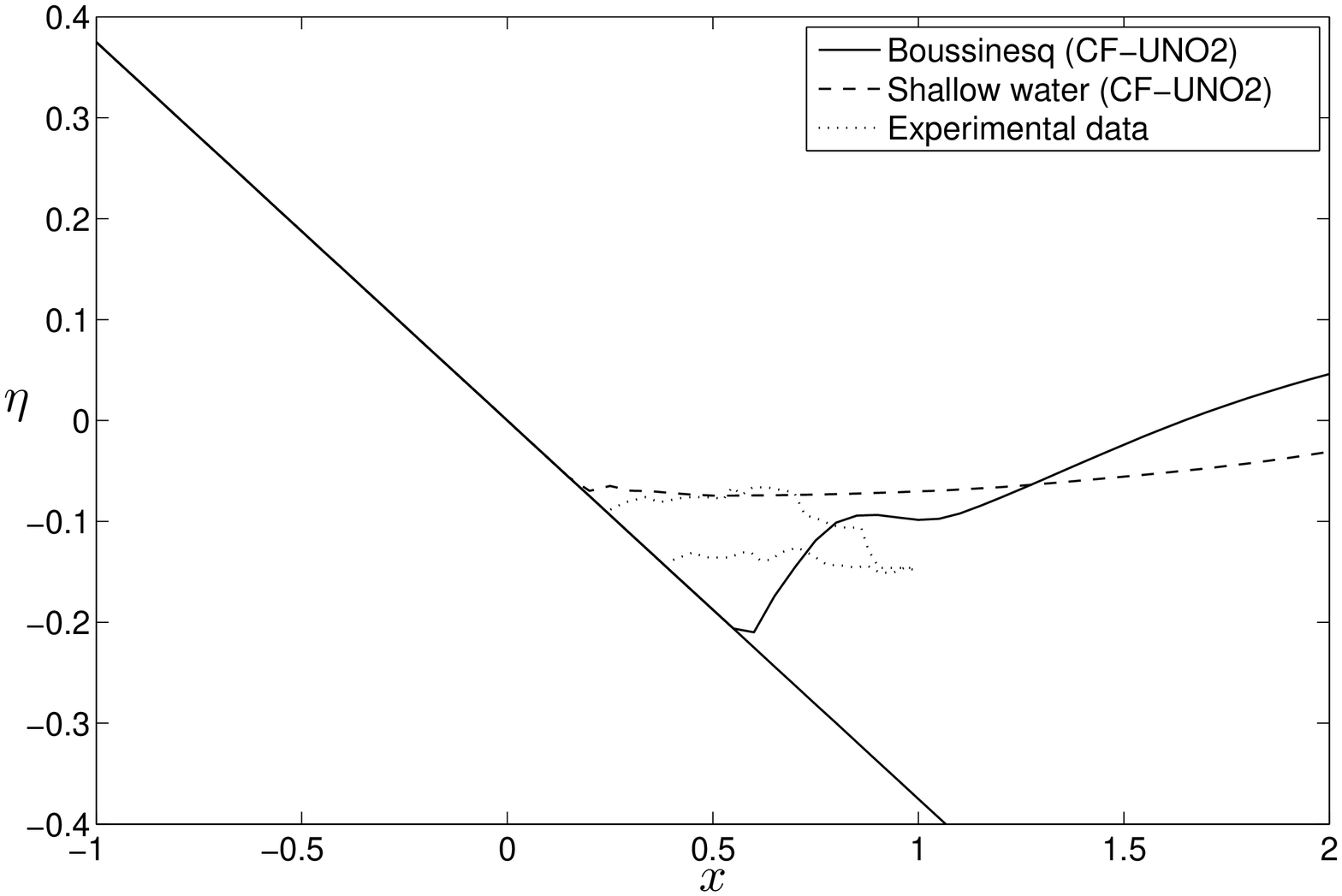}}
\subfigure[$t=22.29$]{\includegraphics[scale=.333]{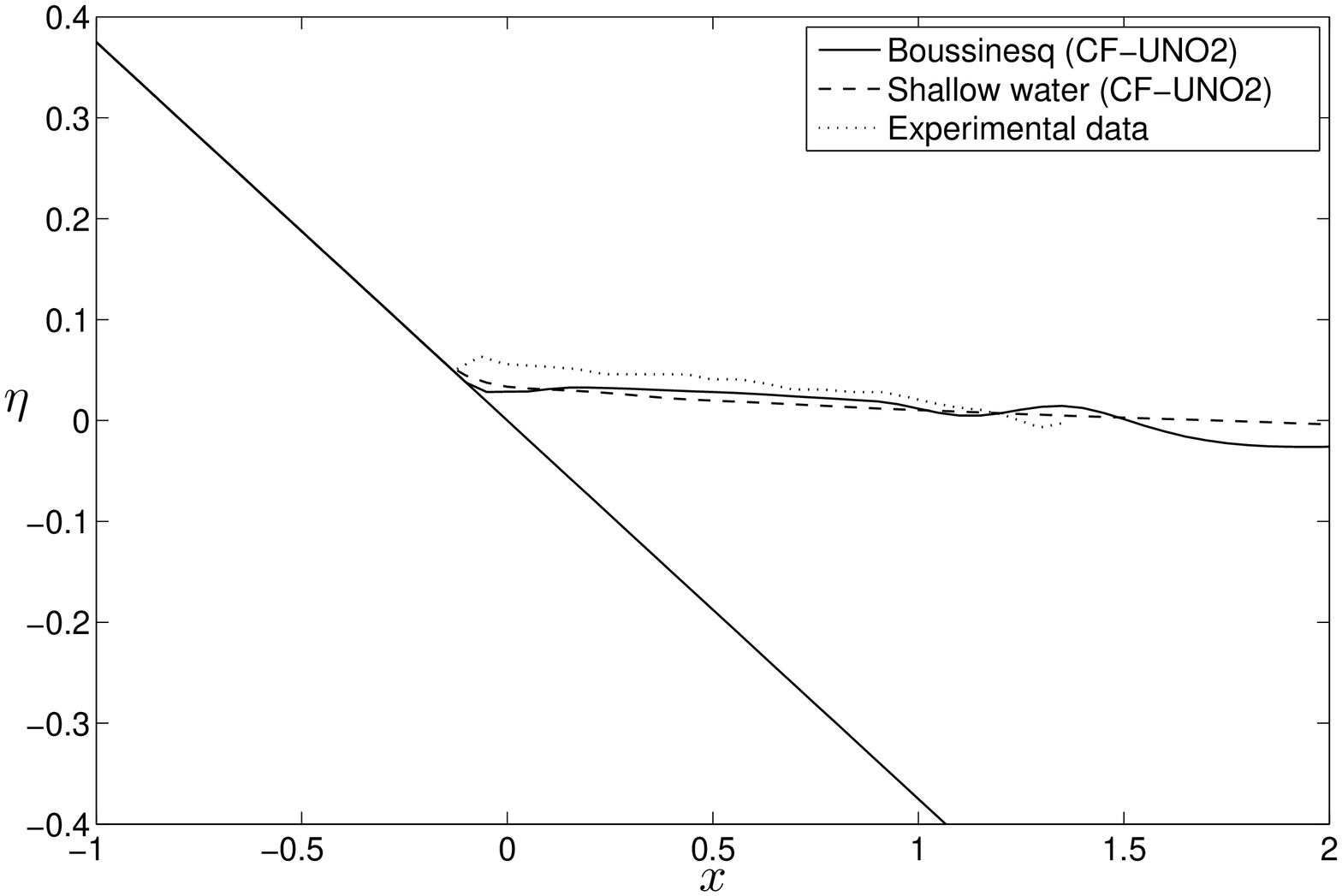}}
\caption{Rundown of the wave with amplitude $A_s = 0.2$.}% 
\label{F6iiia}%
\end{figure}

\begin{figure}%
\centering
\subfigure[$A_s = 0.12$]{\includegraphics[scale=.333]{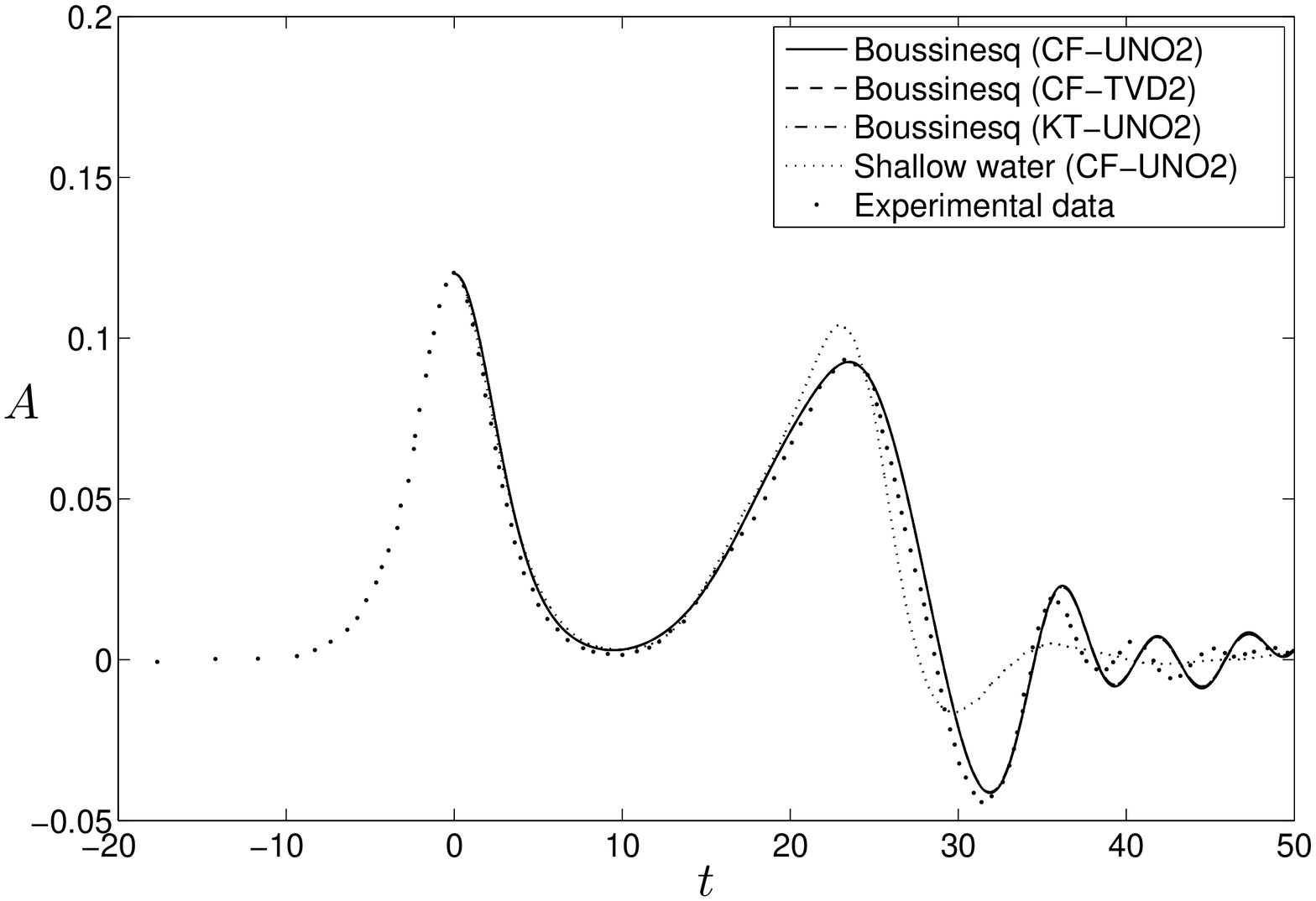}}
\subfigure[$A_s = 0.2$]{\includegraphics[scale=.333]{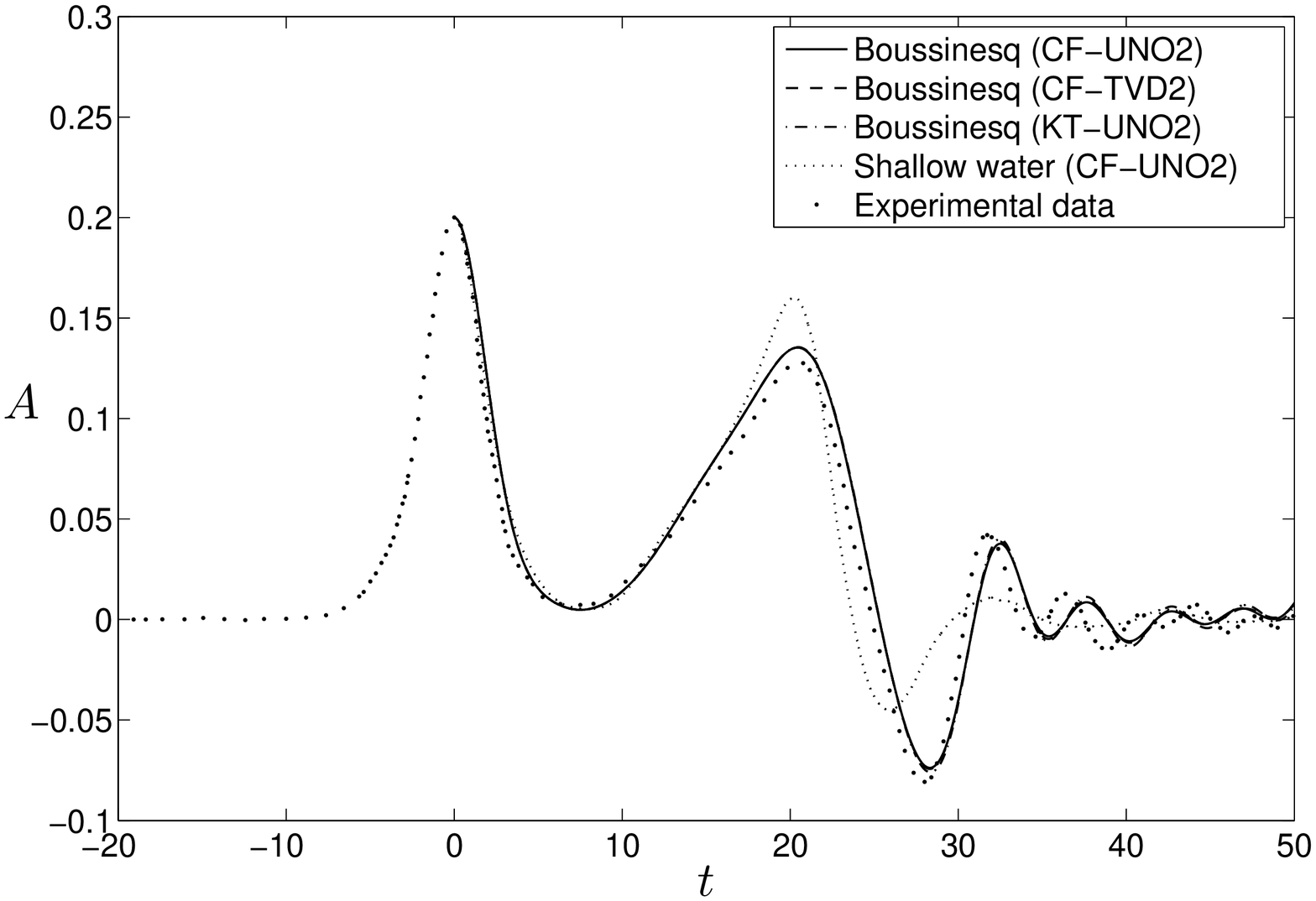}}
\caption{The amplitude at the wave gauge $A$}% 
\label{F6iiii.2}%
\end{figure}

Figure \ref{F6iiia} shows the rundown of the solitary wave of amplitude $A_s = 0.2$ during breaking. One may observe that the experimental data consist of two curves due to the difficulty in measuring the surface elevation of the breaking wave due to 3D effects which become important. On Figure \ref{F6iiii.2} the free surface elevation at a gauge is presented. The  gauge is located $8.85$ meters away from the still shoreline position. The reflected wave appears in both cases to be highly dispersive thus, the Boussinesq model provides a much better approximation. In the case $A_s = 0.2$, the mass remained equal to $I_0^h = 29.770808175$,  while in the case case $A_s = 0.12$,  $I_0^h = 29.4861671693$ conserving the digits shown.

%%%%%%%%%%%%%%%%%%%%%%%%%%%%%%%%%%%%
%%%%%%%%%%%%%%%%%%%%%%%%%%%%%%%%%
%%%%%%%%%%%%%%%%%%%%%%%%%%%%%%%%%
%%%%%%%%%%%%%%%%%%%%%%%%%%%%%%%%%

\subsection{Solitary wave runup on a gradual slope $(\beta = 2.88^{\circ})$ with a pond}

\begin{figure}%
\centering
\subfigure[Wave gauge at $x=-3.4$]{\includegraphics[scale=0.333]{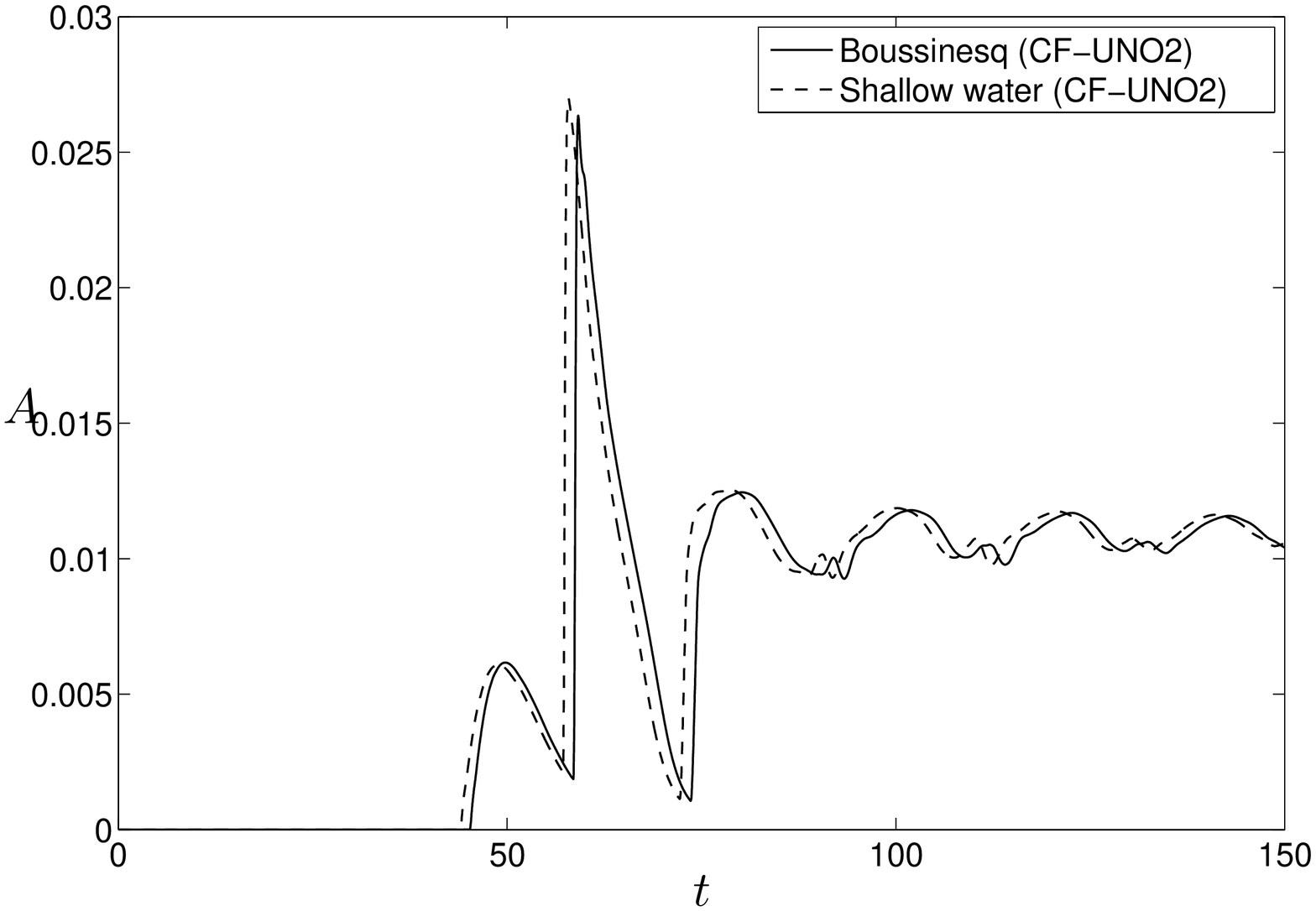}}
\subfigure[Wave gauge at $x=8$]{\includegraphics[scale=0.333]{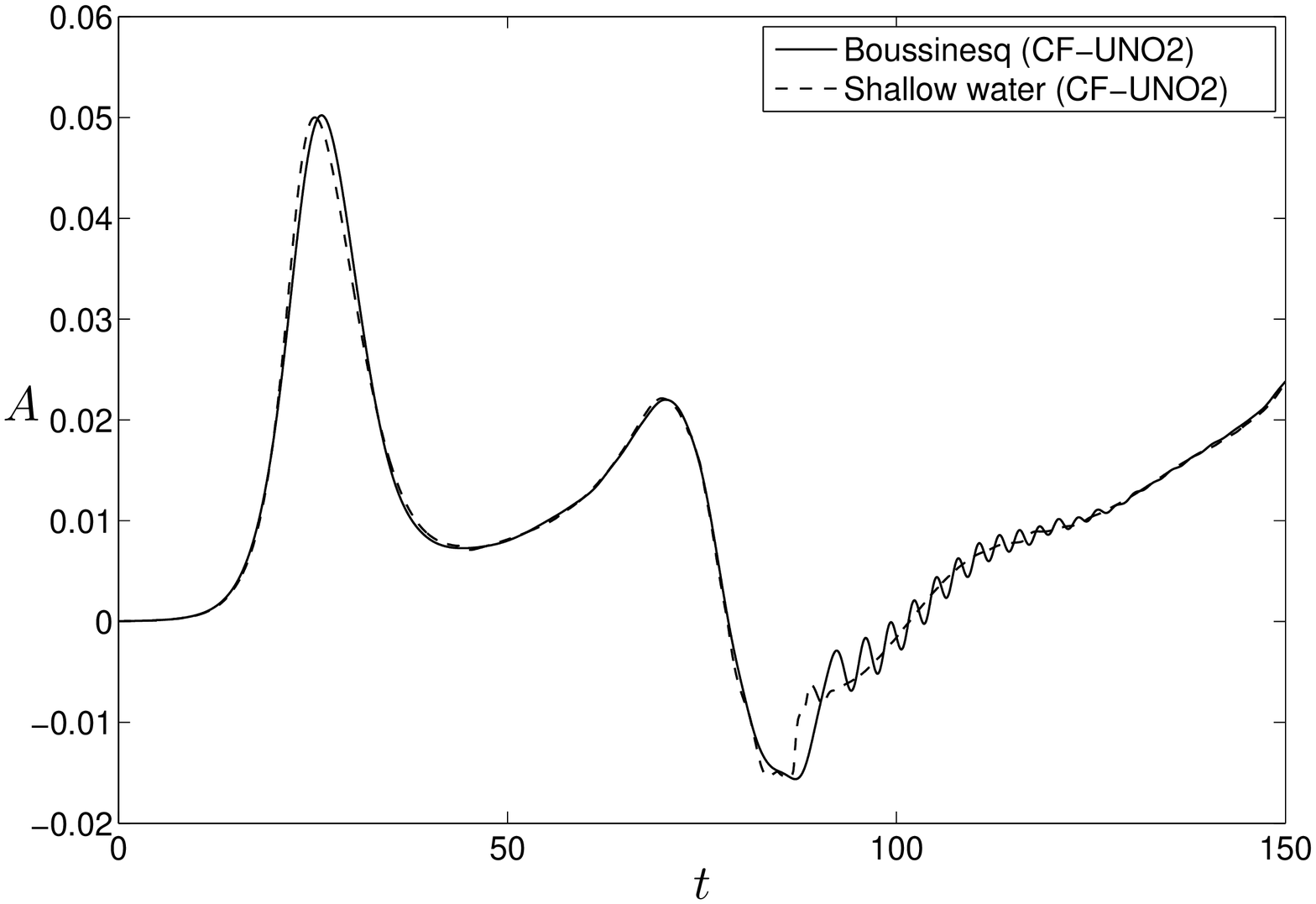}}
\caption{Evolution of the free surface elevation at two wave gauges.}% 
\label{Fb2}%
\end{figure}

We repeat the experiment of Section \ref{nolake} in $I=[-10,50]$ with solitary wave amplitude equal to $A_s = 0.05$. However, we modify the bottom by adding a small pond over the shoreline described by the exponential function $0.1\, e^{-(x+4)^2}$. The Boussinesq system preserves the correct dispersion characteristics for the waves reaching the pond. In Figure \ref{Fb1a} we present the overall process. It is worth noting that after the pond was filled a breaking wave was reflected back. As the wave slides down, a small hydraulic jump appears. In the case of shallow water system  this jump propagates as a shock wave due to hyperbolic character of equations. On the other hand, the Boussinesq system develops  into an Airy type wave according to its dispersive characteristics. In Figure \ref{Fb2} we show the solution at two wave gauges located at $x=-3.4$ and $x=8$ for both the dispersive and nondispersive models. The mass during the simulations is constantly equal to $I_0^h=40.5198087147$.

%%%%%%%%%%%%%%%%%%%%%%%%%%%%%%%%%%%%%%
\begin{figure}%
\centering
\subfigure[$t=37.58$]{\includegraphics[scale=0.333]{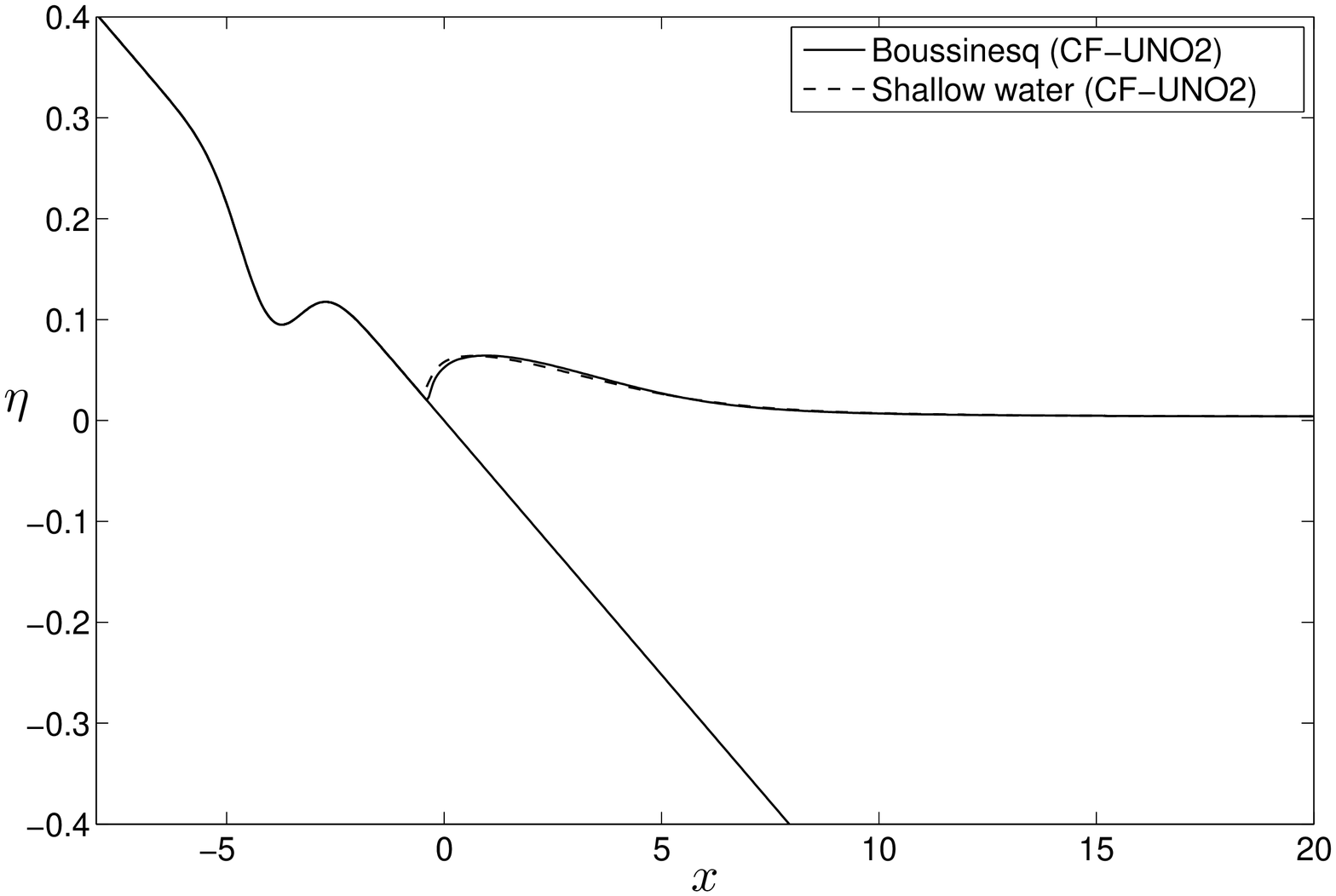}}
\subfigure[$t=48.31$]{\includegraphics[scale=0.333]{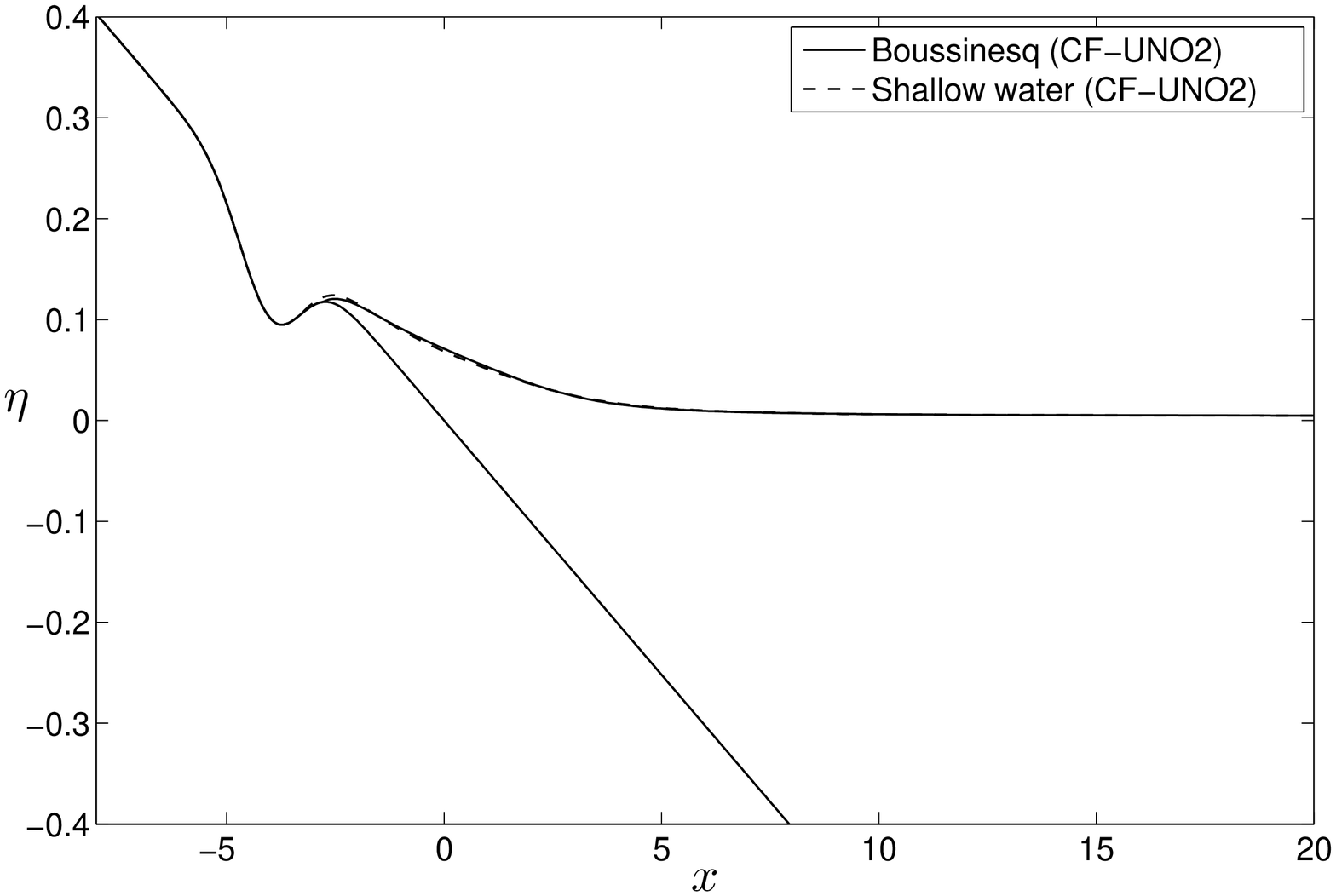}}
\subfigure[$t=55.22$]{\includegraphics[scale=0.333]{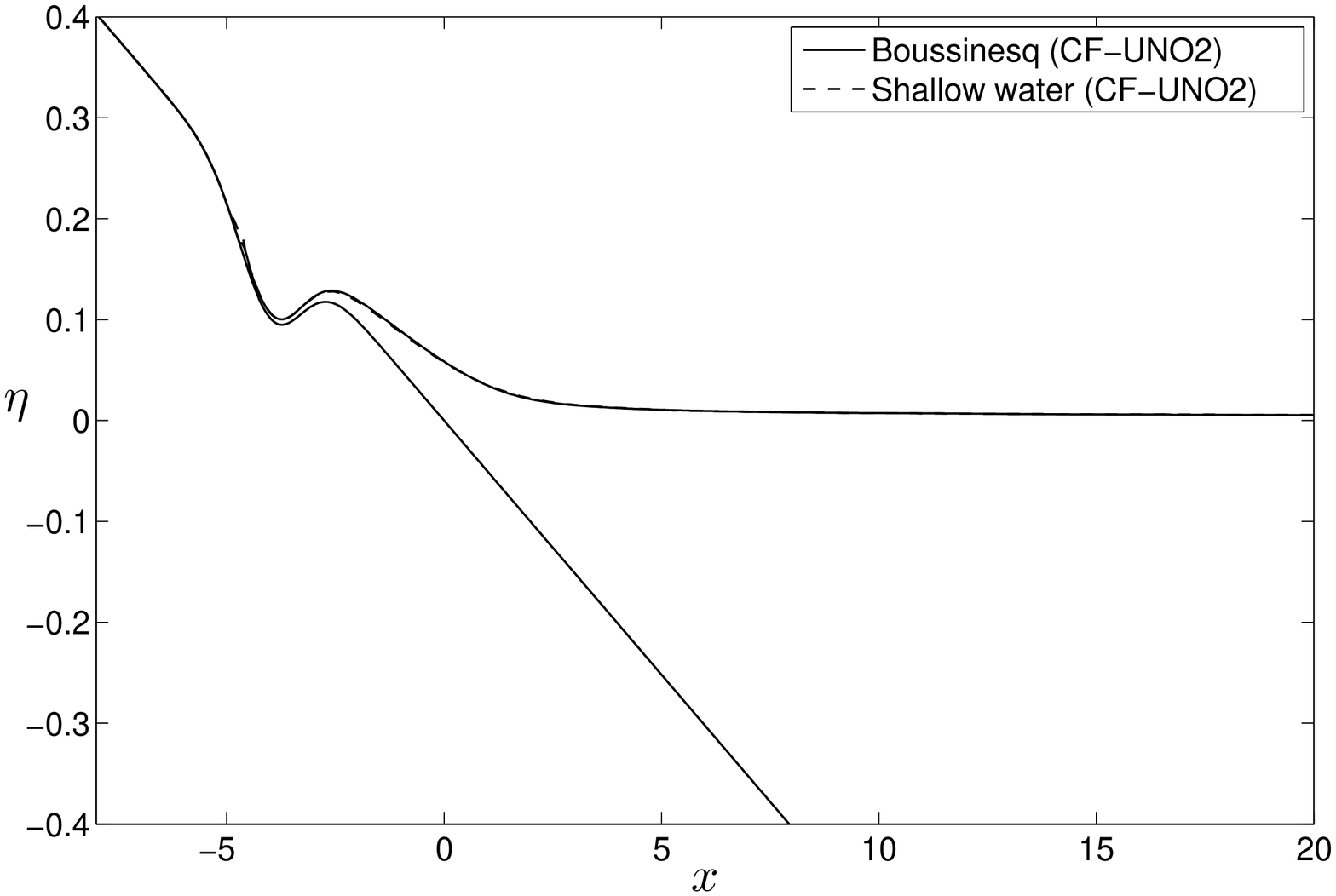}}
\subfigure[$t=58.67$]{\includegraphics[scale=0.333]{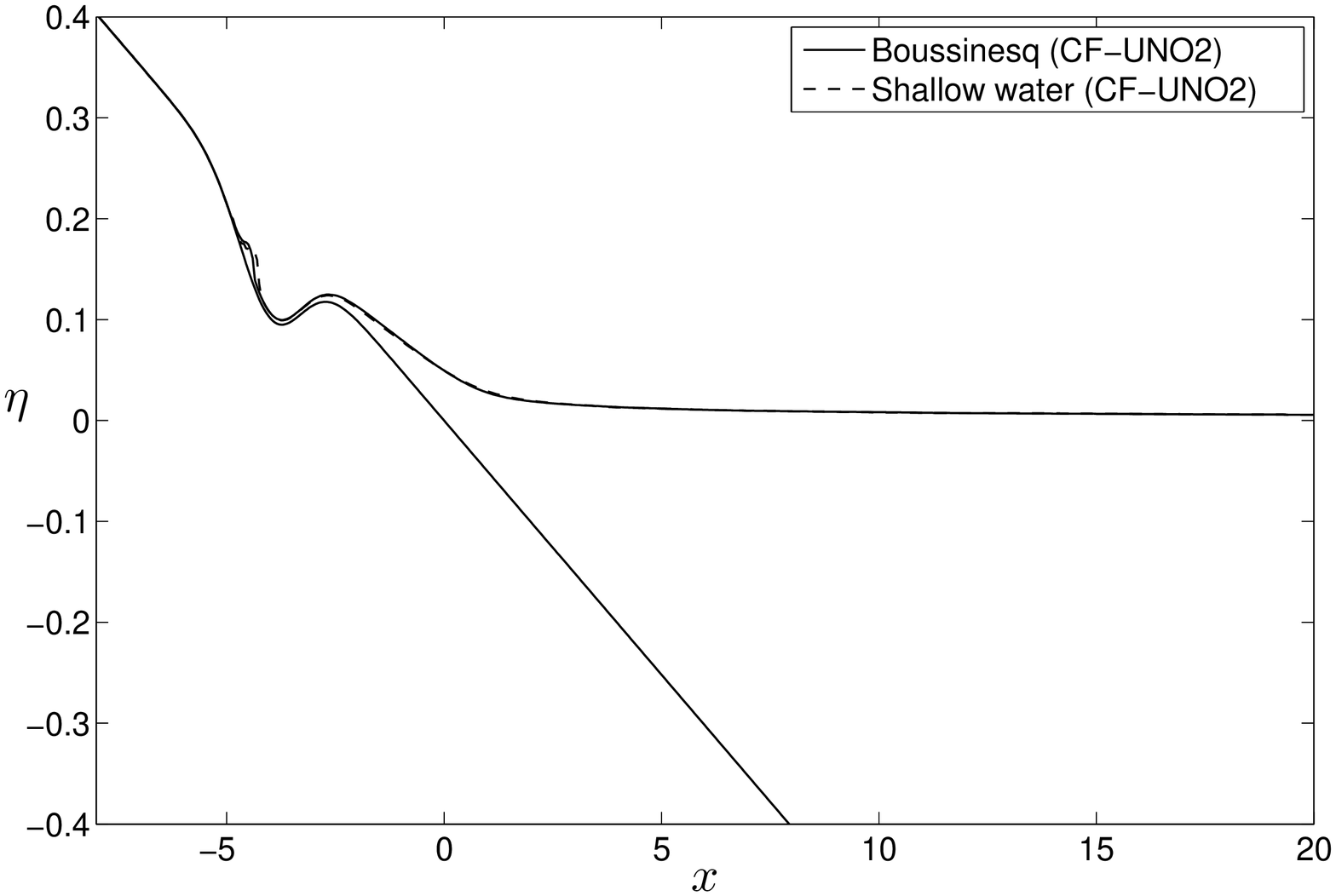}}
\subfigure[$t=65.57$]{\includegraphics[scale=0.333]{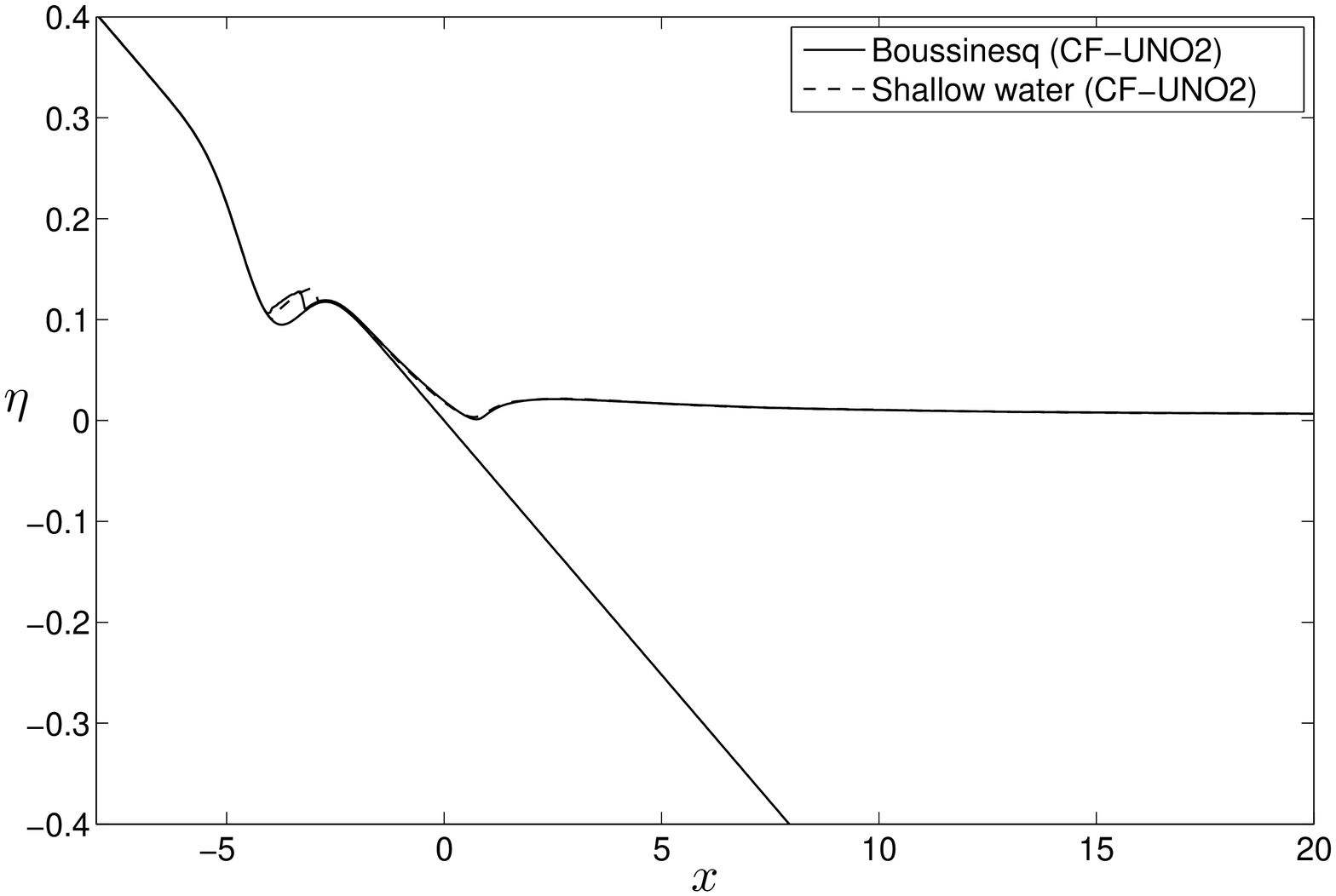}} 
\subfigure[$t=72.47$]{\includegraphics[scale=0.333]{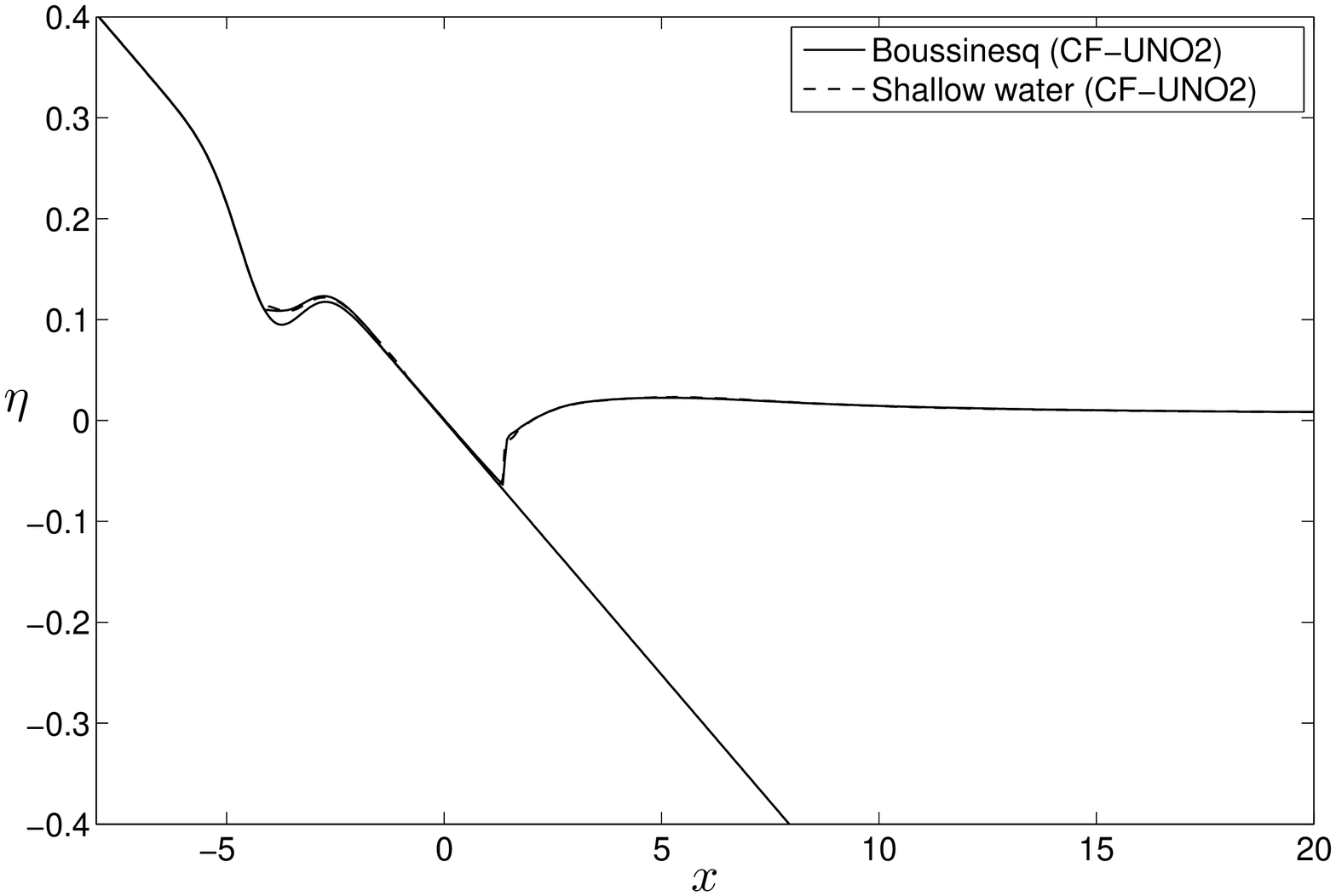}}
\caption{Long wave runup on a shore with a pond.}% 
\label{Fb1a}%
\end{figure}
%%%%%%%%%%%%%%%%%%%%%%%%%%%%%%%%%%%%%%
\begin{figure}%
\ContinuedFloat
\centering
\subfigure[$t=75.93$]{\includegraphics[scale=0.333]{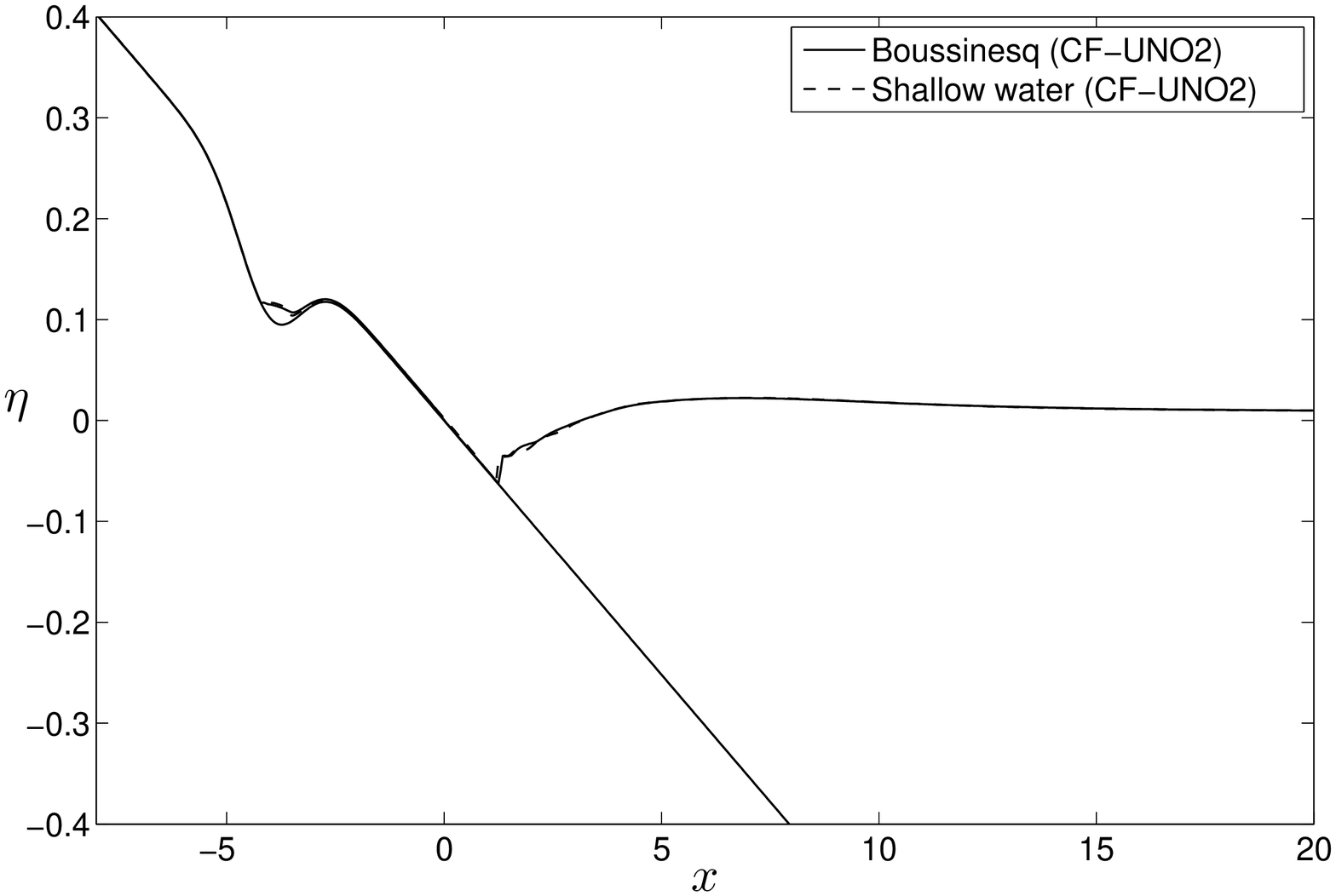}}
\subfigure[$t=79.38$]{\includegraphics[scale=0.333]{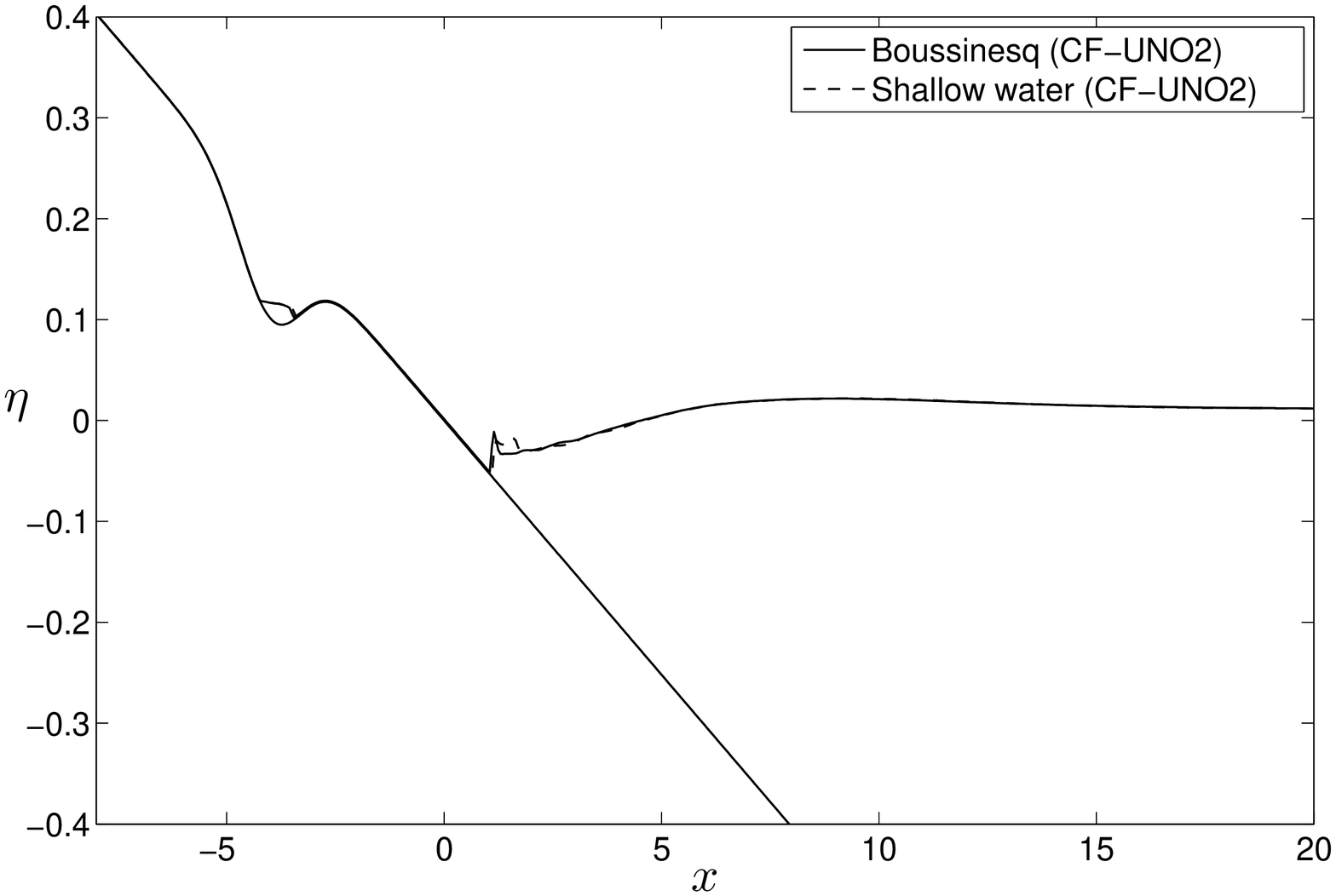}}
\subfigure[$t=86.28$]{\includegraphics[scale=0.333]{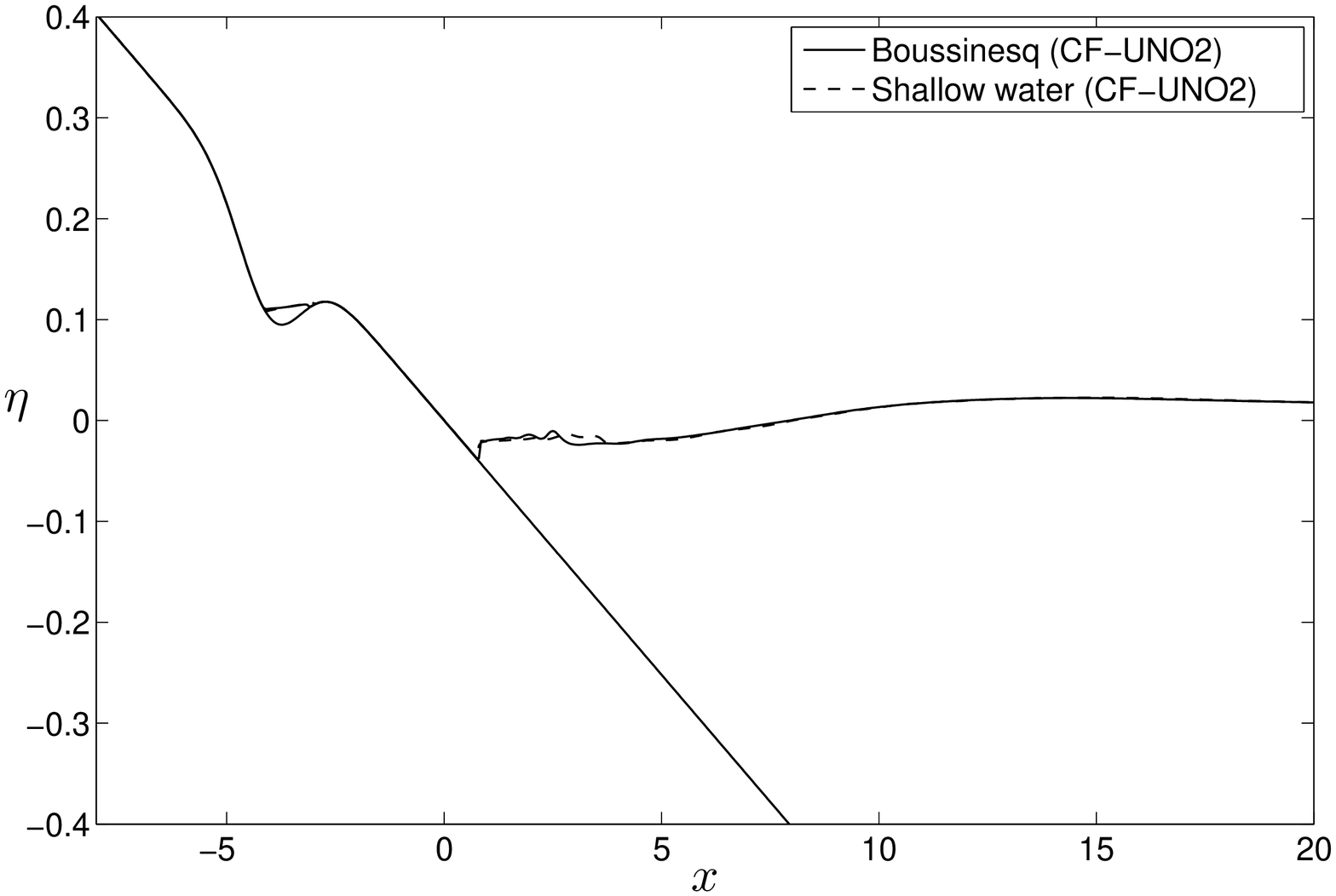}}
\subfigure[$t=103.54$]{\includegraphics[scale=0.333]{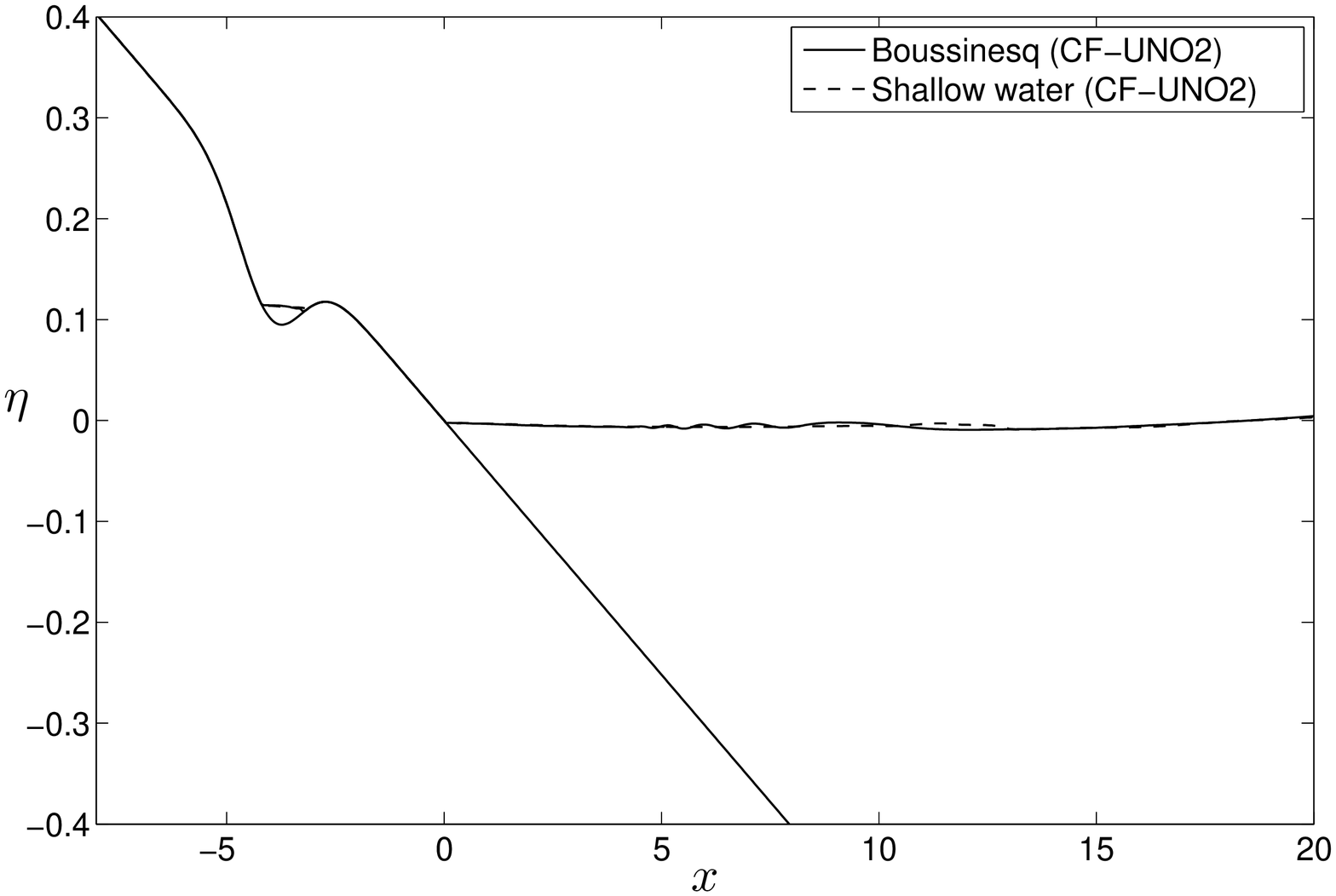}}
\caption{(Cont'd): Long wave runup on a shore with a pond.}%
\label{Fb1b}%
\end{figure}
%%%%%%%%%%%%%%%%%%%%%%%%%%%%%%%%%%%%%%
%%%%%%%%%%%%%%%%%%%%%%%%%%%%%%%%%%%%%%%%%%%%%%%%%%%%%%%%%

\section{Conclusions}

In the present study we extend the finite volume framework, developed for hyperbolic conservations laws, to approximate solutions of dispersive wave equations.  This type of equations arises naturally in many physical problems. In the water wave theory dispersive equations have been well known since the pioneering work of J. Boussinesq \cite{bouss} and Korteweg-de Vries \cite{KdV}. Currently, the so-called Boussinesq-type models become more and more popular as an operational model for coastal hydrodynamics and other fields of engineering. 

We extend the finite volume framework to dispersive models. We tested several choices of numerical fluxes, various reconstruction methods ranging from classical MUSCL type  to modern approaches such as WENO. Various choices of limiters have been also tested out. Advantages of specific methods are discussed and some suggestions are outlined.

For operational modeling of the wave runup we derived a new system which has some advantages over its classical counterpart. The new system together with proposed novel discretization procedure are validated by extensive comparisons with experimental data of C.E.~Synolakis \cite{Synolakis1987} and J.A.~Zelt \cite{Zelt1991}.

We paid a special attention to the comparison of dispersive (Boussinesq) and nondispersive (shallow water) models. Nowadays shallow water equations have become the model of choice for operational tsunami modeling including the inundation zone estimation \cite{Titov1997, Synolakis2008}. The question of dispersive effects importance arises recurrently in the tsunami wave modeling community \cite{Kulikov, Tkalich2007}. Our results show that shallow water equations are sufficient to predict maximum runup values. However, the dispersive effects can be beneficial for more accurate description of long wave propagation, runup and rundown.

%%%%%%%%%%%%%%%%%%%%%%%%%%%%%%%%%%%%%%%%%%%%%%%%%%%%%%%%%

\section*{Acknowledgment}

D.~Dutykh acknowledges the support from French Agence Nationale de la Recherche, project MathOcean (Grant ANR-08-BLAN-0301-01) and Ulysses Program of the French Ministry of Foreign Affairs under the project 23725ZA. The work of Th.~Katsaounis was partially supported by European Union FP7 program Capacities(Regpot 2009-1), through ACMAC (http://acmac.tem.uoc.gr). The work of D.~Mitsotakis was supported by Marie Curie Fellowship No. PIEF-GA-2008-219399 of the European Commission. We would like to thank also professors Diane Henderson and Costas Synolakis for providing us their experimental data and professors Jerry Bona and Vassilios Dougalis for very helpful discussions.

%%%% Bibliography  %%%%%%%%%%
\bibliographystyle{alpha}
\bibliography{biblio}

\end{document}